\documentclass[11pt]{article}

\usepackage[]{acl}

\usepackage{times}
\usepackage{latexsym}

\usepackage[T1]{fontenc}

\usepackage[utf8]{inputenc}

\usepackage{microtype}

\usepackage{inconsolata}

\usepackage{times}
\usepackage{enumitem}
\usepackage{latexsym}
\usepackage{amsmath}
\usepackage{booktabs}
\usepackage{xcolor}
\usepackage{colortbl}
\usepackage{amssymb}
\usepackage[T1]{fontenc}
\usepackage[utf8]{inputenc}
\usepackage{microtype}
\usepackage{inconsolata}
\usepackage{times}
\usepackage{latexsym}
\usepackage{microtype}
\usepackage{inconsolata}
\usepackage{amsmath,amssymb}
\usepackage{booktabs}
\usepackage{multirow}
\usepackage{array}
\usepackage{xcolor}
\usepackage{graphicx}
\usepackage{tikz}
\usepackage{pgfplots}
\pgfplotsset{compat=1.18}
\usetikzlibrary{arrows.meta,positioning,fit,calc}
\usepackage{enumitem}
\usepackage{siunitx}
\usepackage{url}

\usepackage{graphicx}
\usepackage{xcolor}
\usepackage{pifont}
\usepackage[most]{tcolorbox}
\usepackage{tikz}
\usetikzlibrary{shapes.geometric}
\usepackage{pgfplots}
\pgfplotsset{compat=1.18}
\usepgfplotslibrary{groupplots}

\newtcolorbox{examplebox}[1]{
  enhanced,breakable,colback=gray!3,colframe=gray!50,
  boxrule=0.5pt,arc=2pt,left=5pt,right=5pt,top=4pt,bottom=4pt,
  title=\textbf{#1},fonttitle=\small,coltitle=black
}

\newcommand{\cmark}{\textcolor{green!45!black}{\ding{51}}}
\newcommand{\xmark}{\textcolor{red!55!black}{\ding{55}}}

\newcommand{\mptb}{\texttt{MultiPartyTurnBench}}

\newcommand{\eot}{\textsc{EOT}}
\newcommand{\intr}{\textsc{INT}}
\newcommand{\fdstart}{\textsc{Start}}
\newcommand{\fdstop}{\textsc{Stop}}

\definecolor{budgetgray}{RGB}{239,242,246}
\definecolor{bestcyan}{RGB}{214,245,250}

\title{\texttt{MultiPartyTurnBench}: Benchmarking Floor Control in Multi-Party Spoken Dialogue}

\author{Puneet Mathur, Dinesh Manocha\\
\small{University of Maryland College Park, USA}\\
Project Page: \url{https://multiparty-turnbench.github.io/}}

\begin{document}
\maketitle

\begin{abstract}
Full-duplex speech systems must decide when to enter, hold, or relinquish the conversational floor, even as multiple speakers overlap or backchannel without taking it. Existing evaluations study dyadic turn-taking or end-to-end agent behavior, but lack a unified, controlled evaluation of genuine floor transitions against confusable non-events in multi-party settings. We introduce \textbf{\texttt{MultiPartyTurnBench}} (MPTB), a benchmark of 240 synthetically generated, channel-separated conversations spanning two to five speakers and six interaction types. MPTB evaluates two complementary tasks: End-of-Turn (EOT) detection distinguishes genuine turn completion from temporary holds, while Interruption (INT) detection distinguishes successful floor takeovers from non-floor-taking speech. Exact event-conditioned annotations enable causal evaluation using false-positive-constrained recall and signed latency. Across acoustic, semantic, and turn-projection baselines, we find a stark asymmetry: learned systems achieve 82.5--92.6\% EOT recall at 4.3--11.8\% false-positive rates, while INT detection suffers from missed takeovers or excessive false activations. This asymmetry extends to five full-duplex agents which achieve 86.1--95.8\% balanced accuracy on turn entry, yet perform near chance (46.9--54.7\%) on selective interruption handling. Gemini-3.8-Live yields to 97.7\% of backchannels, whereas MiniCPM-o-4.5 yields to only 1.4\% of genuine interruptions. These findings establish selective floor control as a distinct, unresolved challenge for reliable multi-party voice interaction.
\end{abstract}

\begin{figure*}[htbp] 
    \centering
    \includegraphics[width=\textwidth]{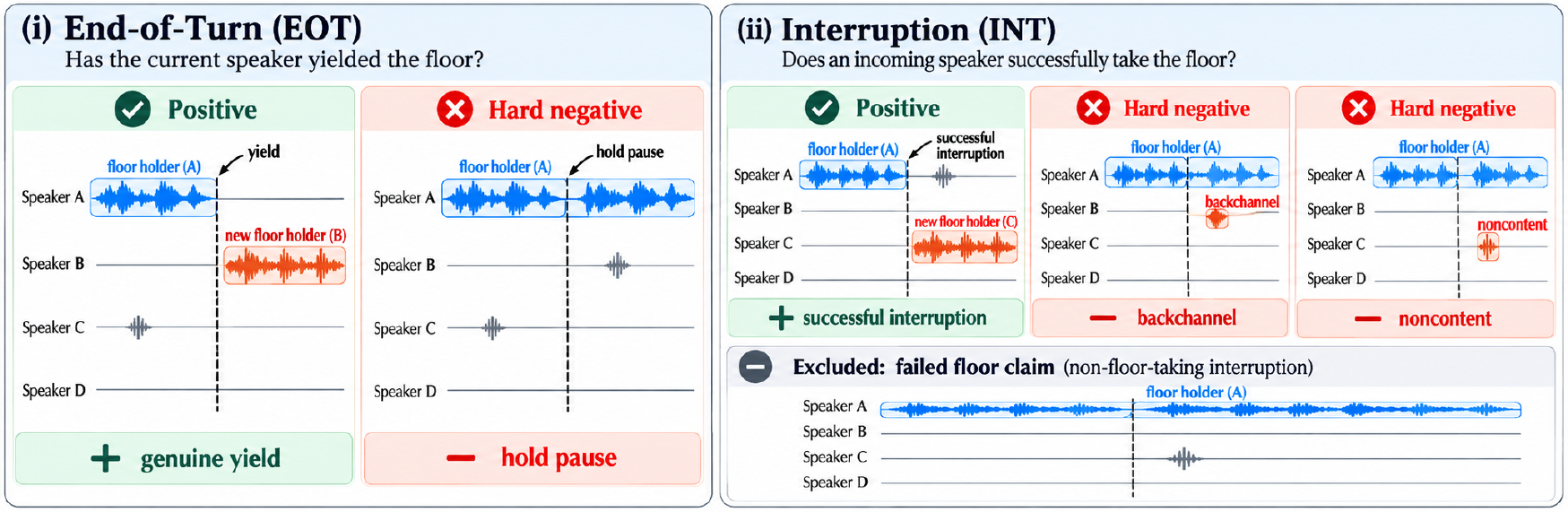}
    \caption{\small{\textbf{MultiPartyTurnBench evaluation tracks.} The End-of-Turn (EOT) track distinguishes genuine floor relinquishment from mid-turn pauses where the speaker retains the floor. The Interruption (INT) track distinguishes successful floor takeovers from backchannels and non-content vocalizations that leave floor ownership unchanged. Non-floor-taking interruption attempts are excluded from causal INT scoring because their ambiguous outcomes become apparent only through subsequent interaction.}}
    \label{fig:intro}
\end{figure*}

\section{Introduction}
\label{sec:introduction}

Full-duplex speech systems must continuously decide when to enter, hold, or relinquish the conversational floor while processing ongoing speech \citep{vap,fdbench}. These decisions require more than detecting speech activity: a speaker may pause without yielding the floor, while a listener may backchannel without attempting to take it \citep{sacks1974,yngve1970}. Misinterpreting either can cause an agent to respond prematurely or unnecessarily abandon its own response, even when its language understanding is otherwise correct \citep{sidbench}.

This distinction becomes particularly important in multi-party conversation, where several listeners may backchannel, overlap, or interject while the current speaker retains the floor. Existing evaluations address complementary aspects of this problem: dyadic benchmarks distinguish turn endings and interruptions in 2-party settings \citep{turnbench,sidbench}, natural multi-party corpora study next-speaker prediction \citep{triadicvap,muvap}, and full-duplex agent benchmarks assess speaking and stopping behavior \citep{fdb1,fdb2,duplexmpe}. However, the joint evaluation of genuine floor relinquishment and successful floor acquisition under controlled multi-party conditions remains underexplored.

We introduce \textbf{\mptb} (Fig.~\ref{fig:intro}), a benchmark for evaluating floor control in multi-party full-duplex speech. MPTB evaluates two complementary decisions: End-of-Turn (\eot) detection distinguishes genuine turn completion from temporary holds, while Interruption (\intr) detection distinguishes successful floor takeovers from backchannels and other non-floor-taking speech. The benchmark comprises 240 synthetically generated, channel-separated conversations balanced across $N\in\{2,3,4,5\}$ speakers and six interaction types. A discrete-event orchestrator establishes floor ownership and event timing before speech synthesis, enabling controlled variation in speaker count, interaction dynamics, and acoustic conditions. We evaluate acoustic, semantic, and turn-projection systems through causal event detection with false-positive-constrained recall and signed latency, complemented by real-speech validation and interactive testing of full-duplex agents.

\noindent\textbf{Main Results.} Our evaluation reveals a pronounced asymmetry between turn completion and successful floor acquisition. At qualified operating points, systems recover up to 92.6\% of turn endings but only 56.2\% of successful interruptions; acoustic detectors achieve higher recall only at excessive false-positive rates. Increasing speaker count produces jagged degradation, while conversational style exposes substantial model-specific interruption failures. Crucially, this asymmetry extends to observable voice agent behavior: four of five full-duplex agents achieve 86.1--95.8\% balanced accuracy on turn entry, yet all five remain near chance (46.9--54.7\%) on selective yielding. Gemini-3.8-Live~\citep{google2026gemini38live} yields almost indiscriminately, whereas MiniCPM-o-4.5~\citep{cui2026minicpm} rarely yields even to genuine interruptions, revealing opposite failures concealed by aggregate accuracy. Complementary real-speech evaluation further shows that interruption detection remains challenging in natural multi-party meetings. These findings establish selective interruption handling as a persistent limitation of multi-party full-duplex interaction.\textbf{Our main contributions are:}
\begin{itemize}[leftmargin=*,nosep]
    \item \textbf{MultiPartyTurnBench}, a controlled benchmark of 240 multi-party conversations spanning two to five speakers and six interaction types, with exact event-state annotations for end-of-turn (EOT) and interruption (INT) detection.
    \item \textbf{A systematic evaluation of floor-transition detection} across acoustic, semantic, and turn-projection systems, revealing an EOT--INT capability gap and model-specific sensitivities to speaker count and conversational style.
    \item \textbf{End-to-end validation across five full-duplex voice agents}, exposing near-chance selective yielding despite strong turn-entry performance, with opposing tendencies to yield indiscriminately or ignore genuine interruptions.
\end{itemize}

\begin{table*}[t]
\centering
\small
\setlength{\tabcolsep}{5pt}
\renewcommand{\arraystretch}{1.14}
\resizebox{\textwidth}{!}{%
\begin{tabular}{@{}lcccccccc@{}}
\toprule
\textbf{Benchmarks}
& \shortstack{\textbf{Multi-}\\\textbf{party}}
& \shortstack{\textbf{EOT event}\\\textbf{detection}}
& \shortstack{\textbf{INT event}\\\textbf{detection}}
& \shortstack{\textbf{Event-specific}\\\textbf{negatives}}
& \shortstack{\textbf{Timestamped}\\\textbf{event scoring}}
& \shortstack{\textbf{FPR}\\\textbf{budget}}
& \shortstack{\textbf{Controlled Speaker}\\\textbf{Comparison}}
& \shortstack{\textbf{Agent behavior}\\\textbf{scoring}} \\
\midrule

\multicolumn{9}{@{}l}{\textit{\textbf{Turn-taking and interruption benchmarks}}} \\
TurnBench~\citep{turnbench} & \xmark & \cmark & \cmark & \cmark & \cmark & \cmark & \xmark & \xmark \\
Talking Turns~\citep{talkingturns} & \xmark & \xmark & \xmark & \xmark & \xmark & \xmark & \xmark & \cmark \\
SID-Bench~\citep{sidbench} & \xmark & \xmark & \cmark & \cmark & \cmark & \xmark & \xmark & \xmark \\
ECHO~\citep{zhao2026echo} & \xmark & \xmark & \xmark & \cmark & \xmark & \xmark & \xmark & \cmark \\

\addlinespace[3pt]
\midrule
\multicolumn{9}{@{}l}{\textit{\textbf{Multi-party turn-taking evaluation resources}}} \\
TEIDAN (Triadic-VAP)~\citep{triadicvap} & \cmark & \xmark & \xmark & \xmark & \xmark & \xmark & \xmark & \xmark \\
AVCC (MuVAP)~\citep{muvap} & \cmark & \xmark & \xmark & \xmark & \xmark & \xmark & \cmark & \xmark \\
Fast When, Careful Who~\citep{fastcareful} & \cmark & \cmark & \xmark & \cmark & \cmark & \xmark & \cmark & \xmark \\

\addlinespace[3pt]
\midrule
\multicolumn{9}{@{}l}{\textit{\textbf{Full-duplex and multi-speaker agent benchmarks}}} \\
Full-Duplex-Bench~\citep{fdb1,fdb2,fdb3} & \xmark & \xmark & \xmark & \cmark & \xmark & \xmark & \xmark & \cmark \\
FD-Bench~\citep{fdbench} & \xmark & \xmark & \xmark & \xmark & \xmark & \xmark & \xmark & \cmark \\
Duplex-MPE~\citep{duplexmpe} & \cmark & \xmark & \xmark & \xmark & \xmark & \xmark & \xmark & \cmark \\
MultiTalkBench~\citep{multitalkbench} & \cmark & \xmark & \xmark & \xmark & \xmark & \xmark & \xmark & \cmark \\
MSI-Bench~\citep{msibench} & \cmark & \xmark & \xmark & \xmark & \xmark & \xmark & \xmark & \cmark \\

\midrule
\rowcolor{cyan!12}
\textbf{MultiPartyTurnBench (ours)} & \cmark & \cmark & \cmark & \cmark & \cmark & \cmark & \cmark & \cmark \\
\bottomrule
\end{tabular}%
}
\caption{\small{\textbf{Comparison of turn-taking and full-duplex evaluation benchmarks.} Existing benchmarks address dyadic turn-taking events, natural multi-party next-speaker predictions, or agent interaction behavior. \texttt{MultiPartyTurnBench} jointly evaluates multi-party EOT and INT event detection with event-specific hard negatives, timestamped causal decisions, FPR-constrained scoring, controlled speaker-count comparisons, and complementary agent behavior validation.}}
\label{tab:landscape}
\end{table*}

\section{Related Work}
\label{sec:related}

\paragraph{Turn-taking and interruption benchmarks.}
Conversational floor management involves turn transitions, interruptions, backchannels, and competing speaker claims~\citep{sacks1974,duncan1972,yngve1970}. Talking Turns~\citep{talkingturns} evaluates turn-taking in audio foundation models, while TurnBench~\citep{turnbench} established causal EOT and interruption detection in dyadic conversational styles. SID-Bench~\citep{sidbench} further studies semantic interruption detection against non-interruptive backchannels. Table \ref{tab:landscape} shows that \textbf{\texttt{MultiPartyTurnBench} addresses the complementary multi-party setting}, where several listeners may produce overlapping, non-floor-taking speech and successful floor acquisition must be distinguished from competing vocal activity under a common event-level evaluation protocol.

\paragraph{Multi-party turn-taking resources.}
Natural multi-speaker corpora support complementary prediction tasks. TEIDAN~\citep{triadicvap} evaluates future speaker activity in triadic Japanese conversations. AVCC~\citep{muvap} evaluates shift/hold and next-speaker prediction in audiovisual dialogue. VoxConverse~\citep{chung2020spot} studies floor-shift detection, boundary timing, and next-speaker identification~\citep{fastcareful}. These resources capture natural conversational dynamics, but do not score end-of-turn and interruption detection.

\paragraph{Full-duplex agent benchmarks.}
Full-Duplex-Bench series~\citep{fdb1,fdb2,fdb3} and FD-Bench~\citep{fdbench} evaluate real-time conversational behavior. Recent multi-party benchmarks broaden this scope in multi-party settings: Duplex-MPE~\citep{duplexmpe} assesses if agents respond, remain silent, or stop speaking in group conversations; MultiTalkBench~\citep{multitalkbench} examines long-form multi-party dialogue, and MSI-Bench~\citep{msibench} evaluates speaker-grounded 
memory, instruction following, and reasoning. These benchmarks primarily assess conversational outcomes rather than standalone floor-event predictions. \textbf{MultiPartyTurnBench isolates the underlying floor-discrimination problem}, while separately validating how these distinctions manifest in full-duplex agent behavior.

\section{\texttt{MultiPartyTurnBench}}
\label{sec:benchmark}

\begin{table}[t]
\centering
\small
\setlength{\tabcolsep}{4.5pt}
\renewcommand{\arraystretch}{1.12}
\resizebox{0.85\linewidth}{!}{
\begin{tabular}{lccc}
\toprule
\textbf{Event} & \textbf{Floor outcome} & \textbf{EOT} & \textbf{INT} \\
\midrule
\textsc{Turn}: yield & Released & $\checkmark$ & -- \\
\hspace{2.8em} hold & Retained & $\times$ & -- \\
\hspace{2.8em} interrupted & Taken over & -- & -- \\
\textsc{Interruption} & Transferred & -- & $\checkmark$ \\
\textsc{Non-Floor-Taking Int.} & Retained & -- & Excl. \\
\textsc{Backchannel} & Retained & -- & $\times$ \\
\textsc{Noncontent} & Unchanged & -- & $\times$ \\
\bottomrule
\end{tabular}
}
\caption{\small{\textbf{Floor-state event taxonomy in \texttt{MultiPartyTurnBench}.} Events are mapped to EOT and INT evaluation targets: $\checkmark$ denotes a positive event, $\times$ a hard negative, and -- an event not scored in that track. Excl.\ denotes events explicitly excluded from scoring.}}
\label{tab:event_taxonomy}
\end{table}

\begin{figure*}[htbp] 
    \centering
    \includegraphics[width=0.9\textwidth]{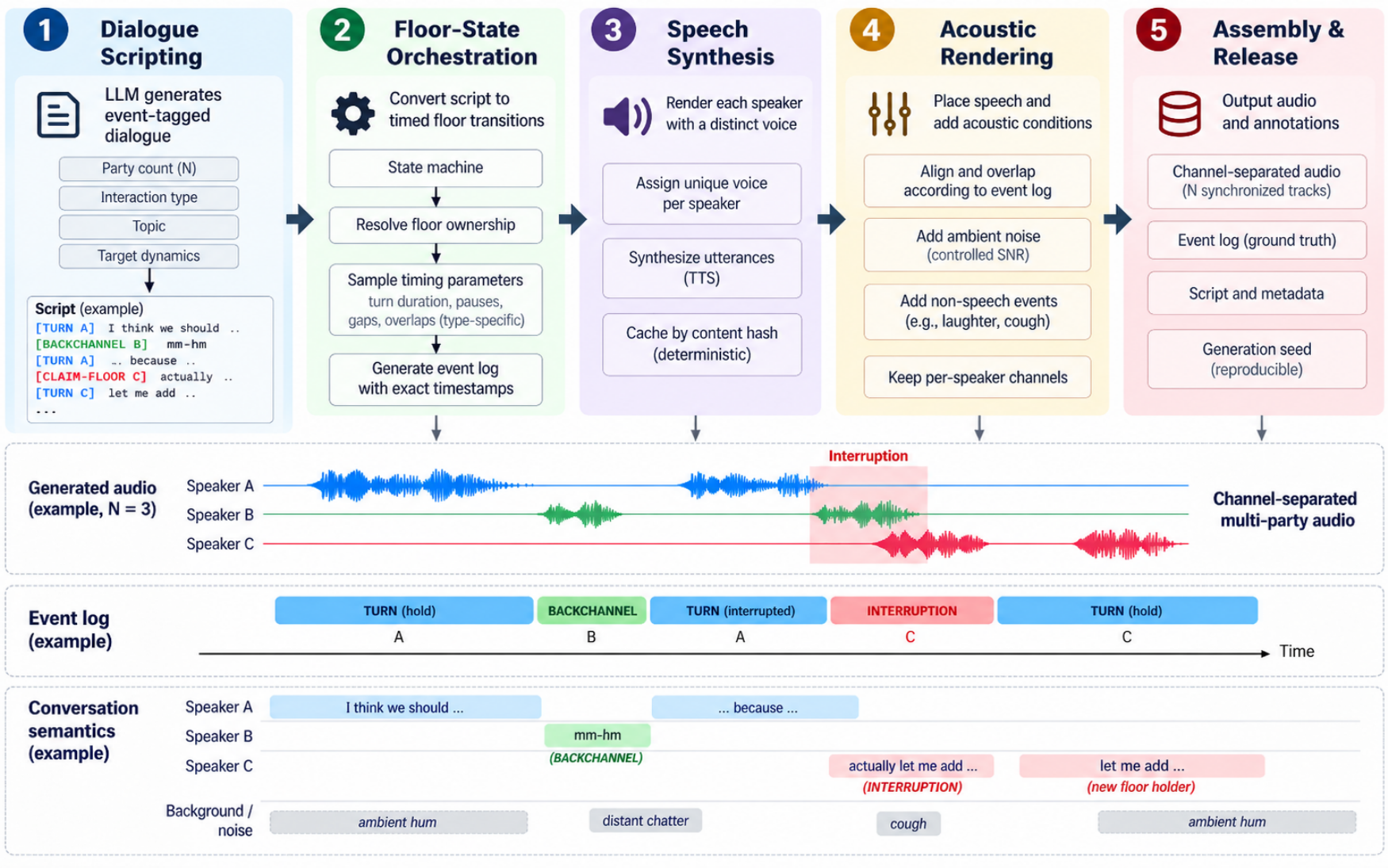}
    \caption{\small{\textbf{MultiPartyTurnBench conversation generation pipeline.} Five-stage pipeline transforms event-conditioned dialogue scripts into synchronized multi-speaker audio with exact floor-transition annotations. A discrete-event orchestrator resolves floor ownership and temporal dynamics before speech synthesis, followed by speaker rendering, acoustic augmentation, and channel assembly. The example illustrates overlapping speech, backchannels, successful interruptions, and their event annotations.}}
    \label{fig:main}
\end{figure*}

\texttt{MultiPartyTurnBench} evaluates whether systems can causally infer changes in \emph{conversational floor ownership} from ongoing multi-speaker audio. It defines two complementary, event-conditioned tasks: \textbf{(i) End-of-Turn (EOT)} asks whether the \emph{current speaker has relinquished the floor} and aims to detect genuine floor relinquishment by the current speaker, anchored at speech offset against hard negatives such as mid-turn pauses where the speaker retains the floor. \textbf{(ii) Interruption (INT)} asks whether \emph{another speaker has successfully acquired the floor} and detects successful floor acquisition by another speaker, anchored at the interrupting speech onset, against hard negatives like backchannels and non-content vocalizations that leave floor ownership unchanged. Unsuccessful interruption attempts are excluded because their outcomes may only become apparent through subsequent interaction, making them ambiguous at onset for a causal detector~\citep{turnbench}. Together, the two tracks evaluate \emph{floor relinquishment and acquisition} against behaviorally confusable non-events, rather than speech activity alone (Table~\ref{tab:event_taxonomy}).


\subsection{Benchmark Construction}
\label{sec:corpus_construction}

MPTB is designed to isolate floor-control decisions under independently controlled interaction conditions with exact event boundaries. In natural multi-party recordings, speaker count, overlap, listener activity, and acoustics vary jointly, complicating controlled analysis. We construct each conversation from an explicit floor-state representation before rendering speech in three steps: (1) define the event taxonomy for floor ownership changes; (2) select the interaction conditions and dialogue scripts; (3) staged generation pipeline renders synchronized multi-speaker audio recordings.

\paragraph{Event Taxonomy.} Each event is defined relative to the current floor holder and whether floor ownership is retained, relinquished, or transferred. Five event types capture these transitions: \textsc{Turn} represents speech by the floor holder, resolved as \texttt{hold}, \texttt{yield}, or \texttt{interrupted}; \textsc{Interruption} denotes a successful listener floor claim, while \textsc{Non-Floor-Taking Interruption} denotes an unsuccessful attempt. \textsc{Backchannel} captures listener speech without claiming the floor, and \textsc{Noncontent} captures non-linguistic activity without changing floor ownership. Table~\ref{tab:event_taxonomy} maps these events to the EOT and INT evaluation targets. 

\vspace{-0.05cm}

\paragraph{Controlled Interaction Axes.} MPTB crosses speaker count ($N\in\{2,3,4,5\}$) with six interaction types derived from TurnBench~\citep{turnbench}: Casual, Task-Oriented, Instructional, Collaborative, Argumentative, and Narrative. These regimes elicit distinct turn-taking dynamics while enabling controlled comparisons across group sizes at fixed interaction style, and vice versa. Generation additionally controls backchannel and interruption occurrences, while acoustic rendering varies ambient noise and nuisance events. See Appendix~\ref{app:interaction_types} for definitions of each interaction type.

\paragraph{Conversation Generation Pipeline}
Figure~\ref{fig:main} illustrates the five-stage pipeline for multi-speaker acoustic realization annotated with conversational floor dynamics. The first two stages establish the canonical event log and interaction timeline before speech synthesis, ensuring that ground-truth labels derive from the underlying floor state rather than retrospective waveform analysis.

\noindent\textbf{1. Event-conditioned dialogue generation.}
Off-the-shelf LLM like GPT-4.1 generates multi-party scripts conditioned on speaker count, interaction type, topic, narrative plot, and target turn-taking statistics. A restricted event vocabulary (\texttt{TURN}, \texttt{CLAIM\_FLOOR}, \texttt{VOCALIZE\_NO\_CLAIM}) specifies conversational actions, with backchannel and interruption frequencies guided by TurnBench~\citep{turnbench}. The LLM determines linguistic content and proposes listener actions but does not assign final timestamps or resolve floor ownership. Scripts are schema-validated, with malformed outputs regenerated or rejected.

\noindent\textbf{2. Floor-state orchestration.}
A discrete-event orchestrator maintains the current floor holder and resolves symbolic actions into realized transitions: \texttt{TURN} becomes a hold, yield, or interruption-forced termination; \texttt{CLAIM\_FLOOR} resolves either to a successful \textsc{Interruption} or a \textsc{Non-Floor-Taking Interruption}; and \texttt{VOCALIZE\_NO\_CLAIM} produces a backchannel or related non-floor-changing event. The orchestrator assigns turn durations, inter-speaker gaps, pauses, and interruption overlap based on real-world timing statistics from AMI corpus~\citep{ami}. Explicit mid-turn pauses provide EOT hard negatives. The resulting placement schedule and canonical event log define exact EOT and INT anchors.

\noindent\textbf{3. Speaker rendering.}
Utterances are independently synthesized using Kokoro-82M\citep{kokoro_tts}, with distinct voices sampled from a gender-balanced pool of 24 voices, preventing voice reuse within a conversation. Development and test splits use disjoint voice inventories (8 and 16 voices, respectively). Word-level timing is retained for alignment and quality control.

\noindent\textbf{4. Acoustic scene rendering.}
Synthesized speech is placed on synchronized speaker channels according to the orchestrated timeline, preserving overlap determined by floor-state transitions. Successful interruptions end with a short fade applied at the truncation boundary, and channels are peak-normalized independently. A shared voice-free ambient bed (filtered pink noise and humming) is mixed at a per-conversation SNR. MUSAN~\citep{musan2015} sound effects and intermittent distant third-party speech introduces nuisance activity to stress false-positive detection; neither creates benchmark-positive floor events.

\noindent\textbf{5. Channel assembly and release artifacts.}
Each conversation is released with its noisy evaluation channels, clean reference channels, tagged script, canonical event log, render metadata, and generation seed. Clean channels are retained for quality control, while benchmark systems operate on the noisy rendition. All non-generative stages are deterministic with recorded seeds and manifests, enabling exact reproduction of the interaction schedule and acoustic assembly (see Appendix~\ref{generation}).

\subsection{Benchmark Composition}
\label{sec:benchmark_composition}

MPTB contains 240 conversations (6.89 hours), balanced across four speaker-count settings ($N\in\{2,3,4,5\}$) crossed with the six interaction types. It contains 48 development and 192 test conversations using disjoint speaker voices. The corpus contains 3,422 \eot\ positives and 241 hold negatives, alongside 367 \intr\ positives and 1,481 non-floor-taking negatives. MPTB closely mimic natural turn-taking distributions. See Appendix~\ref{app:corpus_stats} for event-interaction trends. We additionally validate the generated floor events through independent human annotations in Appendix~\ref{app:human_eval}).



\section{Evaluation}
\label{sec:evaluation_protocol}

\begin{figure}[] 
    \centering
    \includegraphics[width=0.85\linewidth]{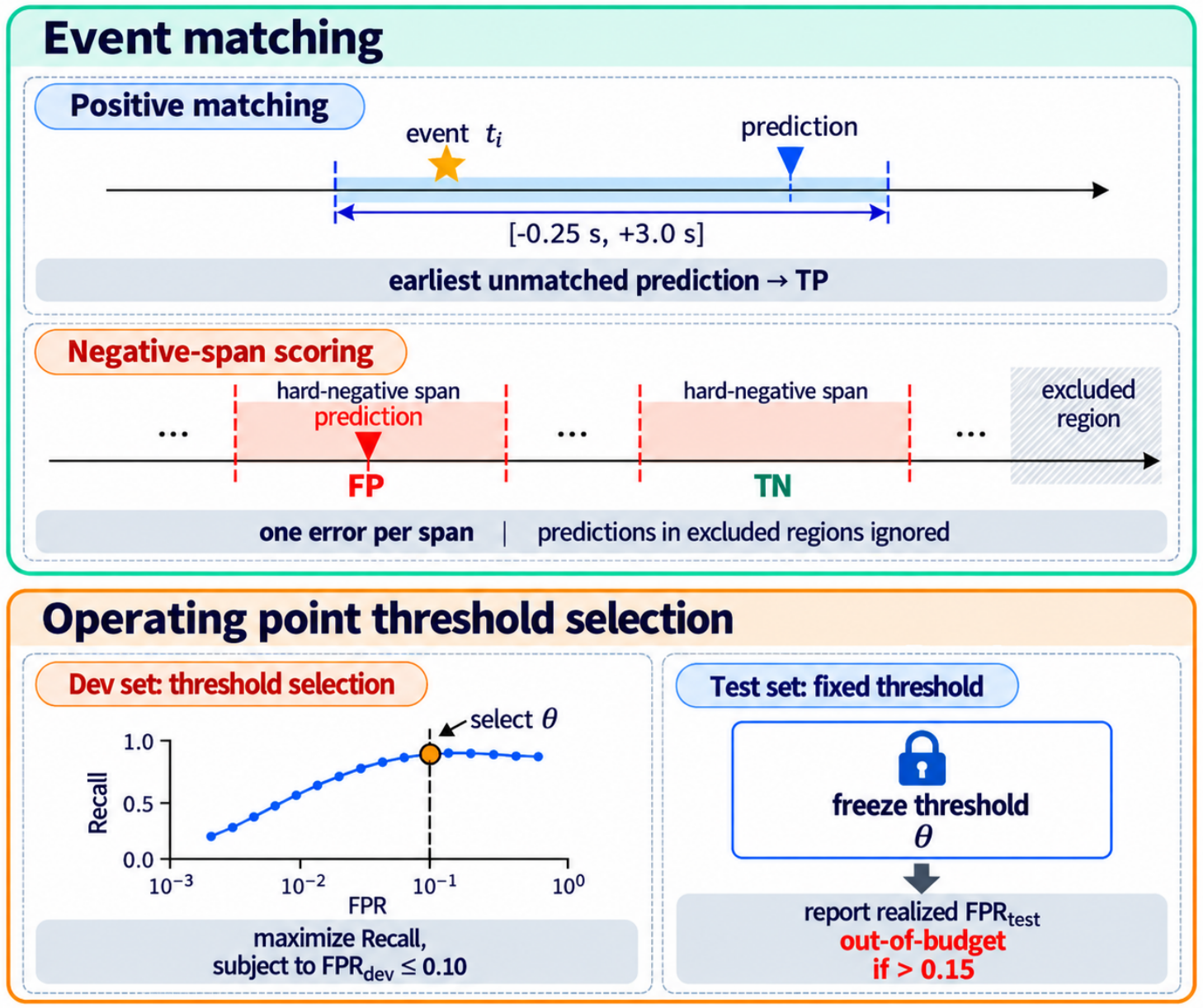}
    \caption{\small{\textbf{\texttt{MultiPartyTurnBench} evaluation protocol.} Causal per-speaker predictions are matched to reference events and scored against hard-negative spans. Development-selected thresholds are frozen for test evaluation, with performance measured through recall, false-positive rate, and latency.}}
    \label{fig:evaluation_protocol}
\end{figure}

Figure~\ref{fig:evaluation_protocol} illustrates the causal evaluation protocol: systems emit timestamped EOT and INT decisions using only audio observed up to prediction time. We evaluate event recall, false-positive rate, and signed decision latency at fixed operating points.

\noindent\textbf{Prediction and Event Matching.}
Per-speaker model scores are converted to discrete events through threshold crossings with a 2\,s refractory period. Following TurnBench~\citep{turnbench}, each gold positive at $t_i$ is matched one-to-one to the earliest unmatched prediction within $[t_i-0.25,\;t_i+3.0]$\,s. Matched events are true positives (TP), while unmatched gold positives are false negatives (FN). Hard negatives are scored as temporal spans: any detection within a span counts as a false positive (FP), otherwise a true negative (TN). Predictions outside scored regions are ignored.

\noindent\textbf{Operating Points and Metrics.}
For each model, we select the development threshold maximizing recall subject to $\mathrm{FPR}_{\mathrm{dev}}\leq0.10$~\citep{turnbench}, then freeze it across all test conversations, speaker counts, and interaction types. Systems exceeding $\mathrm{FPR}_{\mathrm{test}}>0.15$ are marked out-of-budget. We report $\mathrm{Recall}=\mathrm{TP}/(\mathrm{TP}+\mathrm{FN})$ and $\mathrm{FPR}=\mathrm{FP}/(\mathrm{FP}+\mathrm{TN})$, alongside signed latency $\Delta t_i=t_i^{\mathrm{pred}}-t_i^{\mathrm{gold}}$ (p10/p50/p90), where $+$ve / $-$ve values indicate delayed / early commitments.

\noindent\textbf{Full-Duplex Behavioral Diagnostics.}
For full duplex voice agents, recall alone cannot distinguish indiscriminate yielding from selective floor control. We additionally report balanced accuracy $\mathrm{bACC}=(\mathrm{Recall}+1-\mathrm{FPR})/2$ and action bias $A_{\mathrm{bias}}=\mathrm{Recall}+\mathrm{FPR}-1$. Balanced accuracy measures discrimination between genuine floor transitions and hard negatives. Action bias indicates excessive responsiveness ($+$ve) or reluctance to act ($-$ve). More details in Appendix~\ref{app:matching}.

\section{Experiments}
\label{sec:experiments}

\noindent\textbf{Offline EOT and INT Evaluation.}
We evaluate 12 systems across three families: \textbf{acoustic VAD} (RMS Energy, WebRTC VAD~\citep{webrtcvad}, Silero VAD~\citep{silero}); \textbf{semantic endpointers} (Smart Turn v2~\citep{smartturn}, Kyutai SVAD~\citep{kyutai_svad}, Mimi Endpointer~\citep{mimi}); and \textbf{turn-taking models} (VAP~\citep{vap}, MuVAP~\citep{muvap}, DualTurn~\citep{dualturn}, WavLM-Large anchor~\citep{wavlm}, Turn-1-Mini~\citep{turn1mini}, ESPnet Turn-Taking~\citep{espnet}). VAP is adapted to multi-party audio through speaker-versus-rest inference. We report EOT and INT performance across speaker-count and interaction-type breakdowns using development-selected operating points. See Appendix~\ref{app:offline_config} and \ref{app:operating_points} for model configurations and score mappings.

\noindent\textbf{Full Duplex Agent Evaluation.} We test five full-duplex voice agents---\textbf{GPT-Live-1}~\citep{openai2026gptlive}, \textbf{Gemini-3.8-Live}~\citep{google2026gemini38live}, MiniCPM-o-4.5~\citep{cui2026minicpm}, \textbf{Realtime-Venus}~\citep{zhao2026realtime}, and \textbf{Step-Audio3-Realtime}~\citep{lin2026stepaudio}---on 120 interactive floor-control scenarios, repeated three times per system. These experiments distinguish offline event detection from agents' observable decisions to enter or relinquish the conversational floor. See Appendix~\ref{app:experimental_details} for scoring configurations.

\section{Results}
\label{sec:results}


\definecolor{mptbBudgetGray}{RGB}{240,243,246}
\definecolor{mptbBestCyan}{RGB}{217,242,247}

\newcommand{\mptbRec}[3]{#1\,{\scriptsize\color{black!65}[#2,\,#3]}}
\newcommand{\mptbGray}[1]{\cellcolor{mptbBudgetGray}#1}
\newcommand{\mptbCyan}[1]{\cellcolor{mptbBestCyan}#1}

\begin{table*}[t]
\centering
\footnotesize
\setlength{\tabcolsep}{3.0pt}
\renewcommand{\arraystretch}{1.16}
\resizebox{0.8\textwidth}{!}{%
\begin{tabular}{@{}lcccccc@{}}
\toprule
& \multicolumn{3}{c}{\textbf{End-of-Turn (EOT)}} & \multicolumn{3}{c}{\textbf{Interruption (INT)}} \\
\cmidrule(lr){2-4}\cmidrule(l){5-7}
\textbf{System} & \textbf{Recall (\%) $\uparrow$} & \textbf{FPR (\%) $\downarrow$} & \textbf{p50 (ms) $\downarrow$} & \textbf{Recall (\%) $\uparrow$} & \textbf{FPR (\%) $\downarrow$} & \textbf{p50 (ms) $\downarrow$} \\
\midrule

\multicolumn{7}{@{}l}{\textit{Acoustic VAD}} \\
RMS Energy & \mptbRec{49.8}{44.4}{57.3} & 40.9 & 292 & \mptbRec{66.2}{61.5}{72.8} & 25.0 & 859 \\
WebRTC VAD~{\scriptsize\citep{webrtcvad}} & \mptbRec{91.8}{90.8}{92.8} & 90.3 & 332 & \mptbRec{84.3}{80.6}{89.2} & 83.8 & 128 \\
Silero VAD~{\scriptsize\citep{silero}} & \mptbRec{97.2}{96.3}{97.8} & 93.5 & 397 & \mptbRec{85.4}{81.4}{89.7} & 89.2 & 119 \\

\addlinespace[2pt]
\midrule
\multicolumn{7}{@{}l}{\textit{Semantic endpointers}} \\
Smart Turn v2~{\scriptsize\citep{smartturn}} & \mptbRec{82.5}{81.0}{84.2} & 11.8 & -18 & \mptbRec{67.6}{61.5}{73.6} & 64.3 & 87 \\
Kyutai SVAD~{\scriptsize\citep{kyutai_svad}} & \mptbGray{\mptbRec{92.5}{91.1}{93.9}} & \mptbGray{4.3} & \mptbGray{1829} & \mptbCyan{\mptbRec{\textbf{56.2}}{50.5}{62.0}} & \mptbCyan{6.6} & \mptbCyan{689} \\
Mimi Endpointer~{\scriptsize\citep{mimi}} & \mptbGray{\mptbRec{90.5}{89.8}{91.2}} & \mptbGray{3.8} & \mptbGray{1231} & \mptbGray{\mptbRec{13.9}{10.0}{17.9}} & \mptbGray{3.8} & \mptbGray{870} \\

\addlinespace[2pt]
\midrule
\multicolumn{7}{@{}l}{\textit{Turn-taking models}} \\
VAP~{\scriptsize\citep{vap}} & \mptbCyan{\mptbRec{\textbf{92.6}}{92.0}{93.1}} & \mptbCyan{7.5} & \mptbCyan{\textbf{143}} & \mptbGray{\mptbRec{11.7}{8.1}{15.6}} & \mptbGray{2.8} & \mptbGray{492} \\
MuVAP~{\scriptsize\citep{muvap}} & \mptbGray{\mptbRec{88.1}{87.1}{89.1}} & \mptbGray{10.2} & \mptbGray{553} & \mptbGray{\mptbRec{34.5}{29.1}{39.9}} & \mptbGray{8.6} & \mptbGray{618} \\
DualTurn~{\scriptsize\citep{dualturn}} & \mptbGray{\mptbRec{89.4}{88.6}{90.2}} & \mptbGray{10.8} & \mptbGray{777} & \mptbGray{\mptbRec{37.0}{31.6}{42.6}} & \mptbGray{8.2} & \mptbGray{572} \\
WavLM-Large (anchor)~{\scriptsize\citep{wavlm}} & \mptbRec{90.4}{89.5}{91.2} & 22.6 & 1185 & \mptbGray{\mptbRec{12.8}{9.3}{16.4}} & \mptbGray{7.3} & \mptbGray{\textbf{42}} \\
Turn-1-Mini~{\scriptsize\citep{turn1mini}} & \mptbRec{86.9}{85.9}{87.9} & 25.3 & 1087 & \multicolumn{3}{c}{---} \\
ESPnet Turn-Taking (mix)~{\scriptsize\citep{espnet}} & \mptbGray{\mptbRec{86.9}{85.6}{88.0}} & \mptbGray{12.4} & \mptbGray{1084} & \mptbGray{\mptbRec{1.1}{0.0}{2.5}} & \mptbGray{1.9} & \mptbGray{71} \\
\bottomrule
\end{tabular}%
}
\caption{\small\textbf{EOT and INT detection on \texttt{MultiPartyTurnBench}.} Recall is reported with 95\% conversation-bootstrap CI; p50 denotes median signed latency (ms). \colorbox{mptbBudgetGray}{Gray} indicates operating points satisfying dev FPR $\leq10\%$ and test FPR $\leq15\%$; \colorbox{mptbBestCyan}{cyan} highlights the highest qualified recall in each task. Budget-qualified systems achieve up to 92.6\% EOT recall, yet the best INT recall reaches only 56.2\%. Acoustic detectors recover more events but incur substantially higher FPR, highlighting the difficulty of distinguishing genuine floor transitions from speech activity.}
\label{tab:offline_results}
\end{table*}


\definecolor{mptbBehCyanColor}{RGB}{217,242,247}
\definecolor{mptbBehRedColor}{RGB}{252,228,228}

\newcommand{\mptbBehCI}[3]{#1\,{\scriptsize\color{black!65}[#2,\,#3]}}
\newcommand{\mptbBehCyan}[1]{\cellcolor{mptbBehCyanColor}#1}
\newcommand{\mptbBehRed}[1]{\cellcolor{mptbBehRedColor}#1}

\begin{table*}[t]
\centering
\footnotesize
\setlength{\tabcolsep}{3.0pt}
\renewcommand{\arraystretch}{1.13}
\resizebox{0.65\textwidth}{!}{%
\begin{tabular}{@{}lccccccc@{}}
\toprule
& \multicolumn{3}{c}{\textbf{START: enter the floor}} & \multicolumn{4}{c}{\textbf{STOP: yield selectively}} \\
\cmidrule(lr){2-4}\cmidrule(lr){5-8}
\textbf{Voice agent} & \textbf{Start $\uparrow$} & \textbf{False start $\downarrow$} & \textbf{bACC $\uparrow$} & \textbf{Yield $\uparrow$} & \textbf{False yield $\downarrow$} & \textbf{bACC $\uparrow$} & $\mathbf{A}_{\mathrm{Bias}}\rightarrow 0$ \\
\midrule
GPT-Live-1~\citep{openai2026gptlive} & 93.3 & 1.7 & \mptbBehCyan{\mptbBehCI{\textbf{95.8}}{90.8}{99.2}} & 50.0 & 40.6 & \mptbBehRed{\mptbBehCI{54.7}{45.7}{63.5}} & -0.09 \\
Gemini-3.8-Live~\citep{google2026gemini38live} & 88.9 & 0.0 & \mptbBehCyan{\mptbBehCI{94.4}{90.0}{97.8}} & 100.0 & 97.7 & \mptbBehRed{\mptbBehCI{51.1}{50.0}{52.8}} & +0.98 \\
MiniCPM-o-4.5~\citep{cui2026minicpm} & 86.7 & 5.6 & \mptbBehCyan{\mptbBehCI{90.6}{85.0}{95.6}} & 1.4 & 0.0 & \mptbBehRed{\mptbBehCI{50.7}{50.0}{52.2}} & -0.99 \\
Realtime-Venus~\citep{zhao2026realtime} & 81.1 & 8.9 & \mptbBehCyan{\mptbBehCI{86.1}{80.6}{91.1}} & 8.1 & 12.7 & \mptbBehRed{\mptbBehCI{47.7}{43.1}{52.4}} & -0.79 \\
Step-Audio3-Realtime~\citep{lin2026stepaudio} & 0.0 & 0.0 & \mptbBehCI{50.0}{50.0}{50.0} & 9.1 & 15.4 & \mptbBehRed{\mptbBehCI{46.9}{34.6}{60.0}} & -0.76 \\
\bottomrule
\end{tabular}%
}
\caption{\small\textbf{Full-duplex agents know when to start, but not when to stop.} START distinguishes genuine turn endings from hold pauses; STOP distinguishes interruptions from backchannels. Four agents achieve 86.1--95.8\% START balanced accuracy (bACC), yet all five remain near chance on STOP (46.9--54.7\%). Gemini yields to 97.7\% of backchannels, whereas MiniCPM-o yields to only 1.4\% of interruptions, exposing opposite failure behaviors. Brackets denote 95\% scenario-clustered bootstrap CIs; \colorbox{mptbBehCyanColor}{cyan} highlights above-chance START performance; \colorbox{mptbBehRedColor}{red} marks STOP estimates whose CIs include 50\%.}
\label{tab:fd_behavior}
\end{table*}


\definecolor{MPTBblue}{RGB}{41,89,143}
\definecolor{MPTBorange}{RGB}{206,110,42}
\definecolor{MPTBteal}{RGB}{15,143,140}
\definecolor{MPTBviolet}{RGB}{134,88,162}
\definecolor{MPTBgray}{RGB}{99,105,117}

\begin{figure}[t]
\centering
\pgfplotsset{MPTBline/.style={thick,draw=#1,mark options={fill=#1,scale=0.88},error bars/y dir=both,error bars/y explicit,error bars/error bar style={draw=#1!65,line width=0.55pt},error bars/error mark options={draw=#1!65,mark size=1.6pt}}}
\resizebox{\linewidth}{!}{
\begin{tikzpicture}
\begin{groupplot}[
    group style={group size=2 by 1,horizontal sep=1.55cm},
    width=0.424\textwidth,height=5.35cm,
    xmin=1.8,xmax=5.2,xtick={2,3,4,5},
    xlabel={Number of speakers ($N$)},
    xlabel style={font=\small,yshift=2pt},
    ylabel style={font=\small},
    tick label style={font=\scriptsize},
    title style={font=\small\bfseries,yshift=2pt},
    axis lines=left,axis line style={black!65,line width=0.55pt},
    tick style={black!60},grid=major,
    grid style={black!10,line width=0.3pt},
    enlarge y limits=false,
]

\nextgroupplot[title={(a) End-of-Turn},ylabel={EOT recall (\%)},ymin=79,ymax=101,ytick={80,85,90,95,100},legend to name=mptbSpeakerLegend,legend columns=3,legend style={draw=none,fill=none,font=\scriptsize,column sep=3pt},legend cell align=left]

\addplot+[MPTBline=MPTBblue,mark=*] table[x=N,y=rec,y error plus=upper,y error minus=lower,row sep=\\] {
N rec upper lower\\
2 84.6 3.2 3.4\\
3 92.1 2.2 2.3\\
4 96.9 1.5 1.7\\
5 96.6 1.3 1.3\\
};
\addlegendentry{Kyutai SVAD}

\addplot+[MPTBline=MPTBorange,mark=square*] table[x=N,y=rec,y error plus=upper,y error minus=lower,row sep=\\] {
N rec upper lower\\
2 93.2 0.8 0.8\\
3 93.2 1.1 1.1\\
4 92.0 0.9 1.0\\
5 92.0 1.2 1.4\\
};
\addlegendentry{VAP}

\addplot+[MPTBline=MPTBteal,mark=triangle*] table[x=N,y=rec,y error plus=upper,y error minus=lower,row sep=\\] {
N rec upper lower\\
2 88.6 1.7 1.8\\
3 86.8 2.1 2.2\\
4 88.8 2.2 2.6\\
5 88.3 1.8 2.1\\
};
\addlegendentry{MuVAP}

\addplot+[MPTBline=MPTBviolet,mark=diamond*] table[x=N,y=rec,y error plus=upper,y error minus=lower,row sep=\\] {
N rec upper lower\\
2 89.4 1.6 1.8\\
3 89.4 1.7 1.9\\
4 89.3 1.5 1.5\\
5 89.6 1.4 1.6\\
};
\addlegendentry{DualTurn}

\addplot+[MPTBline=MPTBgray,mark=pentagon*] table[x=N,y=rec,y error plus=upper,y error minus=lower,row sep=\\] {
N rec upper lower\\
2 90.2 1.2 1.3\\
3 89.7 1.5 1.7\\
4 92.0 0.9 1.0\\
5 90.1 1.3 1.6\\
};
\addlegendentry{Mimi EP}


\nextgroupplot[title={(b) Interruption},ylabel={INT recall (\%)},ymin=0,ymax=80,ytick={0,20,40,60,80}]
\addplot+[MPTBline=MPTBblue,mark=*] table[x=N,y=rec,y error plus=upper,y error minus=lower,row sep=\\] {
N rec upper lower\\
2 63.3 13.0 11.6\\
3 63.0 11.5 11.2\\
4 51.4 10.4 11.1\\
5 48.7 12.1 12.3\\
};
\addplot+[MPTBline=MPTBorange,mark=square*] table[x=N,y=rec,y error plus=upper,y error minus=lower,row sep=\\] {
N rec upper lower\\
2 6.7 7.1 5.2\\
3 9.6 7.1 5.8\\
4 11.1 7.1 6.6\\
5 18.4 9.6 9.2\\
};
\addplot+[MPTBline=MPTBteal,mark=triangle*] table[x=N,y=rec,y error plus=upper,y error minus=lower,row sep=\\] {
N rec upper lower\\
2 23.3 11.2 10.3\\
3 32.9 10.0 9.6\\
4 38.9 11.1 10.7\\
5 40.8 11.2 11.1\\
};
\addplot+[MPTBline=MPTBviolet,mark=diamond*] table[x=N,y=rec,y error plus=upper,y error minus=lower,row sep=\\] {
N rec upper lower\\
2 33.3 11.0 11.5\\
3 35.6 10.6 10.2\\
4 38.9 10.4 10.3\\
5 39.5 12.5 11.8\\
};
\addplot+[MPTBline=MPTBgray,mark=pentagon*] table[x=N,y=rec,y error plus=upper,y error minus=lower,row sep=\\] {
N rec upper lower\\
2 13.3 7.7 7.5\\
3 19.2 8.8 8.3\\
4 12.5 8.3 6.9\\
5 10.5 7.8 6.6\\
};
\end{groupplot}
\end{tikzpicture}
}
\vspace{1pt}
\pgfplotslegendfromname{mptbSpeakerLegend}
\vspace{-2pt}
\caption{\textbf{Speaker-count sensitivity of EOT and INT detection.} Recall by speaker count ($N=2$--$5$), with 95\% conversation-bootstrap CI. EOT is broadly stable except for Kyutai, whereas INT trends differ. At five speakers, no plotted detector reaches 50\% INT recall, highlighting the challenge of identifying successful floor takeovers among multiple speech claimants.}
\label{fig:speaker_analysis}
\end{figure}
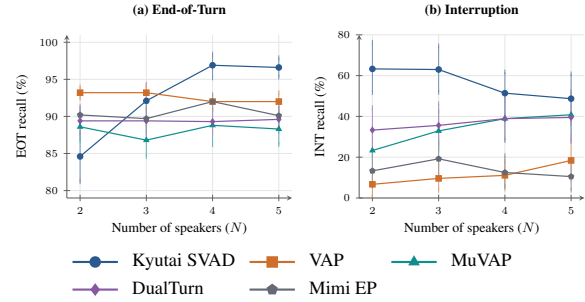


\definecolor{mptbHeatOrange}{RGB}{217,95,2}

\begin{figure*}[t]
\centering
\resizebox{\textwidth}{!}{%
%
}
\caption{\small{\textbf{EOT / INT recall by interaction type on \texttt{MultiPartyTurnBench} test set.} \colorbox{mptbHeatOrange}{Darker} means higher recall. $^\dagger$ denotes over the FPR budget; its recall reflects indiscriminate firing. EOT recall remains comparatively stable across conversation types. INT recall swings up to 30 points. Multi-party floor-control difficulty depends on conversational dynamics (argumentative for VAP and Mimi EP, collaborative for Kyutai, narrative for MuVAP, and task-oriented for DualTurn), not merely on speaker count.}}
\label{fig:eot_int_type_heatmaps_small}
\end{figure*}


\subsection*{\textit{RQ1: Can Models Reliably Identify Floor Transitions?}}

\noindent\textbf{Turn completion is comparatively tractable; successful floor acquisition is not.}
Table~\ref{tab:offline_results} reveals a pronounced asymmetry between the two floor-control decisions. VAP and Kyutai SVAD achieve comparable EOT recall (92.6\% and 92.5\%) at low FPR (7.5\% and 4.3\%), yet recover only 11.7\% and 56.2\% of successful interruptions, respectively. Even the strongest qualified INT detector misses nearly half of genuine takeovers. Thus, recognizing when a speaker has finished does not translate into determining whether incoming speech has actually displaced the current floor holder.

\noindent\textbf{High recall without selectivity is insufficient for conversational control.}
Acoustic detectors illustrate why floor transitions cannot be reduced to speech activity. Silero VAD achieves 97.2\% EOT and 85.4\% INT recall, but incurs prohibitive FPRs of 93.5\% and 89.2\%. Semantic and turn-taking models suppress these false activations, often at the cost of missed takeovers: Mimi Endpointer achieves 90.5\% EOT recall but detects only 13.9\% of interruptions. This exposes the central challenge: distinguishing genuine floor claims from listener activity that leaves floor ownership unchanged.

\noindent\textbf{Detection quality and latency trade-offs.}
Even among strong EOT detectors, comparable recall conceals substantial timing differences. VAP detects turn endings at 143\,ms p50, versus 1,829\,ms for Kyutai SVAD, despite nearly identical recall. Thus, a correct floor transition detected too late may still disrupt natural interaction.

\subsection*{\textit{RQ2: Does Speaker Count Make Floor Control Harder?}}

\noindent\textbf{More speakers do not produce uniform degradation in EOT detection.}
Figure~\ref{fig:speaker_analysis} shows that EOT detection remains comparatively stable across two to five speakers. VAP maintains 92--93\% recall, while MuVAP and DualTurn show similarly small variations. Surprisingly, Kyutai improves from 84.6\% to 96.6\%, suggesting that increasing speaker count does not inherently make turn completion harder under MPTB's controlled conditions.

\noindent\textbf{Interruption detection exhibits opposing speaker-count trends.}
Kyutai's INT recall declines from 63.3\% to 48.7\% as speaker count increases from two to five, whereas MuVAP improves from 23.3\% to 40.8\%, with DualTurn showing a similar upward trend. Yet no model exceeds 50\% INT recall at five speakers. These opposing trajectories suggest that models exploit multi-speaker context differently when interpreting competing floor claims. Relational speaker modeling may offer advantages over channel-local endpointing. Speaker count therefore exposes model-specific vulnerabilities rather than a universal scaling penalty. 
See Appendix~\ref{app:by_n} for recall--FPR breakdowns across speaker counts.

\subsection*{\textit{RQ3: Which Interaction Types Expose Failures?}}

\noindent\textbf{Interruption detection is substantially more sensitive to interaction style.} Figure~\ref{fig:eot_int_type_heatmaps_small} (and Appendix~\ref{app:type_breakdown}) shows that EOT detection remains comparatively stable across interaction types with limited variation. In contrast, INT recall shows large swings under the same threshold (Kyutai SVAD is 47.0\% in collaborative dialogue to 77.5\% in instructional exchanges). This disparity suggests that EOT detection generalizes more consistently across interaction regimes compared to INT detection.

\noindent\textbf{Conversational difficulty is model-dependent.} Kyutai struggles most with collaborative exchanges, while MuVAP performs worst on narrative dialogue (25.0\% INT recall), DualTurn on task-oriented interactions (32.6\%), and VAP and Mimi Endpointer on argumentative exchanges. These contrasting failures suggest that models rely on different conversational cues when interpreting floor claims, rather than a shared hierarchy of interaction difficulty. See Appendix~\ref{app:qualitative} for qualitative examples.

\subsection*{\textit{RQ4: Can Voice Agents Yield Selectively?}}

\paragraph{Knowing when to speak does not imply knowing when to stop.}
Table~\ref{tab:fd_behavior} extends offline detection to observable behavior in five full-duplex agents. Four systems achieve 86.1--95.8\% START balanced accuracy, reliably distinguishing genuine turn endings from mid-turn holds. Yet this competence collapses when deciding whether to yield: all five achieve only 46.9--54.7\% STOP balanced accuracy, with confidence intervals including chance performance. The offline EOT--INT asymmetry thus manifests directly in agent behavior: models recognize end of turn but struggle to determine when incoming speech interruptions.

\paragraph{Similar STOP accuracy conceals opposite behavioral failures.}
Gemini-3.8-Live effectively surrendering the floor indiscriminately (97.7\% of backchannels, bias $+0.98$). Conversely, MiniCPM-o-4.5 almost never yields, ignoring 98.6\% of genuine interruptions (bias $-0.99$). Realtime-Venus similarly resists yielding, while GPT-Live-1 is more balanced but remains unreliable. These opposing behaviors suggest inability of the voice agents to distinguish the conversational significance of incoming audio (see Appendix~\ref{app:fd_details}).

\section{Discussion}
\label{sec:discussion}

\noindent\textbf{Multi-party floor control requires reasoning about speaker roles, not merely speech activity.} MPTB exposes a fundamental limitation of conventional endpointing: incoming speech does not necessarily imply a change in floor ownership. Reliable interruption handling in multi-speaker settings requires distinguishing competing floor claims from backchannels and other non-floor-taking activity, conditioned on the current speaker and conversational context. This motivates event-conditioned modeling of floor transitions rather than independent speech-activity decisions.

\noindent\textbf{Floor-control evaluation must capture the consequences of incorrect decisions.} High event recall can conceal excessive false activations, while conservative yielding policies can appear reliable despite missing genuine interruptions. MPTB demonstrates the importance of evaluation breakdowns across speaker counts and interaction regimes. Such diagnostics expose failure modes obscured by aggregate scores and provide actionable targets for developing more selective full-duplex systems. 

\noindent\textbf{Synthetic to real evaluation gap.} Table \ref{tab:real_vs_synthetic} illustrates that this difficulty is not confined to synthetic speech: VAP recovers only 35.9\% of interruptions on real four-party AMI meetings at 5.3\% FPR, despite in-slice threshold tuning. This supports corroborating learnings from synthetic evaluations to real-world relevance of the interruption challenge.

\section{Conclusion}
\label{sec:conclusion}

We introduced \mptb, a controlled benchmark for turn taking dynamics for end-of-turn and interruption detection in multi-party spoken dialogues. Existing systems detect turn completion reliably but struggle to recognize floor takeovers without excessive false activations, with failures varying across conversational regimes and speaker counts. This asymmetry extends to full-duplex agents where all systems perform near chance on selective yielding. \texttt{MultiPartyTurnBench} establishes selective interruption handling as a critical unresolved challenge for reliable multi-party voice interaction.

\section{Limitations}
\label{sec:limitations}

\noindent\textbf{Synthetic speech and external validity.}
MPTB's controlled construction provides exact floor-state annotations and reproducible interaction conditions, but synthesized speech cannot fully capture the prosody, disfluencies, timing adaptations, and ambiguous floor negotiations of spontaneous conversation. The synthetic-to-real comparison (Table~\ref{tab:real_vs_synthetic}) reveals non-uniform difficulty: learned endpointers generally achieve higher \eot\ recall on MPTB, whereas VAP detects fewer interruptions on MPTB (11.7\%) than on real four-party AMI (35.9\%). Thus, synthetic evaluation does not uniformly overestimate performance, and MPTB scores should be interpreted as controlled diagnostics rather than estimates of natural-conversation error rates.

\noindent\textbf{Scope of real-speech validation.}
Our multi-party validation uses 15 four-speaker AMI meetings with floor events derived from manual speech segments rather than independently annotated under the MPTB taxonomy. The evaluated corpora also differ in speaker count, conversational domain, acoustics, annotation protocols, and threshold calibration, with real-audio slices using in-slice tuning. Cross-corpus differences therefore cannot be attributed causally to speaker count or synthetic speech. AMI provides complementary evidence that interruption detection remains challenging in natural multi-party conversation, but does not establish direct synthetic-to-real performance transfer.

\noindent\textbf{Linguistic and demographic coverage.}
MPTB is restricted to English, with accent diversity limited to the voices sampled during synthesis rather than representative real-world speaker populations. We do not systematically evaluate generalization across languages, regional accents, or speaker demographics. Turn-taking conventions, prosodic cues, and interruption behaviors may vary across linguistic and sociocultural contexts. Consequently, our findings should not be interpreted as universal patterns of conversational floor control; broader multilingual and demographically diverse real-speech evaluation is necessary to establish their generality.

\section*{Ethics Statement}
The released MPTB corpus contains synthetic speech and fictional LLM-generated dialogue rather than newly collected private conversations. The benchmark is designed to diagnose conversational control failures and should not be interpreted as a safety certification for deployment. Real TurnBench material used for scorer validation is not redistributed, and AMI is used under its existing research terms. Because full-duplex systems may be used in multi-user or high-stakes settings, benchmark performance should be considered alongside broader evaluations of recognition quality, safety, fairness, privacy, and task correctness.

\bibliography{references}

@article{sacks1974,
  author  = {Sacks, Harvey and Schegloff, Emanuel A. and Jefferson, Gail},
  title   = {A Simplest Systematics for the Organization of Turn-Taking for Conversation},
  journal = {Language},
  volume  = {50},
  number  = {4},
  pages   = {696--735},
  year    = {1974}
}

@article{duncan1972,
  author  = {Duncan, Starkey},
  title   = {Some Signals and Rules for Taking Speaking Turns in Conversations},
  journal = {Journal of Personality and Social Psychology},
  volume  = {23},
  number  = {2},
  pages   = {283--292},
  year    = {1972}
}

@inproceedings{yngve1970,
  author    = {Yngve, Victor H.},
  title     = {On Getting a Word in Edgewise},
  booktitle = {Papers from the Sixth Regional Meeting, Chicago Linguistic Society},
  pages     = {567--578},
  year      = {1970}
}

@article{turnbench,
  author  = {Jiang, Freeman and others},
  title   = {{TurnBench}: A Multi-Domain Benchmark for Turn-Taking Dynamics in Spoken Dialogue},
  journal = {arXiv preprint arXiv:2608.25218},
  year    = {2026}
}

@inproceedings{talkingturns,
  author    = {Arora, Siddhant and Lu, Zhenyu and Chiu, Chung-Cheng and Pang, Ruoming and Watanabe, Shinji},
  title     = {Talking Turns: Benchmarking Audio Foundation Models on Turn-Taking Dynamics},
  booktitle = {International Conference on Learning Representations (ICLR)},
  year      = {2025}
}

@inproceedings{fdb1,
  author    = {Lin, Guan-Ting and others},
  title     = {{Full-Duplex-Bench}: A Benchmark to Evaluate Full-Duplex Spoken Dialogue Models on Turn-Taking Capabilities},
  booktitle = {IEEE Automatic Speech Recognition and Understanding Workshop (ASRU)},
  year      = {2025}
}

@inproceedings{fdb2,
  author    = {Lin, Guan-Ting and others},
  title     = {{Full-Duplex-Bench-v2}: A Multi-Turn Evaluation Framework for Duplex Dialogue Systems with an Automated Examiner},
  booktitle = {Proceedings of the Annual Meeting of the Association for Computational Linguistics (ACL), Short Papers},
  year      = {2026}
}

@article{fdb3,
  author  = {Lin, Guan-Ting and Chen, Chien-Yu and Chen, Zhehuai and Lee, Hung-yi},
  title   = {{Full-Duplex-Bench-v3}: Benchmarking Tool Use for Full-Duplex Voice Agents under Real-World Disfluency},
  journal = {arXiv preprint arXiv:2604.04847},
  year    = {2026}
}

@inproceedings{triadicvap,
  author    = {{Triadic-VAP}},
  title     = {Triadic Voice Activity Projection for Three-Party Conversation},
  booktitle = {Proceedings of Interspeech},
  note      = {arXiv:2507.07518},
  year      = {2025}
}

@article{fastcareful,
  author  = {{Fast When, Careful Who}},
  title   = {Fast When, Careful Who: Dual-Process End-of-Turn and Next-Speaker Prediction for Multi-Party Dialogue},
  journal = {arXiv preprint arXiv:2606.16568},
  year    = {2026}
}

@inproceedings{ami,
  author    = {Carletta, Jean and others},
  title     = {The {AMI} Meeting Corpus: A Pre-Announcement},
  booktitle = {Machine Learning for Multimodal Interaction},
  series    = {Lecture Notes in Computer Science},
  volume    = {3869},
  pages     = {28--39},
  publisher = {Springer},
  year      = {2006}
}

@misc{silero,
  author       = {{Silero Team}},
  title        = {{Silero VAD}: Pre-trained Enterprise-grade Voice Activity Detector (VAD), Number Detector and Language Classifier},
  year         = {2024},
  publisher    = {GitHub},
  howpublished = {GitHub repository},
  note         = {Available: https://github.com/snakers4/silero-vad}
}

@misc{smartturn,
  author       = {{Pipecat AI}},
  title        = {{Smart Turn v3}: Native-Audio Semantic Turn Detection},
  year         = {2025},
  howpublished = {Open-source model and software},
  note         = {Available: https://github.com/pipecat-ai/smart-turn}
}

@article{wavlm,
  author  = {Sanyuan Chen and Chengyi Wang and Zhengyang Chen and Yu Wu and Shujie Liu and Zhuo Chen and Jinyu Li and Naoyuki Kanda and Takuya Yoshioka and Xiong Xiao and Jian Wu and Long Zhou and Shuo Ren and Yanmin Qian and Yao Qian and Jian Wu and Michael Zeng and Xiangzhan Yu and Furu Wei},
  title   = {{WavLM}: Large-Scale Self-Supervised Pre-Training for Full Stack Speech Processing},
  journal = {IEEE Journal of Selected Topics in Signal Processing},
  volume  = {16},
  number  = {6},
  pages   = {1505--1518},
  year    = {2022},
  month   = oct,
  doi     = {10.1109/JSTSP.2022.3188113}
}

@inproceedings{vap,
  author    = {Erik Ekstedt and Gabriel Skantze},
  title     = {Voice Activity Projection: Self-Supervised Learning of Turn-Taking Events},
  booktitle = {Proc. Interspeech},
  pages     = {5190--5194},
  year      = {2022},
  doi       = {10.21437/Interspeech.2022-10955}
}

@inproceedings{whisper,
  title={Robust speech recognition via large-scale weak supervision},
  author={Radford, Alec and Kim, Jong Wook and Xu, Tao and Brockman, Greg and McLeavey, Christine and Sutskever, Ilya},
  booktitle={International conference on machine learning},
  pages={28492--28518},
  year={2023},
  organization={PMLR}
}

@misc{openai2026gptlive,
  title        = {Introducing {GPT-Live}},
  author       = {{OpenAI}},
  year         = {2026},
  month        = jul,
  howpublished = {\url{https://openai.com/index/introducing-gpt-live/}},
  note         = {Published July 8, 2026}
}

@article{lin2026stepaudio,
  title={StepAudio 3 Realtime Technical Report},
  author={Lin, Bin and Zhao, Bo and Zhang, Boyang and Wu, Boyong and Yan, Chao and Geng, Chen and Wu, Chen and Yi, Cheng and Feng, Chengli and Zhu, Chenglin and others},
  journal={arXiv preprint arXiv:2609.14005},
  year={2026}
}

@misc{google2026gemini38live,
  author = {{Google}},
  title = {Gemini 3.8 Live},
  year = {2026},
  month = sep,
  howpublished = {Google AI for Developers, \url{https://ai.google.dev/gemini-api/docs/models/gemini-3.8-live}},
  note = {Last updated September 15, 2026}
}

@article{cui2026minicpm,
  title={Minicpm-o 4.5: Towards real-time full-duplex omni-modal interaction},
  author={Cui, Junbo and Xu, Bokai and Wang, Chongyi and Yu, Tianyu and Sun, Weiyue and Xu, Yingjing and Wang, Tianran and He, Zhihui and Ma, Wenshuo and Cai, Tianchi and others},
  journal={arXiv preprint arXiv:2604.27393},
  year={2026}
}

@article{zhao2026realtime,
  title={Realtime-Venus: A full-duplex interaction system with asynchronous delegation},
  author={Zhao, Ruixiang and Wang, Hualei and Sun, Renhe and Zhou, Enzhi and Wu, Jincenzi and Song, Xujie and Shi, Kexin and Liu, Zihang and Zhu, Pengcheng and Zhou, Jiayi and others},
  journal={arXiv preprint arXiv:2609.13814},
  year={2026}
}

@misc{webrtcvad,
  author       = {{Google WebRTC Project}},
  title        = {{WebRTC Voice Activity Detector}},
  howpublished = {Open-source software},
  url          = {https://github.com/wiseman/py-webrtcvad},
  note         = {Python interface to the WebRTC VAD implementation}
}

@misc{kyutai_svad,
  author       = {{Kyutai}},
  title        = {{Kyutai Streaming Speech-to-Text with Semantic Voice Activity Detection}},
  howpublished = {Model and software release},
  url          = {https://huggingface.co/kyutai/stt-1b-en_fr-candle},
  note         = {Semantic VAD head of the stt-1b-en\_fr model;
                  implementation: https://github.com/kyutai-labs/delayed-streams-modeling}
}

@misc{duplexmpe,
  title        = {{Duplex-MPE}: Benchmarking Multi-Party Interaction
                  in Full-Duplex Dialogue},
  author       = {Ma, Chengqian and Feng, Wenhao and Jin, Weixuan
                  and Dai, Gaole and Xie, Tianyu and Ma, Yuexiao
                  and Kang, Zhaolu and Zhao, Xiangyu
                  and Zheng, Xiawu and Chao, Fei},
  year         = {2026},
  eprint       = {2609.31948},
  archivePrefix = {arXiv},
  url          = {https://arxiv.org/abs/2609.31948}
}

@misc{multitalkbench,
  title        = {{MultiTalk}: Scaling Full-Duplex Speech Models
                  to Long, Multi-Party, Bilingual Conversation},
  author       = {Wang, Ke and Ren, Houxing and Lu, Zimu
                  and Yang, Yunqiao and Zong, Zhuofan
                  and Zhan, Mingjie and Li, Hongsheng},
  year         = {2026},
  eprint       = {2609.36903},
  archivePrefix = {arXiv},
  url          = {https://arxiv.org/abs/2609.36903}
}

@misc{msibench,
  title        = {{MSI-Bench}: Evaluating Multi-Speaker Voice
                  Interaction for Collaborative AI Agents},
  author       = {Xiong, Chenxu and Shen, Dongming
                  and Tang, Yuzhi and Ma, Wentao
                  and Li, Mu and Smola, Alex},
  year         = {2026},
  eprint       = {2609.24812},
  archivePrefix = {arXiv},
  url          = {https://arxiv.org/abs/2609.24812}
}

@misc{sidbench,
  title        = {Semantic-Aware Interruption Detection in Spoken Dialogue Systems: Benchmark, Metric, and Model},
  author       = {Xia, Kangxiang and Mu, Bingshen and Shi, Xian and Xu, Jin and Xie, Lei},
  year         = {2026},
  eprint       = {2603.24144},
  archivePrefix = {arXiv},
  primaryClass = {cs.SD},
  url          = {https://arxiv.org/abs/2603.24144}
}

@inproceedings{fdbench,
  title     = {{FD-Bench}: A Full-Duplex Benchmarking Pipeline Designed for Full Duplex Spoken Dialogue Systems},
  author    = {Peng, Yizhou and Chao, Yi-Wen and Ng, Dianwen and Ma, Yukun and Ni, Chongjia and Ma, Bin and Chng, Eng Siong},
  booktitle = {Proceedings of Interspeech},
  year      = {2025},
  pages     = {176--180},
  doi       = {10.21437/Interspeech.2025-739}
}

@article{muVAP,
  title={MuVAP: Multimodal Multiparty Voice Activity Projection for Turn-taking Prediction in the Wild},
  author={Haotian Qi and Gabriel Skantze},
  journal={ArXiv},
  year={2026},
  volume={abs/2606.16731},
  url={https://api.semanticscholar.org/CorpusID:289301754}
}

@article{chung2020spot,
  title={Spot the conversation: speaker diarisation in the wild},
  author={Chung, Joon Son and Huh, Jaesung and Nagrani, Arsha and Afouras, Triantafyllos and Zisserman, Andrew},
  booktitle={Interspeech},
  year={2020}
}

@misc{zhao2026echo,
  title = {Same Words, Different Actions: Paired Turn-Taking Evaluation under Rewritten Dialogue Contexts},
  author = {Zhao, Shuofeng and Cai, Hongwei and Fan, Wenke and Guo, Qingxiang and Wang, Zhou and Yang, Dawei and Zhou, Zhiyang and Shang, Yingxin and Wang, Weixu and Yang, Lin and Zhou, Shuran and Song, Yang},
  year = {2026},
  eprint = {2609.17360},
  archivePrefix = {arXiv},
  primaryClass = {cs.CL},
  url = {https://arxiv.org/abs/2609.17360}
}

@misc{dualturn,
      title={DualTurn: Learning Turn-Taking from Dual-Channel Generative Speech Pretraining},
      author={Shangeth Rajaa},
      year={2026},
      eprint={2603.08216},
      archivePrefix={arXiv},
      primaryClass={eess.AS},
      url={https://arxiv.org/abs/2603.08216},
}

@misc{turn1mini,
  title  = {turn-1-mini: a 5M-parameter streaming end-of-turn model},
  author = {p99lab},
  year   = {2026},
  url    = {https://p99lab.com}
}

@inproceedings{espnet,
title={Talking Turns: Benchmarking Audio Foundation Models on Turn-Taking Dynamics},
author={Siddhant Arora and Zhiyun Lu and Chung-Cheng Chiu and Ruoming Pang and Shinji Watanabe},
booktitle={The Thirteenth International Conference on Learning Representations},
year={2025},
url={https://openreview.net/forum?id=2e4ECh0ikn}
}

@inproceedings{mimi,
  title={Streaming endpointer for spoken dialogue using neural audio codecs and label-delayed training},
  author={Udupa, Sathvik and Watanabe, Shinji and Schwarz, Petr and Cernocky, Jan},
  booktitle={2025 IEEE Automatic Speech Recognition and Understanding Workshop (ASRU)},
  pages={1--8},
  year={2025},
  organization={IEEE}
}

@software{kokoro_tts,
  author = {hexgrad},
  title  = {Kokoro-82M: Open-weight TTS Model},
  year   = {2024},
  url    = {https://huggingface.co/hexgrad/Kokoro-82M}
}

@misc{musan2015,
  author    = {David Snyder and Guoguo Chen and Daniel Povey},
  title     = {{MUSAN}: {A} {M}usic, {S}peech, and {N}oise {C}orpus},
  year      = {2015},
  eprint    = {1510.08484},
  note      = {arXiv:1510.08484v1}
}
\clearpage
\appendix

\providecolor{mptbBlue}{RGB}{33,102,172}
\providecolor{mptbOrange}{RGB}{217,95,2}
\providecolor{mptbGreen}{RGB}{27,120,55}
\providecolor{mptbRed}{RGB}{178,24,43}
\providecolor{mptbViolet}{RGB}{118,42,131}
\providecolor{mptbTeal}{RGB}{1,102,94}
\providecolor{mptbSlate}{RGB}{70,80,95}
\providecolor{mptbBudgetGray}{RGB}{240,243,246}
\providecolor{mptbBestCyan}{RGB}{217,242,247}
\providecolor{mptbSoftYellow}{RGB}{255,246,213}
\makeatletter
\@ifpackageloaded{tcolorbox}{%
  \newtcolorbox{mptbbox}[2][mptbBlue]{colback=#1!4!white,colframe=#1!80!black,
    colbacktitle=#1!80!black,coltitle=white,fonttitle=\bfseries\footnotesize,
    title={#2},fontupper=\footnotesize,boxrule=0.6pt,arc=1.2mm,
    left=1.4mm,right=1.4mm,top=1mm,bottom=1mm,toptitle=0.6mm,bottomtitle=0.6mm,
    before skip=7pt,after skip=7pt}%
}{%
  \newsavebox{\mptb@sbox}%
  \newenvironment{mptbbox}[2][mptbBlue]{%
    \def\mptb@col{#1}%
    \begin{lrbox}{\mptb@sbox}\begin{minipage}{\dimexpr\linewidth-2\fboxsep-2\fboxrule\relax}%
    \footnotesize\colorbox{#1!80!black}{\parbox{\dimexpr\linewidth-2\fboxsep\relax}{\color{white}\bfseries #2}}\par\smallskip}%
    {\end{minipage}\end{lrbox}\par\smallskip\noindent
     \fcolorbox{\mptb@col!80!black}{\mptb@col!4!white}{\usebox{\mptb@sbox}}\par\smallskip}%
}
\makeatother
\newcommand{\mptbU}[4]{\par\noindent\hangindent=1.4em\hangafter=1
  \textbf{\textcolor{#1}{#2}}\,{\sffamily\tiny\textcolor{black!55}{[#3]}}~#4}
\newcommand{\mptbT}[4]{\par\noindent\hangindent=1.4em\hangafter=1
  \colorbox{mptbSoftYellow}{\textcolor{mptbRed}{$\triangleright$}\,\textbf{\textcolor{#1}{#2}}\,{\sffamily\tiny\textcolor{black!70}{[#3]}}}~\textit{#4}}
\newcommand{\mptbMeta}[1]{\par\noindent{\sffamily\scriptsize\textcolor{black!60}{#1}}\par\smallskip}
\newcommand{\mptbOK}{\tikz[baseline=-0.55ex]\fill[mptbGreen] (0,0) circle (0.55ex);}
\newcommand{\mptbBad}{\tikz[baseline=-0.55ex]\draw[mptbRed,line width=0.9pt] (-0.5ex,-0.5ex)--(0.5ex,0.5ex) (-0.5ex,0.5ex)--(0.5ex,-0.5ex);}
\newcommand{\mptbOut}[3]{\mbox{#1\,#2{\,\scriptsize\textcolor{black!55}{#3}}}}
\newcommand{\mptbTake}[1]{\par\smallskip\noindent\textcolor{black!75}{\textbf{Reading.}~#1}}
\newcommand{\mptbXX}{\textbf{[X]}}

\section{Human Evaluation Study}
\label{app:human_eval}

\subsection{Human Annotation}
\label{sec:human_annotation}

Although MPTB derives floor-state labels before speech synthesis, a procedurally valid event may not be perceived as intended in the rendered audio. We therefore conduct a human validation study assessing whether listeners recognize the intended event categories and their temporal realization. The study covers all 240 conversations and 6,116 candidate events, with three independent annotations per event (18,348 primary categorical judgments, excluding blind repeats).

\paragraph{Annotators and qualification.}
Annotators have native or near-native English comprehension and complete written guidelines, worked audio examples, and a held-out calibration set emphasizing turn completion versus continuation, backchannels versus floor claims, and successful versus unsuccessful interruptions. Qualification requires at least 75\% agreement with expert labels across 10 calibration examples. Annotators are blinded to generator labels, event logs, and model predictions.

\paragraph{Annotation procedure.}
Annotators inspect candidate regions with 10\,s of preceding and following context using synchronized multi-speaker audio, independently controllable speaker channels, and a shared timeline. For each region, they assign an event category, identify the perceived floor holder before and after the event, and mark onset and offset timestamps. An \textsc{ambiguous} label is permitted when the floor outcome cannot be determined reliably. Automatic transcripts are available as references, but audio takes precedence. Table~\ref{tab:annotation_schema} summarizes the annotation fields.

Annotations are collected independently without discussion or adjudication. Examples are presented in randomized order across interaction conditions, speaker counts, and voices. A subset of 25\% of examples is repeated blindly to assess intra-annotator consistency. Since listeners receive subsequent conversational context, this study validates the perceptual realization of floor transitions rather than their causal predictability at event onset.

\begin{table}[t]
\centering
\scriptsize
\setlength{\tabcolsep}{3.5pt}
\renewcommand{\arraystretch}{1.12}
\begin{tabular}{@{}p{2.15cm}p{5.25cm}@{}}
\toprule
\textbf{Field} & \textbf{Values recorded per candidate region} \\
\midrule
Event category & \textsc{turn-yield}, \textsc{turn-hold}, \textsc{interruption}, \textsc{non-floor-taking int.}, \textsc{backchannel}, \textsc{noncontent}, \textsc{ambiguous} \\
Floor holder (pre) & Perceived speaker ID immediately before the event ($1,\ldots,N$) \\
Floor holder (post) & Perceived speaker ID immediately after the event ($1,\ldots,N$) \\
Onset / offset & Timestamps (s) on the shared timeline \\
Stimulus & Noisy speaker channels (24\,kHz), solo/mix controls, speaker timeline, reference transcript \\
Hidden information & Generator label, event log, model predictions, condition metadata \\
\bottomrule
\end{tabular}
\caption{\small\textbf{Human annotation schema.} Annotators independently identify event categories, floor holders, and temporal boundaries using the role-relative taxonomy (Table~\ref{tab:event_taxonomy_app}). \textsc{Noncontent} corresponds to listener laughter in the released corpus.}
\label{tab:annotation_schema}
\end{table}

\subsection{Consensus and Agreement Measures}
\label{sec:agreement}

We construct human consensus independently of the generator event log. A candidate receives a majority-consensus label when at least two of three annotators assign the same event category. Candidates without categorical consensus are treated as unresolved for human-validation analysis. For agreed events, the consensus boundary is the median timestamp of the agreeing annotators.

We distinguish three complementary measures. \textbf{Categorical consensus} is the fraction of candidate events receiving a majority label, while Fleiss' $\kappa$ measures inter-annotator categorical agreement corrected for chance. \textbf{Boundary F1} measures temporal agreement using a 500\,ms boundary tolerance. Finally, \textbf{generator--human agreement} measures whether the procedurally assigned event label matches the independent human consensus. Fleiss' $\kappa$ is calculated using a one-versus-rest measure over the complete corpus. Boundary F1 temporal agreement is calculated relative to consensus boundaries. 

Let $\mathcal{C}$ denote events with categorical consensus, and let $y_i^{\mathrm{gen}}$ and $y_i^{\mathrm{human}}$ be the generator and consensus labels. Generator--human agreement is
\begin{equation}
A_{\mathrm{gen}} =
\frac{1}{|\mathcal{C}|}
\sum_{i\in\mathcal{C}}
\mathbf{1}\left[y_i^{\mathrm{gen}}=y_i^{\mathrm{human}}\right].
\label{eq:gen_agreement}
\end{equation}
Thus, generator--human agreement is conditional on human consensus and should not be interpreted as accuracy over all generated events.

\subsection{Human Validation Results}

\paragraph{Floor-event labels are perceptually distinguishable, but not equally unambiguous.}
Table~\ref{tab:annotation_agreement} reports agreement across the six generated event categories. Backchannels show the strongest consistency, with 93.6\% categorical consensus, Fleiss' $\kappa=0.92$, and boundary F1 of 0.95. Successful interruptions exhibit lower consensus (81.4\%) and temporal agreement (F1 $=0.89$), indicating greater uncertainty in identifying the realization and timing of floor takeovers. Turn endings and mid-turn holds also exhibit lower categorical consensus than backchannels, supporting the distinction between acoustic speech cessation and perceived floor relinquishment.

\paragraph{Generator--human agreement supports label plausibility, with important qualifications.}
Among events receiving human consensus, generator--human agreement ranges from 86.5\% for turn endings to 97.8\% for backchannels. Successful interruptions achieve 91.2\% generator--human agreement despite having the lowest categorical consensus (81.4\%). These measures capture different properties: consensus reflects consistency across annotators, whereas generator--human agreement evaluates label correspondence only on retained events. High conditional agreement therefore supports the plausibility of recognized synthetic events but does not establish that every generated event is perceptually unambiguous.

\begin{table*}[t]
\centering
\small
\setlength{\tabcolsep}{5.0pt}
\renewcommand{\arraystretch}{1.10}
\resizebox{\textwidth}{!}{%
\begin{tabular}{lccccc}
\toprule
\textbf{Event Class} &
\textbf{\# Events} &
\textbf{Consensus (\%)} &
\textbf{Fleiss' $\kappa$} &
\textbf{Boundary F1} &
\textbf{Gen.--Human Match (\%)} \\
\midrule
TURN / Yield & 3,422 & 83.4 & 0.80 & 0.94 & 86.5 \\
TURN / Hold & 241 & 85.6 & 0.84 & 0.93 & 87.3 \\
INTERRUPTION & 367 & 81.4 & 0.78 & 0.89 & 91.2 \\
NON-FLOOR-TAKING INT. & 605 & 88.6 & 0.82 & 0.91 & 92.5 \\
BACKCHANNEL & 1,176 & 93.6 & 0.92 & 0.95 & 97.8 \\
NONCONTENT & 305 & 90.5 & 0.83 & 0.90 & 94.5 \\
\bottomrule
\end{tabular}%
}
\caption{\small\textbf{Human validation of MPTB floor-state events.} Event counts cover all 240 conversations, with three independent annotator judgments per candidate. Consensus denotes majority categorical agreement; Fleiss' $\kappa$ measures chance-corrected inter-annotator agreement; boundary F1 measures temporal agreement within 500\,ms; and generator--human match is conditional on consensus (Eq.~\ref{eq:gen_agreement}). \textsc{Noncontent} denotes listener laughter. Non-floor-taking interruption attempts are included in human validation but excluded from causal INT scoring. Pooled agreement statistics require event-level recomputation and are not estimated by averaging class-level coefficients.}
\label{tab:annotation_agreement}
\end{table*}

\paragraph{Human agreement decreases with speaker count.}
Table~\ref{tab:agreement_by_n} shows a consistent decline across all three agreement measures as the number of speakers increases. Categorical consensus falls from 87.8\% at $N=2$ to 80.1\% at $N=5$, while Fleiss' $\kappa$ decreases from 0.84 to 0.76 and boundary F1 from 0.96 to 0.90. The simultaneous reduction in categorical and temporal agreement suggests that resolving floor transitions becomes less consistent across annotators in larger groups. This is a descriptive association rather than an isolated causal effect of speaker count, since realized event composition and acoustic conditions may also differ.

\begin{table}[t]
\centering
\small
\setlength{\tabcolsep}{4.2pt}
\renewcommand{\arraystretch}{1.10}
\begin{tabular}{@{}crrccc@{}}
\toprule
& \multicolumn{2}{c}{\textbf{Corpus}} & \multicolumn{3}{c}{\textbf{Human Validation}} \\
\cmidrule(lr){2-3}\cmidrule(l){4-6}
\textbf{$N$} & \textbf{Conv.} & \textbf{Events} & \textbf{Cons. (\%)} & \textbf{Fleiss' $\kappa$} & \textbf{Bnd. F1} \\
\midrule
2 & 60 & 1,490 & 89.8 & 0.84 & 0.96 \\
3 & 60 & 1,527 & 88.6 & 0.82 & 0.93 \\
4 & 60 & 1,536 & 84.3 & 0.81 & 0.91 \\
5 & 60 & 1,563 & 82.1 & 0.76 & 0.90 \\
\bottomrule
\end{tabular}
\caption{\small\textbf{Human agreement across speaker counts.} Event totals include the six generator-defined categories in Table~\ref{tab:annotation_agreement}. Categorical consensus, Fleiss' $\kappa$, and boundary F1 consistently decrease from two to five speakers.}
\label{tab:agreement_by_n}
\end{table}

\paragraph{Interpretation and scope.}
The human study provides complementary evidence that MPTB's generated floor-state categories have recognizable acoustic and conversational realizations, while identifying residual ambiguity in floor transitions, particularly successful interruptions. Since candidate regions are selected from the generator event log and annotators may inspect future context, these findings validate event-label plausibility rather than independent event discovery or real-time detectability. The study also does not establish that synthetic speech captures the full variability of natural multi-party interaction.


\begin{table*}[t]
\centering
\small
\setlength{\tabcolsep}{4.0pt}
\renewcommand{\arraystretch}{1.10}
\resizebox{\textwidth}{!}{%
\begin{tabular}{@{}lrrrrrrr@{}}
\toprule
\textbf{Interaction Type} &
\textbf{\# Conv.} &
\textbf{Hours} &
\textbf{Turns/min} &
\textbf{BC/min} &
\textbf{INT/min} &
\textbf{Overlap (\%)} &
\textbf{Mean Turn (s)} \\
\midrule
Casual & 40 & 1.10 & 12.7 & 4.24 & 0.92 & 11.4 & 4.81 \\
Task-Oriented & 40 & 1.03 & 11.4 & 3.80 & 0.89 & 9.9 & 5.27 \\
Instructional & 40 & 1.17 & 7.9 & 3.81 & 0.70 & 10.7 & 7.79 \\
Collaborative & 40 & 1.07 & 11.4 & 3.43 & 1.33 & 11.0 & 5.32 \\
Argumentative & 40 & 1.11 & 9.6 & 2.95 & 1.08 & 10.7 & 6.29 \\
Narrative & 40 & 1.41 & 6.6 & 3.34 & 0.53 & 9.1 & 9.22 \\
\midrule
\textbf{All} & \textbf{240} & \textbf{6.9} & \textbf{9.8} & \textbf{3.58} & \textbf{0.89} & \textbf{10.4} & \textbf{6.45} \\
\bottomrule
\end{tabular}%
}
\caption{\small\textbf{Corpus statistics by interaction type.} Turns/min counts floor-holding \textsc{turn} segments; BC/min includes backchannels and laughter; INT/min counts successful interruptions; overlap denotes the fraction of conversation time with simultaneous vocal activity from at least two speakers. Interaction types induce distinct conversational regimes despite balanced conversation counts.}
\label{tab:corpus_stats_type}
\end{table*}

\begin{table}[t]
\centering
\small
\setlength{\tabcolsep}{4.0pt}
\renewcommand{\arraystretch}{1.10}
\begin{tabular}{@{}lrrrrr@{}}
\toprule
\textbf{$N$} &
\textbf{\# Conv.} &
\textbf{EOT} &
\textbf{INT} &
\textbf{BC/min} &
\textbf{Overlap (\%)} \\
\midrule
2 & 60 & 853 & 80 & 3.12 & 8.9 \\
3 & 60 & 838 & 96 & 3.61 & 10.9 \\
4 & 60 & 875 & 94 & 3.60 & 10.9 \\
5 & 60 & 856 & 97 & 4.06 & 11.1 \\
\midrule
\textbf{All} & \textbf{240} & \textbf{3,422} & \textbf{367} & \textbf{3.58} & \textbf{10.4} \\
\bottomrule
\end{tabular}
\caption{\small\textbf{Corpus composition by speaker count.} Each party-count condition contains 60 conversations. EOT and INT denote positive event anchors; overlap measures the percentage of conversation time containing simultaneous vocal activity.}
\label{tab:corpus_stats_n}
\end{table}

\begin{table}[t]
\centering
\small
\setlength{\tabcolsep}{4.5pt}
\renewcommand{\arraystretch}{1.10}
\begin{tabular}{@{}lrrrr@{}}
\toprule
\textbf{Split} &
\textbf{\# Conv.} &
\textbf{Hours} &
\textbf{\# Voices} &
\textbf{EOT / INT} \\
\midrule
Dev & 48 & 1.42 & 8 & 706 / 86 \\
Test & 192 & 5.47 & 16 & 2,716 / 281 \\
\midrule
\textbf{Total} & \textbf{240} & \textbf{6.89} & \textbf{24} & \textbf{3,422 / 367} \\
\bottomrule
\end{tabular}
\caption{\small\textbf{MPTB development and test partitions.} The partitions use disjoint TTS voices and retain the complete speaker-count $\times$ interaction-type design, with two development and eight test conversations per condition. EOT and INT denote positive event counts.}
\label{tab:data_split}
\end{table}

\section{MultiPartyTurnBench Evaluation Tracks}
\label{app:tracks}

Table~\ref{tab:benchmark_tracks} summarizes the two benchmark tracks and their positive and hard-negative events. This section gives the exact anchor derivation and scoring rules behind every number in the paper.

\begin{table*}[t]
\centering
\small
\setlength{\tabcolsep}{5pt}
\renewcommand{\arraystretch}{1.12}
\begin{tabular}{p{1.4cm} p{4.4cm} p{4.4cm} p{2.6cm}}
\toprule
\textbf{Track} & \textbf{Positive: floor changes} & \textbf{Hard negative: floor unchanged} & \textbf{Prediction anchor} \\
\midrule
\textbf{EOT} & Current floor holder completes a turn and yields the floor. & Mid-turn pause after which the same speaker resumes while retaining the floor. & End of floor-holding speech \\
\textbf{INT} & Listener begins speaking while another speaker holds the floor and successfully takes ownership. & Backchannel or non-content listener activity that does not change floor ownership. & Onset of floor-taking speech \\
\bottomrule
\end{tabular}
\caption{\small{\textbf{\texttt{MultiPartyTurnBench} evaluation tracks.} Each track contrasts a floor-changing event with a conversationally similar event that leaves floor ownership unchanged. This paired formulation prevents systems from succeeding through indiscriminate silence or speech-activity detection.}}
\label{tab:benchmark_tracks}
\end{table*}

\subsection{Anchor Derivation}
\label{app:anchors}
Scoring anchors are derived deterministically from the canonical event log, separately for every speaker channel $s$:
\begin{itemize}\setlength{\itemsep}{1pt}\setlength{\parskip}{0pt}
\item \textbf{\eot\ positives}: \textsc{turn} events of $s$ with outcome \textit{yield}; anchor $t$ = segment offset. The last turn of a conversation is a yield.
\item \textbf{\eot\ hard negatives}: \textsc{turn} events of $s$ with outcome \textit{hold}; scored span = [pause onset, same-speaker resume onset].
\item \textbf{\eot\ excluded}: \textsc{turn} events of $s$ with outcome \textit{interrupted}; a $\pm0.25$\,s neighborhood around the forced stop is neither rewarded nor penalized.
\item \textbf{\intr\ positives}: \textsc{claim\_floor} events of $s$ resolved as \textsc{interruption}; anchor $t$ = claim onset on the interrupter's own channel.
\item \textbf{\intr\ hard negatives}: \textsc{vocalize\_no\_claim} events of $s$ (backchannel or laughter); scored span = [onset, offset].
\item \textbf{\intr\ excluded}: \textsc{claim\_floor} events of $s$ resolved as \textsc{non-floor-taking interruption}; predictions inside [onset, offset] are discarded before matching.
\end{itemize}
Acoustic nuisance events (ambient bed, MUSAN sound effects, distant third-party speech) never generate anchors. Predictions they trigger outside scored spans are ignored, so they affect a system only when they fall inside a negative span or pre-empt a positive.

\subsection{Matching, Metrics and Operating Points}
\label{app:matching}
Each system submits, per speaker channel, causal decision times for each track. Decision times are rising edges of a score crossing threshold $\theta$, followed by a 2\,s refractory period. Matching follows TurnBench~\citep{turnbench}: (1) predictions inside excluded spans are dropped; (2) gold positives are visited in time order and each is matched to the earliest unclaimed prediction in $[t-0.25,\,t+3.0]$\,s (TP, else FN); (3) each hard-negative span counts as one FP if any remaining unclaimed prediction falls inside it, otherwise one TN. Signed latency is $t_{\text{pred}}-t_{\text{gold}}$ for matched pairs. Metrics are pooled over all speakers and conversations of a split.

For every system and track, $\theta$ is chosen on the development split as the value maximizing recall subject to $\mathrm{FPR}_{\text{dev}}\le0.10$. When no candidate qualifies, the candidate with the lowest development FPR is selected (ties broken by recall). The selected $\theta$ is frozen for the whole test split and for all subgroup analyses. A system is \emph{in budget} when $\mathrm{FPR}_{\text{dev}}\le0.10$ and $\mathrm{FPR}_{\text{test}}\le0.15$. Confidence intervals are percentile intervals from a conversation-clustered bootstrap ($B=10{,}000$, seed 20261010); paired system differences resample the same conversations for both systems. Table~\ref{tab:scored_denominators} lists the scored denominators.

\begin{table}[t]
\centering
\small
\setlength{\tabcolsep}{4pt}
\renewcommand{\arraystretch}{1.1}
\begin{tabular}{@{}lrrrrr@{}}
\toprule
& & \multicolumn{2}{c}{\textbf{\eot}} & \multicolumn{2}{c}{\textbf{\intr}} \\
\cmidrule(lr){3-4}\cmidrule(l){5-6}
\textbf{Split} & \textbf{Conv.} & \textbf{Pos.} & \textbf{Neg.} & \textbf{Pos.} & \textbf{Neg.} \\
\midrule
Dev & 48 & 706 & 55 & 86 & 309 \\
Test & 192 & 2,716 & 186 & 281 & 1,172 \\
\midrule
Total & 240 & 3,422 & 241 & 367 & 1,481 \\
\bottomrule
\end{tabular}
\caption{\small\textbf{Scored denominators} (positive anchors and hard-negative spans). Every test-split number in the paper is computed over these anchors. \intr\ negatives comprise 1,176 backchannels and 305 laughter vocalizations (test: 919 and 253); the 605 non-floor-taking claims are excluded intervals.}
\label{tab:scored_denominators}
\end{table}

\section{Benchmark Construction Details}
\label{app:construction}

\subsection{Role-relative event taxonomy}
\label{app:taxonomy}

Every event is defined relative to the current floor holder. Table~\ref{tab:event_taxonomy_app} gives the operational meaning of each class, its scoring role, and its realized frequency in the released corpus. Labels are produced by the generator (Stage~1 proposes the symbolic act; Stage~2 resolves its outcome) and are never inferred from audio.

\begin{table}[t]
\centering
\scriptsize
\setlength{\tabcolsep}{3pt}
\renewcommand{\arraystretch}{1.12}
\begin{tabular}{@{}>{\raggedright\arraybackslash}p{2.05cm}p{3.0cm}p{1.15cm}r@{}}
\toprule
\textbf{Event} & \textbf{Operational meaning} & \textbf{Role} & \textbf{Count} \\
\midrule
\textsc{turn}--yield & Holder releases the floor to another speaker (or the conversation ends). & \eot\ $+$ & 3,422 \\
\textsc{turn}--hold & Holder pauses mid-turn and resumes. & \eot\ $-$ & 241 \\
\textsc{turn}--interrupted & Holder is cut off by a successful interruption. & \eot\ excl. & 367 \\
\textsc{interruption} & Listener takes the floor before the current turn naturally ends. & \intr\ $+$ & 367 \\
\textsc{non-floor-taking int.} & Listener attempts a claim but the holder keeps the floor. & \intr\ excl. & 605 \\
\textsc{backchannel} & Listener vocalizes ($\le$3 words) without claiming the floor. & \intr\ $-$ & 1,176 \\
\textsc{noncontent} & Listener laughter without claiming the floor. & \intr\ $-$ & 305 \\
\bottomrule
\end{tabular}
\caption{\small\textbf{Role-relative event taxonomy} used by the generator and scorer, with counts over all 240 conversations. Acoustic nuisance (ambient bed, sound effects, distant speech) is rendered in Stage~4 and carries no label.}
\label{tab:event_taxonomy_app}
\end{table}


\subsection{Interaction-Type Definitions}
\label{app:interaction_types}

\begin{figure*}[t]
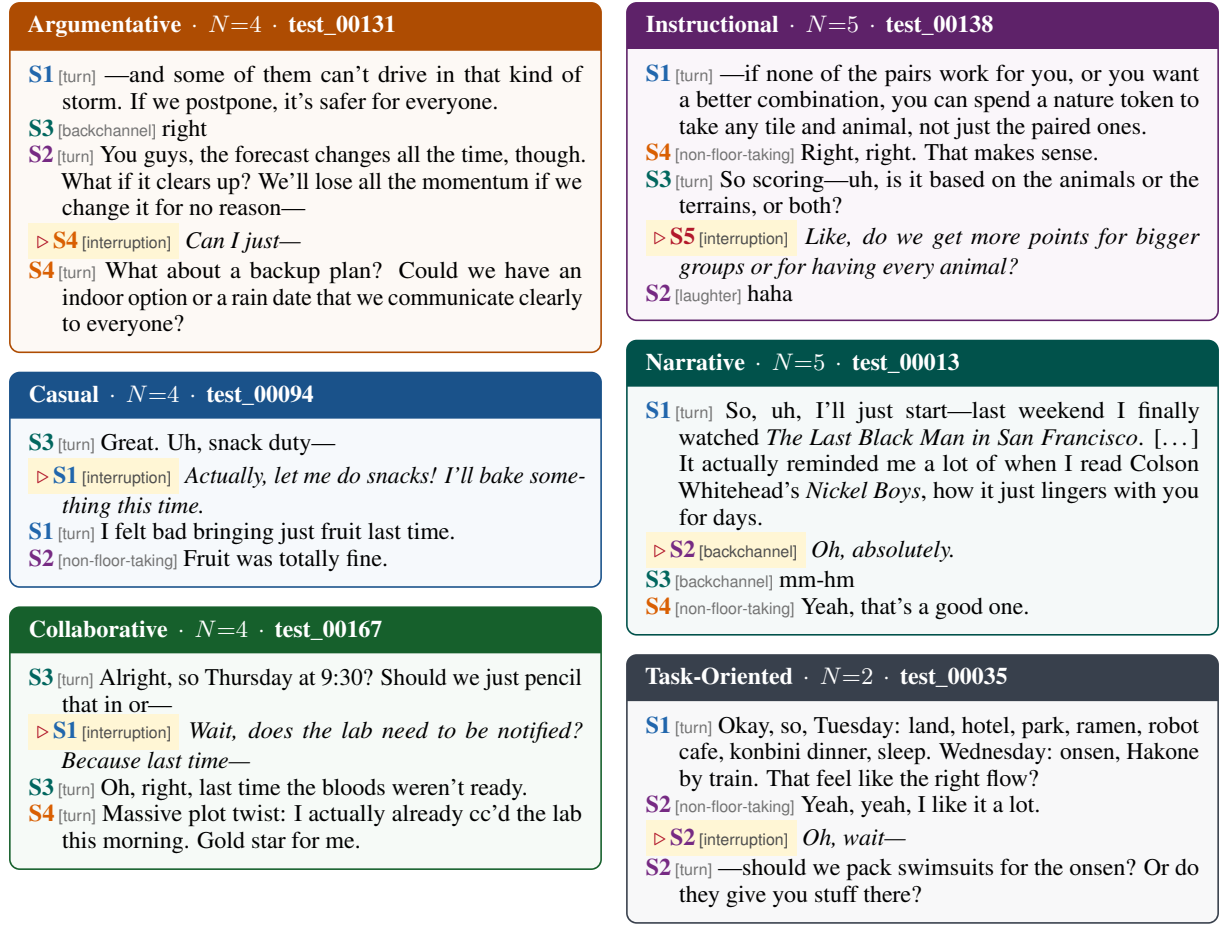

\centering
\begin{minipage}[t]{0.49\textwidth}
\vspace{0pt}
\begin{mptbbox}[mptbOrange]{Argumentative \,$\cdot$\, $N{=}4$ \,$\cdot$\, test\_00131}
\mptbU{mptbBlue}{S1}{turn}{---and some of them can't drive in that kind of storm. If we postpone, it's safer for everyone.}
\mptbU{mptbTeal}{S3}{backchannel}{right}
\mptbU{mptbViolet}{S2}{turn}{You guys, the forecast changes all the time, though. What if it clears up? We'll lose all the momentum if we change it for no reason---}
\mptbT{mptbOrange}{S4}{interruption}{Can I just---}
\mptbU{mptbOrange}{S4}{turn}{What about a backup plan? Could we have an indoor option or a rain date that we communicate clearly to everyone?}
\end{mptbbox}
\begin{mptbbox}[mptbBlue]{Casual \,$\cdot$\, $N{=}4$ \,$\cdot$\, test\_00094}
\mptbU{mptbTeal}{S3}{turn}{Great. Uh, snack duty---}
\mptbT{mptbBlue}{S1}{interruption}{Actually, let me do snacks! I'll bake something this time.}
\mptbU{mptbBlue}{S1}{turn}{I felt bad bringing just fruit last time.}
\mptbU{mptbViolet}{S2}{non-floor-taking}{Fruit was totally fine.}
\end{mptbbox}
\begin{mptbbox}[mptbGreen]{Collaborative \,$\cdot$\, $N{=}4$ \,$\cdot$\, test\_00167}
\mptbU{mptbTeal}{S3}{turn}{Alright, so Thursday at 9:30? Should we just pencil that in or---}
\mptbT{mptbBlue}{S1}{interruption}{Wait, does the lab need to be notified? Because last time---}
\mptbU{mptbTeal}{S3}{turn}{Oh, right, last time the bloods weren't ready.}
\mptbU{mptbOrange}{S4}{turn}{Massive plot twist: I actually already cc'd the lab this morning. Gold star for me.}
\end{mptbbox}
\end{minipage}\hfill
\begin{minipage}[t]{0.49\textwidth}
\vspace{0pt}
\begin{mptbbox}[mptbViolet]{Instructional \,$\cdot$\, $N{=}5$ \,$\cdot$\, test\_00138}
\mptbU{mptbBlue}{S1}{turn}{---if none of the pairs work for you, or you want a better combination, you can spend a nature token to take any tile and animal, not just the paired ones.}
\mptbU{mptbOrange}{S4}{non-floor-taking}{Right, right. That makes sense.}
\mptbU{mptbTeal}{S3}{turn}{So scoring---uh, is it based on the animals or the terrains, or both?}
\mptbT{mptbRed}{S5}{interruption}{Like, do we get more points for bigger groups or for having every animal?}
\mptbU{mptbViolet}{S2}{laughter}{haha}
\end{mptbbox}
\begin{mptbbox}[mptbTeal]{Narrative \,$\cdot$\, $N{=}5$ \,$\cdot$\, test\_00013}
\mptbU{mptbBlue}{S1}{turn}{So, uh, I'll just start---last weekend I finally watched \textit{The Last Black Man in San Francisco}. [\ldots] It actually reminded me a lot of when I read Colson Whitehead's \textit{Nickel Boys}, how it just lingers with you for days.}
\mptbT{mptbViolet}{S2}{backchannel}{Oh, absolutely.}
\mptbU{mptbTeal}{S3}{backchannel}{mm-hm}
\mptbU{mptbOrange}{S4}{non-floor-taking}{Yeah, that's a good one.}
\end{mptbbox}
\begin{mptbbox}[mptbSlate]{Task-Oriented \,$\cdot$\, $N{=}2$ \,$\cdot$\, test\_00035}
\mptbU{mptbBlue}{S1}{turn}{Okay, so, Tuesday: land, hotel, park, ramen, robot cafe, konbini dinner, sleep. Wednesday: onsen, Hakone by train. That feel like the right flow?}
\mptbU{mptbViolet}{S2}{non-floor-taking}{Yeah, yeah, I like it a lot.}
\mptbT{mptbViolet}{S2}{interruption}{Oh, wait---}
\mptbU{mptbViolet}{S2}{turn}{---should we pack swimsuits for the onsen? Or do they give you stuff there?}
\end{mptbbox}
\end{minipage}
\caption{\small\textbf{One generated excerpt per interaction type} (test split, verbatim script text; tags are the generator's event labels, \colorbox{mptbSoftYellow}{highlighted} lines are scored anchors). Argumentative and collaborative claims cut into unfinished clauses; casual and task-oriented claims are short coordination moves; instructional claims are listener questions; narrative regions are dominated by stacked listener backchannels that must not be scored as takeovers.}
\label{fig:type_examples}
\end{figure*}

MPTB adopts the six conversation types used in TurnBench~\citep{turnbench}, which were designed to induce distinct patterns of turn-taking behavior. The categories are retained to provide a common interaction-style axis with prior work, while MPTB uses them within a balanced multi-party factorial design over $N\in\{2,3,4,5\}$. Table~\ref{tab:interaction_types} summarizes the intended interaction characteristics used to condition synthetic dialogue generation. These descriptions specify the dominant conversational regime rather than enforcing deterministic event patterns; realized backchannel, interruption, overlap, and turn-duration statistics are reported separately in Appendix~\ref{sec:data_stats}. Table~\ref{tab:type_conditioning} gives the exact style string and soft targets passed to the generator for each type, and Figure~\ref{fig:type_examples} shows one excerpt per type from the test split.

\begin{table*}[t]
\centering
\small
\setlength{\tabcolsep}{5.5pt}
\renewcommand{\arraystretch}{1.12}
\begin{tabular}{p{2.5cm} p{5.2cm} p{6.3cm}}
\toprule
\textbf{Interaction Type} & \textbf{Conversation Setting} & \textbf{Expected Floor-Management Characteristics} \\
\midrule

\textbf{Casual} &
Unstructured social conversation with flexible topic progression and informal exchange. &
Frequent acknowledgements and backchannels, relatively short turns, topic shifts, and occasional overlap. \\

\textbf{Task-Oriented} &
Goal-directed interaction in which participants jointly complete or resolve a concrete task. &
Frequent clarification and confirmation, comparatively structured floor transfer, and short coordination turns. \\

\textbf{Instructional} &
Asymmetric interaction in which one participant primarily explains, teaches, or guides others. &
Longer floor holds by the primary speaker, question-driven transitions, and acknowledgement backchannels from listeners. \\

\textbf{Collaborative} &
Shared reasoning or problem solving without a single dominant speaker. &
More frequent floor exchange, cooperative overlap, short claims to the floor, and relatively balanced participation. \\

\textbf{Argumentative} &
Structured disagreement or debate in which participants defend competing positions. &
Longer contested turns, more competitive floor claims, interruptions, and overlapping speech. \\

\textbf{Narrative} &
One participant primarily recounts an event or story while others act as listeners. &
Longer sustained turns, frequent listener backchannels, and comparatively limited floor competition. \\

\bottomrule
\end{tabular}
\caption{\small{\textbf{Interaction types used in MultiPartyTurnBench.} MPTB adopts the six conversation categories used in TurnBench~\citep{turnbench} and instantiates each across $N\in\{2,3,4,5\}$ speakers. The final column summarizes the intended floor-management regime used to condition dialogue generation; realized event statistics are measured from the generated corpus rather than assumed from these descriptions.}}
\label{tab:interaction_types}
\end{table*}

\begin{table*}[t]
\centering
\scriptsize
\setlength{\tabcolsep}{4pt}
\renewcommand{\arraystretch}{1.14}
\begin{tabular}{@{}l p{7.6cm} rrrr@{}}
\toprule
& & \multicolumn{4}{c}{\textbf{Soft targets (TurnBench dyadic statistics)}} \\
\cmidrule(l){3-6}
\textbf{Type} & \textbf{Style string inserted into the prompt} & \textbf{Turn (s)} & \textbf{BC/min} & \textbf{INT/min} & \textbf{Words/min} \\
\midrule
Argumentative & \textit{a heated debate with frequent disagreement, interruptions, and people talking over each other} & 10.0 & 3.09 & 2.48 & 206 \\
Casual & \textit{a relaxed casual chat with frequent overlaps, backchannels, and light interruptions} & 8.1 & 5.20 & 2.05 & 211 \\
Collaborative & \textit{participants actively building on each other's ideas, with overlaps and frequent floor contests} & 7.9 & 3.97 & 2.64 & 207 \\
Instructional & \textit{one participant explaining/teaching others who ask clarifying questions; mostly clean turns, few interruptions} & 10.5 & 4.30 & 1.52 & 202 \\
Narrative & \textit{storytelling with personal anecdotes; longer turns, listeners backchanneling, rare interruptions} & 10.2 & 4.76 & 1.50 & 198 \\
Task-Oriented & \textit{a focused task-oriented working session; mostly orderly turns with occasional quick interjections} & 8.9 & 5.00 & 1.75 & 205 \\
\bottomrule
\end{tabular}
\caption{\small\textbf{Generator conditioning per interaction type.} For a conversation of target duration $D$ (min) with $N$ speakers, the prompt requests about $\max(2N,\mathrm{round}(60D/\text{turn}))$ floor-holding turns, $\mathrm{round}(\text{INT/min}\cdot D\cdot s_N)$ floor claims and $\mathrm{round}(\text{BC/min}\cdot D\cdot s_N)$ backchannels/laughter, with $s_N=1+0.1(N-2)$. Targets are guardrails; realized rates are reported in Tables~\ref{tab:corpus_stats} and~\ref{tab:corpus_stats_type}.}
\label{tab:type_conditioning}
\end{table*}

\subsection{Generation pipeline}
\label{generation}

Generation proceeds in five stages. (1) An LLM writes a role-tagged dialogue conditioned on party count, interaction type, and target rates for backchannels and interruptions. (2) A deterministic state machine resolves floor claims, holds, yields, and overlaps into a timed event graph. (3) Kokoro-82M renders each speaker using a disjoint voice pool. (4) A controlled noise stage adds procedural ambient sound, discrete non-content events, and intermittent distant speech. (5) The assembler emits synchronized per-speaker audio and the exact event log. Scripts, seeds, audio manifests, and scorer inputs are recorded for reproducibility. The remainder of this section specifies every stage.

\paragraph{Conversation specifications.}
Specifications are sampled with exactly 10 conversations per $(N,\text{type})$ cell (2 dev, 8 test). Each specification independently draws a domain (20 domains, Table~\ref{tab:domains}), a tone (5 tones), a target duration $D\sim\mathcal{U}(150,240)$\,s, $N$ distinct voices from the split's voice pool, and a noise profile (Stage~4). Each conversation carries an integer seed that drives every downstream random draw.

\begin{table}[t]
\centering
\scriptsize
\setlength{\tabcolsep}{3pt}
\renewcommand{\arraystretch}{1.08}
\begin{tabular}{@{}p{7.5cm}@{}}
\toprule
\textbf{Domains (20).} tech support call; study group / exam prep; family logistics planning; product design standup; podcast panel / current-events discussion; board game night; job interview; travel itinerary planning; classroom discussion; relationship/interpersonal conflict; restaurant/group dining decision; sports commentary; book club; home renovation planning; medical appointment scheduling (logistics only, non-clinical); roommate household planning; startup pitch practice; book/movie recommendation swap; volunteer event coordination; group trivia night. \\
\midrule
\textbf{Tones (5; realized count).} professional and business-like (57); curious and inquisitive (54); enthusiastic and energetic (53); casual and friendly (42); thoughtful and contemplative (34). \\
\bottomrule
\end{tabular}
\caption{\small\textbf{Scenario conditioning vocabulary.} All 20 domains occur in the corpus (most frequent: volunteer event coordination, 19 conversations).}
\label{tab:domains}
\end{table}

\begin{figure*}[t]
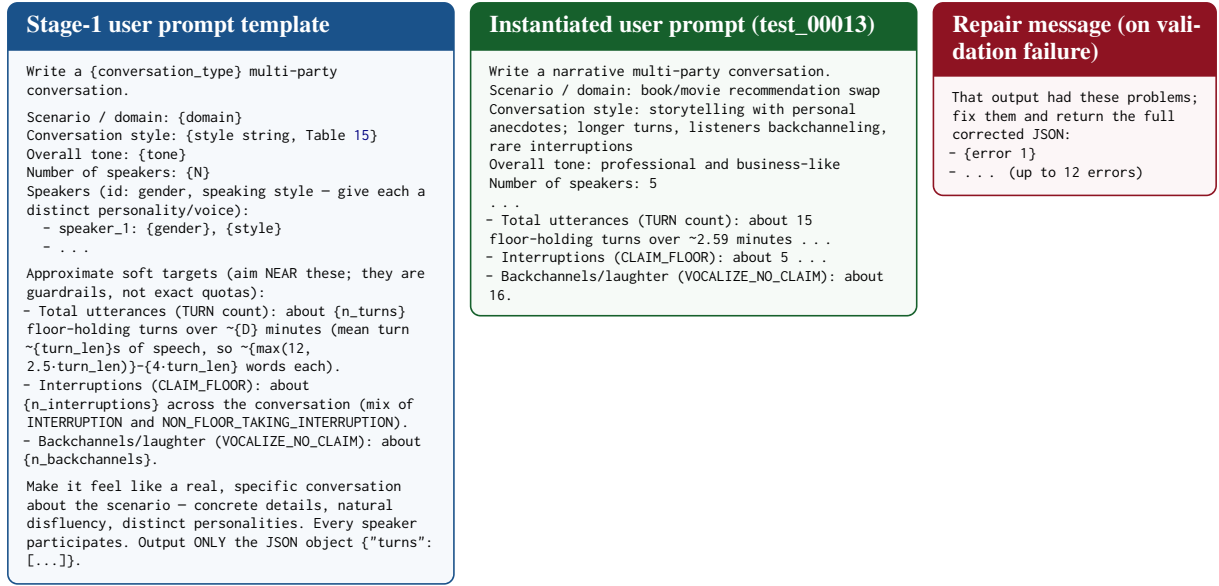

\centering
\begin{minipage}[t]{0.37\textwidth}
\vspace{0pt}
\begin{mptbbox}[mptbBlue]{Stage-1 user prompt template}
{\ttfamily\tiny\raggedright
Write a \{conversation\_type\} multi-party conversation.\par\smallskip
Scenario / domain: \{domain\}\par
Conversation style: \{style string, Table~\ref{tab:type_conditioning}\}\par
Overall tone: \{tone\}\par
Number of speakers: \{N\}\par
Speakers (id: gender, speaking style --- give each a distinct personality/voice):\par
\hspace*{1em}- speaker\_1: \{gender\}, \{style\}\par
\hspace*{1em}- \ldots\par\smallskip
Approximate soft targets (aim NEAR these; they are guardrails, not exact quotas):\par
- Total utterances (TURN count): about \{n\_turns\} floor-holding turns over \textasciitilde\{D\} minutes (mean turn \textasciitilde\{turn\_len\}s of speech, so \textasciitilde\{max(12, 2.5$\cdot$turn\_len)\}-\{4$\cdot$turn\_len\} words each).\par
- Interruptions (CLAIM\_FLOOR): about \{n\_interruptions\} across the conversation (mix of INTERRUPTION and NON\_FLOOR\_TAKING\_INTERRUPTION).\par
- Backchannels/laughter (VOCALIZE\_NO\_CLAIM): about \{n\_backchannels\}.\par\smallskip
Make it feel like a real, specific conversation about the scenario --- concrete details, natural disfluency, distinct personalities. Every speaker participates. Output ONLY the JSON object \{"turns": [...]\}.\par}
\end{mptbbox}
\end{minipage}\hfill
\begin{minipage}[t]{0.37\textwidth}
\vspace{0pt}
\begin{mptbbox}[mptbGreen]{Instantiated user prompt (test\_00013)}
{\ttfamily\tiny\raggedright
Write a narrative multi-party conversation.\par
Scenario / domain: book/movie recommendation swap\par
Conversation style: storytelling with personal anecdotes; longer turns, listeners backchanneling, rare interruptions\par
Overall tone: professional and business-like\par
Number of speakers: 5\par
\ldots\par
- Total utterances (TURN count): about 15 floor-holding turns over \textasciitilde2.59 minutes \ldots\par
- Interruptions (CLAIM\_FLOOR): about 5 \ldots\par
- Backchannels/laughter (VOCALIZE\_NO\_CLAIM): about 16.\par}
\end{mptbbox}
\end{minipage}\hfill
\begin{minipage}[t]{0.235\textwidth}
\vspace{0pt}
\begin{mptbbox}[mptbRed]{Repair message (on validation failure)}
{\ttfamily\tiny\raggedright
That output had these problems; fix them and return the full corrected JSON:\par
- \{error 1\}\par
- \ldots\ (up to 12 errors)\par}
\end{mptbbox}
\end{minipage}
\caption{\small\textbf{Stage-1 user prompt.} The template is filled from the conversation specification and Table~\ref{tab:type_conditioning}; the middle box shows the instantiated prompt of the narrative excerpt in Figure~\ref{fig:type_examples}. On validation failure the repair message is appended and the model is queried again.}
\label{fig:gen_prompts_user}
\end{figure*}

\paragraph{Stage 1: event-conditioned dialogue generation.}
Scripts are generated by \texttt{gpt-4.1} (all 240 conversations) with the system prompt and user template of Figures~\ref{fig:gen_prompts} and~\ref{fig:gen_prompts_user}, temperature 0.9, a 8,000-token output limit and JSON output. Each utterance carries exactly one event type (\textsc{turn}, \textsc{claim\_floor}, \textsc{vocalize\_no\_claim}), a qualitative timing hint and, where applicable, a target holder and a resolution hint. Outputs are validated against structural rules (first utterance is a \textsc{turn}; claims and vocalizations require a current holder and target it; every speaker holds at least one \textsc{turn}; at least $\max(2N,8)$ turns; closed tag and hint vocabularies). A failed validation is returned to the model with the error list for repair, up to three attempts with seeds $\text{seed}+k$. Before validation, any backchannel longer than three words is deterministically relabelled as a non-floor-taking claim, so that multi-word listener speech is never scored as an \intr\ negative. Realized scripts contain 6,800 utterances (28.3 per conversation): 4,347 \textsc{turn} (median 16 words), 367 successful claims (median 9 words), 605 failed claims (median 5 words), 1,176 backchannels (median 1 word) and 305 laughter vocalizations (median 2 words).

\begin{figure*}[t]
\centering
\begin{minipage}[t]{0.495\textwidth}
\vspace{0pt}
\begin{mptbbox}[mptbSlate]{Stage-1 system prompt, verbatim (1/2): output schema}
{\ttfamily\tiny\raggedright
You are a meticulous dialogue data generator that writes REALISTIC multi-party spoken conversation transcripts for a turn-taking benchmark. You output STRICT JSON only.\par\smallskip
The transcript is an ordered list of utterances. Each utterance is one JSON object with these fields:\par
- "speaker": one of the given speaker ids (e.g. "speaker\_1").\par
- "text": what is actually said aloud. Natural spoken English (disfluencies, "um", "like", trailing "---" for cut-offs are welcome). No stage directions, no narration, no lists/bullets.\par
- "event\_type": exactly one of:\par
\hspace*{1em}- "TURN": a normal floor-holding utterance (the speaker "has the floor"). This is the default.\par
\hspace*{1em}- "CLAIM\_FLOOR": this speaker jumps in to try to take the floor FROM the current floor-holder while that holder is still speaking. REQUIRES:\par
\hspace*{2em}- "resolved\_hint": "INTERRUPTION" (the jump-in SUCCEEDS --- the holder stops and this speaker takes over; usually followed by a TURN from this same speaker) OR "NON\_FLOOR\_TAKING\_INTERRUPTION" (the jump-in FAILS --- it is a brief overlap and the ORIGINAL holder keeps going; usually the holder's next TURN continues their sentence).\par
\hspace*{2em}- "target": the speaker id being interrupted (the current holder).\par
\hspace*{1em}- "VOCALIZE\_NO\_CLAIM": a backchannel or laughter over the current holder; it NEVER takes the floor. REQUIRES:\par
\hspace*{2em}- "sublabel": "BACKCHANNEL" or "LAUGHTER".\par
\hspace*{2em}- "target": the current holder id being backchanneled.\par
\hspace*{2em}- A "BACKCHANNEL" MUST be an extremely brief vocalization of at most THREE words --- ideally ONE --- e.g. "mm-hm", "yeah", "right", "uh-huh", "exactly", "oh really", "no way". It is a listening noise, not a comment. If a listener would say a whole phrase or sentence, that is NOT a backchannel: make it a "CLAIM\_FLOOR" instead (resolved\_hint "NON\_FLOOR\_TAKING\_INTERRUPTION" if the holder keeps talking, or "INTERRUPTION" if the holder stops). "LAUGHTER" should also be very short (e.g. "haha", "haha right").\par
- "timing\_hint": exactly one of ["after\_pause",\allowbreak "latched",\allowbreak "long\_pause",\allowbreak "mid\_utterance",\allowbreak "overlap\_onset",\allowbreak "during\_holder"] --- a QUALITATIVE placement cue, never a number:\par
\hspace*{1em}- TURN utterances use "after\_pause" (normal gap), "latched" (right after, \textasciitilde no gap), or "long\_pause".\par
\hspace*{1em}- CLAIM\_FLOOR uses "mid\_utterance" or "overlap\_onset".\par
\hspace*{1em}- VOCALIZE\_NO\_CLAIM uses "during\_holder".\par}
\end{mptbbox}
\end{minipage}\hfill
\begin{minipage}[t]{0.495\textwidth}
\vspace{0pt}
\begin{mptbbox}[mptbSlate]{Stage-1 system prompt, verbatim (2/2): rules and exemplar}
{\ttfamily\tiny\raggedright
Hard rules:\par
1. The FIRST utterance must be a TURN (someone starts speaking).\par
2. A CLAIM\_FLOOR or VOCALIZE\_NO\_CLAIM can only occur while someone holds the floor (never first, and its "target" must be the current holder --- the most recent TURN speaker).\par
3. After a successful CLAIM\_FLOOR (INTERRUPTION), the interrupter normally continues with one or more TURN utterances (they now hold the floor).\par
4. After a failed CLAIM\_FLOOR (NON\_FLOOR\_TAKING\_INTERRUPTION), the ORIGINAL holder normally continues with a TURN (often finishing their interrupted sentence, starting with "---").\par
5. EVERY listed speaker must speak at least one TURN.\par
6. Do NOT emit any tag other than the three above (no HOLDER\_YIELDS etc.) --- clean turn hand-offs are implicit (a TURN by a new speaker after a TURN means the previous speaker yielded the floor).\par
7. BACKCHANNELS must be <=3 words (ideally 1). A listener saying a full phrase/sentence is a CLAIM\_FLOOR, never a BACKCHANNEL.\par
8. Output ONLY a JSON object: \{"turns": [ ... ]\}. No prose before or after.\par\smallskip
Here is a short example of the exact output shape (a 3-speaker snippet exercising every tag):\par
\{"turns": [\par
\hspace*{0.5em}\{"speaker": "speaker\_1", "text": "Okay so for the mitosis question, the key thing is the chromosome number stays the same.", "event\_type": "TURN", "timing\_hint": "after\_pause"\},\par
\hspace*{0.5em}\{"speaker": "speaker\_2", "text": "Well hang on---", "event\_type": "CLAIM\_FLOOR", "resolved\_hint": "NON\_FLOOR\_TAKING\_INTERRUPTION", "target": "speaker\_1", "timing\_hint": "mid\_utterance"\},\par
\hspace*{0.5em}\{"speaker": "speaker\_1", "text": "---right, whereas meiosis halves it, that's the whole point.", "event\_type": "TURN", "timing\_hint": "latched"\},\par
\hspace*{0.5em}\{"speaker": "speaker\_3", "text": "mm-hm", "event\_type": "VOCALIZE\_NO\_CLAIM", "sublabel": "BACKCHANNEL", "target": "speaker\_1", "timing\_hint": "during\_holder"\},\par
\hspace*{0.5em}\{"speaker": "speaker\_2", "text": "Actually wait, I think you've got prophase backwards, can we go back?", "event\_type": "CLAIM\_FLOOR", "resolved\_hint": "INTERRUPTION", "target": "speaker\_1", "timing\_hint": "overlap\_onset"\},\par
\hspace*{0.5em}\{"speaker": "speaker\_2", "text": "In prophase one the homologs pair up, that's what makes it different.", "event\_type": "TURN", "timing\_hint": "latched"\},\par
\hspace*{0.5em}\{"speaker": "speaker\_3", "text": "haha right", "event\_type": "VOCALIZE\_NO\_CLAIM", "sublabel": "LAUGHTER", "target": "speaker\_2", "timing\_hint": "during\_holder"\},\par
\hspace*{0.5em}\{"speaker": "speaker\_1", "text": "Fair enough, good catch. So then the diagram on slide four is wrong too.", "event\_type": "TURN", "timing\_hint": "after\_pause"\}\par
]\}\par}
\end{mptbbox}
\end{minipage}
\caption{\small\textbf{Stage-1 system prompt}, shared by all 240 conversations. The LLM proposes symbolic acts only; it never assigns timestamps or final floor outcomes.}
\label{fig:gen_prompts}
\end{figure*}

\paragraph{Stage 2: floor-state orchestration.}
A discrete-event state machine consumes the script in order and maintains the floor holder. A \textsc{turn} by a new speaker closes the previous holder's segment as a \emph{yield} and starts the new segment after a gap drawn from the timing hint. A \textsc{turn} by the current holder continues the segment; if the open segment is at least 4\,s long, a mid-turn pause is inserted with probability 0.45, closing the segment as a \emph{hold} and producing an \eot\ hard negative. A successful \textsc{claim\_floor} starts inside the holder's ongoing utterance, the holder keeps talking briefly in overlap and is then truncated (\emph{interrupted}), and the floor transfers. A failed claim and a vocalization overlap the holder without changing the floor. Utterance durations come from the synthesized audio, so timing is resolved against real speech lengths. Table~\ref{tab:orchestrator_params} lists every distribution; all draws use the conversation seed.

\begin{table}[t]
\centering
\scriptsize
\setlength{\tabcolsep}{3pt}
\renewcommand{\arraystretch}{1.1}
\begin{tabular}{@{}p{4.6cm}l@{}}
\toprule
\textbf{Parameter} & \textbf{Value} \\
\midrule
Lead-in silence & 0.3\,s \\
Yield gap, \texttt{after\_pause} & $\mathcal{U}(0.20,0.60)$\,s \\
Yield gap, \texttt{latched} & $\mathcal{U}(-0.12,0.06)$\,s \\
Yield gap, \texttt{long\_pause} & $\mathcal{U}(0.80,2.00)$\,s \\
Same-holder continuation gap & $\mathcal{U}(0.02,0.18)$\,s \\
Mid-turn pause: probability / min.\ segment & 0.45 / 4.0\,s \\
Mid-turn pause duration & $\mathcal{U}(0.40,1.20)$\,s \\
Interruption onset (fraction of holder utterance) & $\mathcal{U}(0.45,0.85)$ \\
Holder overlap before cut-off & $\mathcal{U}(0.40,1.20)$\,s \\
Failed-claim onset (fraction) & $\mathcal{U}(0.30,0.80)$ \\
Backchannel/laughter onset (fraction) & $\mathcal{U}(0.20,0.80)$ \\
Self-overlap guard (same channel) & 0.08\,s \\
Truncation fade-out & 30\,ms \\
\bottomrule
\end{tabular}
\caption{\small\textbf{Stage-2 orchestrator parameters.} Realized timing distributions are given in Table~\ref{tab:timing_realized}.}
\label{tab:orchestrator_params}
\end{table}

\paragraph{Stage 3: speaker rendering.}
Each utterance is synthesized independently with Kokoro-82M~\citep{kokoro_tts} (\texttt{hexgrad/Kokoro-82M}) at 24\,kHz, speed 1.0, with a per-utterance seed derived from the text and voice. Leading and trailing silence below $-40$\,dBFS is trimmed with a 30\,ms margin, and clips are cached by content hash so re-rendering is byte-reproducible. Abstract voice ids map to Kokoro voices as in Table~\ref{tab:voice_pool}. The 24 voices are split gender-balanced into 8 development and 16 test voices; no voice occurs in both splits, and no voice is reused within a conversation.

\begin{table}[t]
\centering
\tiny
\setlength{\tabcolsep}{2.4pt}
\renewcommand{\arraystretch}{1.05}
\resizebox{\columnwidth}{!}{%
\begin{tabular}{@{}llcl|llcl@{}}
\toprule
\textbf{ID} & \textbf{Kokoro} & \textbf{G} & \textbf{Style} & \textbf{ID} & \textbf{Kokoro} & \textbf{G} & \textbf{Style} \\
\midrule
v01 & af\_alloy & F & clipped/efficient & v13 & am\_onyx & M & upbeat/expressive \\
v02 & af\_aoede & F & energetic/upbeat & v14 & af\_nicole & F & dry-humor/deadpan \\
v03 & am\_adam & M & calm/measured & v15 & am\_puck & M & laid-back/casual \\
\cellcolor{mptbBestCyan}v04 & am\_echo & M & forceful/assertive & \cellcolor{mptbBestCyan}v16 & am\_santa & M & deadpan/flat \\
\cellcolor{mptbBestCyan}v05 & am\_eric & M & warm/friendly & \cellcolor{mptbBestCyan}v17 & af\_nova & F & skeptical/probing \\
v06 & am\_fenrir & M & hesitant/soft-spoken & v18 & bm\_daniel & M & nervous/tentative \\
v07 & am\_liam & M & thoughtful/slow & v19 & af\_river & F & gentle/reassuring \\
v08 & af\_bella & F & measured/rebuttal & \cellcolor{mptbBestCyan}v20 & af\_sarah & F & authoritative/instructor \\
v09 & af\_heart & F & playful/animated & v21 & bm\_fable & M & gravelly/mature \\
v10 & af\_jessica & F & direct/no-nonsense & \cellcolor{mptbBestCyan}v22 & af\_sky & F & bright/youthful \\
\cellcolor{mptbBestCyan}v11 & af\_kore & F & brisk/professional & \cellcolor{mptbBestCyan}v23 & bm\_george & M & analytical/precise \\
v12 & am\_michael & M & moderator/neutral & v24 & bf\_alice & F & empathetic/attentive \\
\bottomrule
\end{tabular}%
}
\caption{\small\textbf{Voice pool and split.} \colorbox{mptbBestCyan}{Shaded} ids form the development pool (4F/4M); the remaining 16 form the test pool. Prefix \texttt{a}/\texttt{b} = American/British English. The style descriptor is passed to the script generator.}
\label{tab:voice_pool}
\end{table}

\paragraph{Stage 4: acoustic scene rendering.}
Clips are placed on per-speaker channels at the orchestrated times; truncated holder clips receive a 30\,ms fade, and each channel is peak-normalized to $-1$\,dBFS. A single noise realization is added to all channels of a conversation (one room, $N$ microphones), so nuisance sounds are cross-channel. Table~\ref{tab:noise_params} gives the noise model and its realized distribution; sound effects and distant speech are drawn from MUSAN~\citep{musan2015}. If the mixture exceeds $-1$\,dBFS, one common gain is applied to all channels, preserving inter-channel levels.

\begin{table}[t]
\centering
\scriptsize
\setlength{\tabcolsep}{3pt}
\renewcommand{\arraystretch}{1.1}
\begin{tabular}{@{}>{\raggedright\arraybackslash}p{1.9cm}>{\raggedright\arraybackslash}p{3.15cm}>{\raggedright\arraybackslash}p{2.05cm}@{}}
\toprule
\textbf{Component} & \textbf{Model} & \textbf{Realized (240 conv.)} \\
\midrule
Ambient bed & Procedural pink noise, scene-specific low-pass (200--2200\,Hz), mains hum (100/200\,Hz) for office scenes, slow ($<$0.2\,Hz) drift; 10 scenes & SNR $\mathcal{U}(20,38)$\,dB: range 20.2--37.8, median 29.7 \\
Discrete events & One random MUSAN sound-bible clip ($\le$4\,s) per sampled event slot, 12\,dB below active speech, random time; $k\sim\{0,1,1,2\}$ slots & 0 / 1 / 2 slots: 49 / 124 / 67 conv. \\
Distant speech & 2--4 faded MUSAN speech snippets of 1.5--3\,s at 26\,dB below active speech; $p=0.2$ & 51 conversations \\
\bottomrule
\end{tabular}
\caption{\small\textbf{Stage-4 noise model.} SNR is measured against the active-speech RMS pooled over channels. Event-slot labels: keyboard clatter, notification chime, dishes, off-mic laughter, phone ring, door knock, chairs, school bell, alarm beep, paper shuffle (23--30 occurrences each). Scenes: outdoor park 33, living room TV 32, school hallway 27, office HVAC 25, studio room tone 24, cafe 22, office hum 22, dorm 20, quiet office 19, quiet apartment 16.}
\label{tab:noise_params}
\end{table}

\paragraph{Stage 5: channel assembly, quality control and release.}
Each conversation is released as $N$ noisy evaluation channels and $N$ clean channels (FLAC, 24\,kHz), the tagged script, the canonical event log, a render sidecar listing every clip with its placement and truncation, the noise manifest, and the seed. An automatic acoustic check runs on the clean channels: (Q1) audio length matches the event log; (Q2) per-channel speech ratio in $[0.03,0.99]$; (Q3) at least 50\% of each scheduled clip span is voiced; (Q4) at most 30\% voiced time outside scheduled clips; (Q5) at least 60\% of overlap events (interruptions, failed claims, vocalizations) have both mover and holder voiced. All 240 conversations pass. Across the corpus, 2,369 of 2,436 overlap events (97.2\%) are acoustically realized as simultaneous speech, median clip coverage is 0.81, and no channel shows speech leakage outside its clips.

\begin{mptbbox}[mptbSlate]{From script to scoring anchors \,$\cdot$\, test\_00013 ($N{=}5$, narrative)}
\mptbMeta{Event log (first 78\,s) $\rightarrow$ anchors on each speaker channel.}
{\scriptsize
\setlength{\tabcolsep}{2.2pt}
\resizebox{\linewidth}{!}{%
\begin{tabular}{@{}l l l l@{}}
\textbf{Time (s)} & \textbf{Spk} & \textbf{Generator event} & \textbf{Anchor} \\
0.30--19.19 & S1 & \textsc{turn}, yield$\rightarrow$S2 & \eot$+$ @19.19 (S1) \\
7.97--9.18 & S2 & \textsc{backchannel} ``Oh, absolutely.'' & \intr$-$ [7.97, 9.18] (S2) \\
9.45--9.90 & S3 & \textsc{backchannel} ``mm-hm'' & \intr$-$ [9.45, 9.90] (S3) \\
14.88--16.21 & S4 & failed claim ``Yeah, that's a good one.'' & \intr\ excl. (S4) \\
19.21--44.33 & S2 & \textsc{turn}, hold & \eot$-$ [44.33, 45.29] (S2) \\
34.70--35.38 & S3 & failed claim ``Right, I---'' & \intr\ excl. (S3) \\
45.29--51.98 & S2 & \textsc{turn}, yield$\rightarrow$S3 & \eot$+$ @51.98 (S2) \\
51.94--77.85 & S3 & \textsc{turn}, interrupted & \eot\ excl. @77.85$\pm$0.25 \\
77.03 & S5 & \textsc{interruption} ``I, uh---can I just say \ldots'' & \intr$+$ @77.03 (S5) \\
\end{tabular}}}
\mptbTake{S2 takes the floor 22\,ms after S1 stops (a \texttt{latched} hand-over). The S2 hold is created by a same-speaker continuation after S3's failed claim; its scored span is the 0.96\,s silence before S2 resumes. S5's claim starts 0.82\,s before S3 is cut off, so the interruption onset, not S3's forced stop, is the scored event.}
\end{mptbbox}

\section{Corpus Composition and Turn-Taking Dynamics}
\label{app:corpus_stats}

\subsection{Benchmark Analysis}
\label{sec:data_stats}

\begin{table*}[t]
\centering
\scriptsize
\setlength{\tabcolsep}{4pt}
\renewcommand{\arraystretch}{1.12}
\resizebox{\textwidth}{!}{%
\begin{tabular}{lrrrrrrrrr}
\toprule
& \multicolumn{2}{c}{\textbf{Corpus Size}}
& \multicolumn{5}{c}{\textbf{Interaction Dynamics}}
& \multicolumn{2}{c}{\textbf{Benchmark Events}} \\
\cmidrule(lr){2-3}
\cmidrule(lr){4-8}
\cmidrule(lr){9-10}
\textbf{Condition} &
\textbf{Dialogues} &
\textbf{Duration (h)} &
\textbf{Mean Turn (s)} &
\textbf{Events/min} &
\textbf{Backchannels/min} &
\textbf{Interruptions/min} &
\textbf{Overlap/dialogue (s)} &
\textbf{EOT Pos./Neg.} &
\textbf{INT Pos./Neg.} \\
\midrule
\multicolumn{10}{l}{\textit{By speaker count}} \\
$N=2$ & 60 & 1.90 & 7.07 & 13.8 & 3.12 & 0.70 & 10.2 & 853/74 & 80/356 \\
$N=3$ & 60 & 1.66 & 6.38 & 16.3 & 3.61 & 0.96 & 10.8 & 838/64 & 96/359 \\
$N=4$ & 60 & 1.67 & 6.10 & 16.3 & 3.60 & 0.94 & 10.9 & 875/55 & 94/361 \\
$N=5$ & 60 & 1.66 & 6.26 & 16.6 & 4.06 & 0.97 & 11.0 & 856/48 & 97/405 \\
\midrule
\multicolumn{10}{l}{\textit{By interaction type}} \\
Argumentative & 40 & 1.11 & 6.29 & 15.2 & 2.95 & 1.08 & 10.7 & 520/49 & 72/196 \\
Casual & 40 & 1.10 & 4.81 & 19.4 & 4.24 & 0.92 & 11.3 & 739/37 & 61/280 \\
Collaborative & 40 & 1.07 & 5.32 & 17.8 & 3.43 & 1.33 & 10.6 & 610/37 & 85/220 \\
Instructional & 40 & 1.17 & 7.79 & 13.8 & 3.81 & 0.70 & 11.3 & 458/45 & 49/268 \\
Narrative & 40 & 1.41 & 9.22 & 11.8 & 3.34 & 0.53 & 11.5 & 471/45 & 45/282 \\
Task-Oriented & 40 & 1.03 & 5.27 & 17.5 & 3.80 & 0.89 & 9.1 & 624/28 & 55/235 \\
\midrule
\textbf{All} & \textbf{240} & \textbf{6.89} & \textbf{6.45} & \textbf{15.7} & \textbf{3.58} & \textbf{0.89} & \textbf{10.7} & \textbf{3422/241} & \textbf{367/1481} \\
\bottomrule
\end{tabular}%
}
\caption{\textbf{Corpus composition and turn-taking dynamics across speaker counts and interaction types.} MPTB maintains balanced coverage across $N=2$--$5$ while exhibiting relatively stable conversational activity as speaker count increases. Interaction types induce more distinct floor-management patterns: Casual conversations exhibit faster exchanges and shorter turns, Collaborative conversations have the highest interruption rate, and Narrative conversations sustain the longest turns with the fewest interruptions. EOT and INT $+/-$ indicate positive anchors and hard-negative spans, respectively.}
\label{tab:corpus_stats}
\end{table*}

\paragraph{Dataset composition.}
MPTB crosses four speaker counts ($N\in\{2,3,4,5\}$) with six interaction types, yielding 24 conditions with 10 conversations each. The resulting 240 conversations comprise 6.89 hours of speech, partitioned into 48 development and 192 test conversations with disjoint speaker voices. Table~\ref{tab:corpus_stats} provides event counts and conversational statistics for each condition. Although conversation counts are balanced, realized positive and negative event counts vary across conditions.

\paragraph{Turn-taking dynamics.}
Speaker count has a comparatively limited effect on aggregate conversational dynamics: overlap remains between 10.2 and 11.0 seconds per dialogue, while interruption rates increase from 0.70/min at two speakers to approximately 0.97/min at five. Interaction type induces larger differences in conversational behavior. Casual conversations have the fastest exchanges (19.4 events/min) and shortest turns (4.81\,s), whereas Narrative conversations exhibit longer turns (9.22\,s) and fewer interruptions (0.53/min). Collaborative dialogue has the highest interruption rate (1.33/min). These differences establish distinct floor-management regimes within the balanced design, enabling evaluation across conversational styles without conflating them with speaker count.

\paragraph{Floor distribution.}
Adding speakers spreads the floor rather than speeding up exchanges. The share of floor-holding time taken by the most active speaker falls from 60.4\% at $N{=}2$ to 51.4\%, 41.0\% and 36.4\% at $N{=}3,4,5$, while speaker changes per minute stay between 7.6 and 9.1 (Table~\ref{tab:corpus_stats_n}). Hold negatives become rarer with $N$ (74 at $N{=}2$ vs.\ 48 at $N{=}5$) as mean floor segments shorten (7.07 to 6.26\,s); mid-turn pauses are only inserted into segments of at least 4\,s (Table~\ref{tab:orchestrator_params}). Listener activity grows over the same range (3.12 to 4.06 backchannels/min).

\begin{table}[t]
\centering
\scriptsize
\setlength{\tabcolsep}{3.4pt}
\renewcommand{\arraystretch}{1.1}
\resizebox{\columnwidth}{!}{%
\begin{tabular}{@{}lrrrr@{}}
\toprule
\textbf{Quantity (s)} & \textbf{$n$} & \textbf{p10} & \textbf{p50} & \textbf{p90} \\
\midrule
Floor-holding segment & 4,030 & 2.19 & 5.18 & 10.65 \\
Yield gap (next onset $-$ offset) & 3,182 & $-$0.09 & 0.01 & 0.48 \\
Hold pause span (\eot$-$) & 241 & 0.49 & 0.83 & 1.12 \\
Interruption overlap (onset$\rightarrow$cut-off) & 367 & 0.43 & 0.74 & 1.09 \\
Failed claim duration & 605 & 0.80 & 1.63 & 3.72 \\
Backchannel duration (\intr$-$) & 1,176 & 0.44 & 0.64 & 1.04 \\
Laughter duration (\intr$-$) & 305 & 0.53 & 0.97 & 1.82 \\
\bottomrule
\end{tabular}%
}
\caption{\small\textbf{Realized timing distributions} over all 240 conversations. Conversations last 48.2--208.2\,s (median 99.9\,s). Negative yield gaps are \texttt{latched} hand-overs that begin before the previous holder finishes.}
\label{tab:timing_realized}
\end{table}


\section{Experimental Details}
\label{app:experimental_details}

\subsection{Offline Baseline Configuration}
\label{app:offline_config}

The offline suite covers twelve systems in three families of floor-control systems. Acoustic detectors include RMS energy, WebRTC VAD~\citep{webrtcvad}, and Silero VAD~\citep{silero}. Semantic endpoint detectors include Smart Turn v2~\citep{smartturn}, Kyutai Semantic-VAD~\citep{kyutai_svad}, and the Mimi Endpointer~\citep{mimi}. Turn-taking models include Voice Activity Projection (VAP)~\citep{vap}, MuVAP~\citep{muvap}, DualTurn~\citep{dualturn}, the WavLM-Large anchor~\citep{wavlm}, Turn-1-Mini~\citep{turn1mini}, and the ESPnet turn-taking model~\citep{espnet}. Table~\ref{tab:model_configs} gives the exact checkpoint, input configuration and score mapping of every system.

For multi-party evaluation, two-channel models (VAP, MuVAP, Mimi Endpointer, DualTurn) operate in a speaker-versus-rest configuration, treating each participant as the candidate speaker and the mean of the remaining channels as the conversational context. Single-channel models run independently on every speaker channel; ESPnet is reported on the mono mixture of all channels. Systems are evaluated on the noisy, channel-separated MPTB recordings, and all inference is causal: each score is time-stamped at the end of the audio it has consumed. Except for VAP, which gates on its own voice-activity head, every learned \intr\ score is gated by ``another speaker is active'' from WebRTC VAD (aggressiveness 2) on the other channels. Turn-1-Mini releases only its end-of-turn head and is evaluated only on \eot; no \intr\ output is inferred from its predictions. All tunable detectors use thresholds selected on the 48-conversation development split and held fixed on the 192-conversation test split, following Section~\ref{sec:evaluation_protocol}. Thresholds are not adjusted for individual speaker counts or interaction types. The threshold grids, selected thresholds and development scores are listed in Table~\ref{tab:operating_points}.

\begin{table*}[t]
\centering
\scriptsize
\setlength{\tabcolsep}{2.5pt}
\renewcommand{\arraystretch}{1.18}
\begin{tabular}{@{}>{\raggedright\arraybackslash}p{1.8cm}>{\raggedright\arraybackslash}p{3.5cm}>{\raggedright\arraybackslash}p{3.5cm}>{\raggedright\arraybackslash}p{2.8cm}>{\raggedright\arraybackslash}p{3.6cm}@{}}
\toprule
\textbf{System} & \textbf{Checkpoint / implementation} & \textbf{Input and context} & \textbf{\eot\ score} & \textbf{\intr\ score} \\
\midrule
RMS Energy & frame RMS (dBFS) & per channel, 20\,ms frames & speech$\rightarrow$silence offset after 0.3\,s silent hangover & speech onset while another channel is voiced, confirmed after 0.1\,s \\
WebRTC VAD & \texttt{py-\allowbreak webrtcvad} & per channel, 16\,kHz, 20\,ms frames; $\theta$ = aggressiveness & as RMS, on WebRTC masks & as RMS, on WebRTC masks \\
Silero VAD & Silero VAD & per channel, 16\,kHz; $\theta$ = speech prob. & as RMS, on Silero masks & as RMS, on Silero masks \\
\midrule
Smart Turn v2 & \texttt{pipecat-\allowbreak ai/\allowbreak smart-\allowbreak turn-\allowbreak v2} (wav2vec2-base, attention pooling) & per channel, 16\,kHz, $\le$8\,s window ending at each energy-VAD offset ($-40$\,dBFS, 0.3\,s hangover) & $P(\text{complete})$ & $1-P(\text{complete})$ in TurnBench's Smart Turn pipeline (Silero VAD, 512-sample chunks) \\
Kyutai SVAD & \texttt{kyutai/\allowbreak stt-\allowbreak 1b-\allowbreak en\_\allowbreak fr-\allowbreak candle} & per channel, streaming Mimi, 12.5\,Hz & $P(\text{end of turn})$ (VAD head 2) & $1-P(\text{end of turn})$ \\
Mimi Endpointer & \texttt{viks66/\allowbreak mimi-\allowbreak endpointer} (\texttt{oto\_\allowbreak d1f}), two-stream LSTM over Mimi & $s$ vs.\ mean of rest, 24\,kHz, 25\,Hz & $1-P(\text{user})$ & $P(\text{user})$ \\
\midrule
VAP & TurnBench fine-tuned VAP (\texttt{viks66/\allowbreak VAP\_\allowbreak checkpoints}, \texttt{oto}) & $s$ vs.\ mean of rest, 16\,kHz stereo, 50\,Hz; 20\,s context, 5\,s step & $1-p_{\text{now}}[s]$ & $p_{\text{now}}[s]$, gated by VAP $\text{vad}[\text{rest}]>0.5$ \\
MuVAP & \texttt{Haotian-\allowbreak Qi/\allowbreak MuVAP} \texttt{vap-\allowbreak speaker-\allowbreak mimi} (audio-only release) & $s$ vs.\ mean of rest, 16\,kHz, 25\,Hz; 20\,s context, 5\,s step & $1-p_{\text{now}}[s]$ (0--0.6\,s) & $p_{\text{now}}[s]$ \\
DualTurn & \texttt{anyreach-\allowbreak ai/\allowbreak dualturn-\allowbreak qwen2.5-\allowbreak mimi-\allowbreak 0.5B} & $s$ vs.\ mean of rest, 24\,kHz, 12.5\,Hz & \texttt{eot\_\allowbreak probs}[$s$] & \texttt{hold\_\allowbreak probs}[$s$] (mapping chosen on dev) \\
WavLM Anchor & \texttt{ZhuoyanTao/\allowbreak causal-\allowbreak wavlm-\allowbreak turn-\allowbreak taking} (frozen WavLM-Large, AR decoder) & per channel, 16\,kHz, 4\,s window, 40\,ms stride & $P(\textsc{na})+P(\textsc{t})$ & $1-P(\textsc{c})$ \\
Turn-1-Mini & \texttt{p99lab/\allowbreak turn-\allowbreak 1-\allowbreak mini} (released end-of-turn head) & per channel, 16\,kHz, 4\,kHz low-pass, 160\,ms streaming steps & $P(\text{end of turn})$ & not released \\
ESPnet & \texttt{espnet/\allowbreak Turn\_\allowbreak taking\_\allowbreak prediction\_\allowbreak SWBD} (Whisper-medium encoder~\citep{whisper}, 5-class head) & mono mix of all channels, 16\,kHz, $\le$30\,s window, 40\,ms hop & $P(\textsc{t})\times$ energy share of $s$ & $P(\textsc{i})\times$ onset of $s\times$ others' activity \\
\bottomrule
\end{tabular}
\caption{\small\textbf{Offline system configurations.} Classes \textsc{c}/\textsc{na}/\textsc{i}/\textsc{bc}/\textsc{t} = continuation / no-activity / interruption / backchannel / turn-shift. Every score is converted to decisions by rising edges above $\theta$ with a 2\,s refractory period. Mappings follow TurnBench's reference adapters wherever one exists.}
\label{tab:model_configs}
\end{table*}

\subsection{Real-Audio Evaluation}
\label{app:real_audio_eval}

To examine the transferability of results obtained from synthetic speech, we evaluate compatible offline detectors on two real-audio datasets: a dyadic subset of TurnBench~\citep{turnbench} comprising 12 conversations, and a four-party subset of AMI~\citep{ami} comprising 15 meetings (approximately 7.1 hours). The TurnBench subset uses consensus annotations, while AMI floor events are derived from its manual segment and word annotations using a documented floor-state mapping.

The same event definitions and scoring procedure are applied across datasets, with dataset-specific operating points selected where development data are available. Since the datasets differ in recording conditions, interaction structure, and speaker composition, cross-corpus comparisons characterize domain sensitivity rather than isolating a causal effect of speaker count.

\paragraph{TurnBench slice.} Consensus anchors are rebuilt from the three released annotator tracks with $\pm200$\,ms endpoint agreement, following TurnBench's protocol. The slice contains 287 \eot\ positives, 571 hold spans, 107 \intr\ positives and 1,272 negatives.

\paragraph{AMI floor-state mapping.} Each AMI headset channel is one speaker. A single floor holder is tracked over the manual transcriber segments with the thresholds in Table~\ref{tab:ami_mapping}; the resulting events pass through the same anchor derivation and scorer as MPTB. Over the 15 meetings this yields 1,458 \eot\ positives, 862 hold spans, 1,295 \intr\ positives and 1,255 \intr\ negatives (17.3 events, 2.95 backchannels and 3.04 interruptions per minute).

\begin{table}[t]
\centering
\scriptsize
\setlength{\tabcolsep}{3pt}
\renewcommand{\arraystretch}{1.1}
\resizebox{\columnwidth}{!}{%
\begin{tabular}{@{}p{5.3cm}r@{}}
\toprule
\textbf{Rule} & \textbf{Threshold} \\
\midrule
Same-speaker segments merged (contiguous) & gap $<0.15$\,s \\
Mid-turn pause $\rightarrow$ \textsc{turn}--hold (\eot$-$) & gap $\ge0.20$\,s \\
Overlap counted as ``during'' the holder & $\ge0.10$\,s \\
Backchannel (\intr$-$): duration / words & $\le1.20$\,s / $\le3$ \\
Interruption (\intr$+$): interrupter speaks at least & 1.00\,s \\
\quad and holder falls silent within & 2.50\,s of onset \\
Other overlapping listener speech & non-floor-taking (excluded) \\
Non-holder starts after holder silence & holder yields (\eot$+$) \\
\bottomrule
\end{tabular}%
}
\caption{\small\textbf{AMI floor-state mapping} from manual segments to MPTB events.}
\label{tab:ami_mapping}
\end{table}

\subsection{Full-Duplex Behavioral Evaluation}
\label{app:fd_behavioral_eval}

\paragraph{Scenario Design.}
We construct 120 interactive scenarios spanning four event conditions: true end-of-turn, mid-turn hold pause, successful interruption, and backchannel (30 each). Scenarios are balanced across $N\in\{2,3,4\}$ and the six interaction types. Each scenario is evaluated three times on five full-duplex systems: GPT-Live-1~\citep{openai2026gptlive}, Gemini-3.8-Live~\citep{google2026gemini38live}, MiniCPM-o-4.5~\citep{cui2026minicpm}, Realtime-Venus~\citep{zhao2026realtime}, and Step-Audio-3-Realtime~\citep{lin2026stepaudio}, yielding 1,800 attempted trials. Agents receive streaming speech without access to scenario labels or reference event times.

\paragraph{START and STOP Probes.}
The START probe measures whether an agent correctly enters the conversational floor. A response following a genuine EOT is counted as a correct start, while speaking during a mid-turn hold pause constitutes a false start. We denote the corresponding rates by $S$ and $FS$.

The STOP probe measures whether an agent relinquishes the floor appropriately. The agent is first induced to produce a multi-second response, during which an incoming speaker either successfully interrupts or produces a backchannel. Yielding to a genuine interruption constitutes a correct stop; yielding to a backchannel constitutes a false stop. We denote these rates by $Y$ and $FY$.

Both probes use the balanced-accuracy and action-bias definitions in Section~\ref{sec:evaluation_protocol}, with $(\mathrm{Recall},\mathrm{FPR})=(S,FS)$ for START and $(Y,FY)$ for STOP.

\paragraph{Instrumentation and Trial Validity.}
Provider-specific adapters record incoming audio timestamps, assistant speech intervals, emitted audio, and available interruption or cancellation events on a common monotonic clock. START trials require an observable opportunity for the agent to respond; STOP trials additionally require the agent to be speaking when the probe is injected. Trials failing the relevant observability conditions are excluded from behavioral scoring.

\section{Additional Results}
\label{app:additional_results}

\subsection{Synthetic vs Real-world Turn Taking in Multi-party Setting}
\label{app:sim2real_results}

Table~\ref{tab:real_vs_synthetic} places MPTB next to three real-speech sources. \emph{TurnBench official} reports the published TurnBench leaderboard (real dyadic studio speech; 116 test conversations; TurnBench's own runs and operating points). \emph{TurnBench dev slice} applies our scorer and our detector implementations to 12 real dyadic development conversations with consensus anchors (287 \eot\ positives / 571 hold spans; 107 \intr\ positives / 1,272 negatives). \emph{AMI meetings} uses 15 four-party headset meetings (7.09\,h; ES2002a--d, ES2003a, ES2004a, ES2005a, IS1000a--b, IS1001a--b, TS3003a--b, TS3004a--b) with floor events derived from the manual segment annotations (1,458 \eot\ positives / 862 hold spans; 1,295 \intr\ positives / 1,255 negatives; mapping in Appendix~\ref{app:real_audio_eval}). On both real slices thresholds are tuned in-slice.

The single system evaluated on all four sources is VAP. Its \eot\ recall decreases from 92.6\% (MPTB) to 84.5\% (TurnBench official), 79.1\% (dev slice) and 52.1\% (AMI). Its \intr\ recall is 94.5\% on official dyadic TurnBench, 61.7\% on the dev slice, 35.9\% on four-party AMI and 11.7\% on MPTB, with FPR between 2.8\% and 10.7\% in every case. AMI is also denser in interruptions than MPTB (3.04 vs.\ 0.89 per minute).

\providecolor{mptbBudgetGray}{RGB}{240,243,246}
\begin{table*}[t]
\centering
\footnotesize
\setlength{\tabcolsep}{3.0pt}
\renewcommand{\arraystretch}{1.14}
\resizebox{\textwidth}{!}{%
\begin{tabular}{@{}lcccccccc@{}}
\toprule
& \multicolumn{2}{c}{\textbf{MPTB (ours)}} & \multicolumn{2}{c}{\textbf{TurnBench official}} & \multicolumn{2}{c}{\textbf{TurnBench dev slice}} & \multicolumn{2}{c}{\textbf{AMI meetings}} \\
\cmidrule(lr){2-3}\cmidrule(lr){4-5}\cmidrule(lr){6-7}\cmidrule(l){8-9}
\textit{Audio} & \multicolumn{2}{c}{\scriptsize synthetic TTS + noise} & \multicolumn{2}{c}{\scriptsize real, studio} & \multicolumn{2}{c}{\scriptsize real, studio} & \multicolumn{2}{c}{\scriptsize real, headset} \\
\textit{Speakers} & \multicolumn{2}{c}{\scriptsize $N=2$--$5$} & \multicolumn{2}{c}{\scriptsize $N=2$} & \multicolumn{2}{c}{\scriptsize $N=2$} & \multicolumn{2}{c}{\scriptsize $N=4$} \\
\textit{Labels} & \multicolumn{2}{c}{\scriptsize generator script (exact)} & \multicolumn{2}{c}{\scriptsize 3-annotator consensus} & \multicolumn{2}{c}{\scriptsize 3-annotator consensus} & \multicolumn{2}{c}{\scriptsize derived from manual segments} \\
\textit{Threshold} & \multicolumn{2}{c}{\scriptsize MPTB dev $\rightarrow$ test} & \multicolumn{2}{c}{\scriptsize TB dev $\rightarrow$ test} & \multicolumn{2}{c}{\scriptsize tuned in-slice} & \multicolumn{2}{c}{\scriptsize tuned in-slice} \\
\textit{Conversations} & \multicolumn{2}{c}{\scriptsize 192 test} & \multicolumn{2}{c}{\scriptsize 116 test} & \multicolumn{2}{c}{\scriptsize 12} & \multicolumn{2}{c}{\scriptsize 15} \\
\midrule
\textbf{System} & \textbf{EOT} & \textbf{INT} & \textbf{EOT} & \textbf{INT} & \textbf{EOT} & \textbf{INT} & \textbf{EOT} & \textbf{INT} \\
\midrule
RMS Energy & 49.8\,/\,40.9 & 66.2\,/\,25.0 & 71.8\,/\,63.2 & 99.6\,/\,44.5 & --- & --- & 28.9\,/\,17.4 & \cellcolor{mptbBudgetGray}30.0\,/\,8.2 \\
Smart Turn & 82.5\,/\,11.8 & 67.6\,/\,64.3 & \cellcolor{mptbBudgetGray}75.2\,/\,4.7 & \cellcolor{mptbBudgetGray}10.7\,/\,9.3 & \cellcolor{mptbBudgetGray}28.6\,/\,7.2 & --- & \cellcolor{mptbBudgetGray}27.3\,/\,4.2 & --- \\
Kyutai SVAD & \cellcolor{mptbBudgetGray}92.5\,/\,4.3 & \cellcolor{mptbBudgetGray}56.2\,/\,6.6 & \cellcolor{mptbBudgetGray}77.3\,/\,5.9 & \cellcolor{mptbBudgetGray}89.8\,/\,8.1 & 86.1\,/\,31.3 & --- & 62.2\,/\,41.1 & --- \\
VAP & \cellcolor{mptbBudgetGray}92.6\,/\,7.5 & \cellcolor{mptbBudgetGray}11.7\,/\,2.8 & \cellcolor{mptbBudgetGray}84.5\,/\,5.5 & \cellcolor{mptbBudgetGray}94.5\,/\,10.7 & 79.1\,/\,26.4 & \cellcolor{mptbBudgetGray}61.7\,/\,5.1 & \cellcolor{mptbBudgetGray}52.1\,/\,8.2 & \cellcolor{mptbBudgetGray}35.9\,/\,5.3 \\
Mimi Endpointer & \cellcolor{mptbBudgetGray}90.5\,/\,3.8 & \cellcolor{mptbBudgetGray}13.9\,/\,3.8 & \cellcolor{mptbBudgetGray}78.2\,/\,7.8 & \cellcolor{mptbBudgetGray}89.9\,/\,10.6 & --- & --- & --- & --- \\
WavLM Anchor & 90.4\,/\,22.6 & \cellcolor{mptbBudgetGray}12.8\,/\,7.3 & \cellcolor{mptbBudgetGray}80.0\,/\,5.4 & \cellcolor{mptbBudgetGray}86.8\,/\,5.4 & --- & --- & --- & --- \\
ESPnet Turn-Taking & \cellcolor{mptbBudgetGray}86.9\,/\,12.4 & \cellcolor{mptbBudgetGray}1.1\,/\,1.9 & \cellcolor{mptbBudgetGray}82.6\,/\,7.8 & \cellcolor{mptbBudgetGray}57.3\,/\,8.0 & --- & --- & --- & --- \\
\bottomrule
\end{tabular}%
}
\caption{
\textbf{Interruption detection remains challenging in natural multi-party speech.} Recall\,/\,FPR (\%) for systems evaluated on synthetic MPTB, the official TurnBench leaderboard (real dyadic speech; TurnBench's own runs), our TurnBench dev slice, and real four-party AMI meetings. \colorbox{mptbBudgetGray}{Gray}: within that source's FPR budget. Sources differ in audio, labels and operating-point selection (header rows), and both real slices tune thresholds on the evaluated data. Absolute numbers are therefore not comparable across columns but only in directions. Every learned system has lower EOT recall on real audio than on MPTB. Interruptions, however, collapse in multi-party speech on both real and synthetic audio. VAP, the one learned system evaluated on every source, detects 94.5\% of interruptions in real dyadic TurnBench, but only 35.9\% in real four-party AMI, even with in-slice tuning, and 11.7\% on MPTB. VAP, Mimi Endpointer and WavLM Anchor are trained on TurnBench data and are in-domain there.}
\label{tab:real_vs_synthetic}
\end{table*}

\subsection{EOT and INT detection recall-FPR breakdown by interaction type}
\label{app:type_breakdown}

Figure~\ref{fig:eot_int_type_heatmaps} reports recall and FPR for all twelve systems in each interaction type, and Figure~\ref{fig:type_analysis} isolates the five systems that are within budget on both tracks. Each cell uses the pooled operating point; per-type cells contain 32 test conversations, 364--601 \eot\ positives, 21--38 hold spans, 32--66 \intr\ positives and 154--225 \intr\ negatives.

\providecolor{mptbHeatBlue}{RGB}{33,102,172}
\providecolor{mptbHeatOrange}{RGB}{217,95,2}
\begin{figure*}[t]
\centering
\resizebox{\textwidth}{!}{%
%
}
\caption{\textbf{EOT and INT performance across interaction types.} Recall (blue) and false-positive rate (orange) per conversation type on the test split (32 conversations per type), at each system's pooled operating point. Darker means higher; scales are shared across panels. EOT recall barely depends on the regime: every learned system stays within 6.5 points across types (VAP 91.2--93.7\%). INT does depend on it. The two strongest INT systems both peak on asymmetric \emph{instructional} talk (Kyutai 77.5\%, DualTurn 47.5\%), and Kyutai drops to 47.0\% in overlap-heavy \emph{collaborative} talk. Learned INT false alarms stay at or below 11.5\% in every type (excluding the Smart Turn inversion). For Kyutai, Mimi-EP and DualTurn they peak in backchannel-dense \emph{casual} conversation, as TurnBench reports for dyadic speech. Per-type EOT FPR rests on only 21--38 negative spans, so differences between types there are not reliable. Kyutai and Smart Turn INT are inversions of end-of-turn scores; --- marks no INT output.}
\label{fig:eot_int_type_heatmaps}
\end{figure*}


\providecolor{mptbT1}{RGB}{31,119,180}
\providecolor{mptbT2}{RGB}{255,127,14}
\providecolor{mptbT3}{RGB}{44,160,44}
\providecolor{mptbT4}{RGB}{214,39,40}
\providecolor{mptbT5}{RGB}{148,103,189}
\providecolor{mptbT6}{RGB}{140,86,75}
\begin{figure}[t]
\centering
\resizebox{\linewidth}{!}{%
\begin{tikzpicture}
\begin{groupplot}[
    group style={group size=1 by 2,vertical sep=1.35cm},
    width=8.6cm,height=3.8cm,
    axis lines=left,axis line style={black!65,line width=0.55pt},
    tick label style={font=\scriptsize},title style={font=\small\bfseries,yshift=-3pt},
    xmajorgrids,grid style={black!10,line width=0.3pt},
    ytick style={draw=none},clip=false,
]
\nextgroupplot[title={(a) End-of-Turn recall (\%)},xmin=80,xmax=100,xtick={80,85,90,95,100},ymin=0.4,ymax=5.6,ytick={1,...,5},yticklabels={{DualTurn},{MuVAP},{VAP},{Mimi EP},{Kyutai SVAD}},legend to name=mptbTypeLegend,legend columns=3,legend style={draw=none,fill=none,font=\scriptsize,column sep=4pt},legend cell align=left]
\draw[black!14,line width=5pt,line cap=round] (axis cs:90.5,5) -- (axis cs:96.0,5);
\draw[black,line width=1.1pt] (axis cs:92.5,4.66) -- (axis cs:92.5,5.34);
\draw[black!14,line width=5pt,line cap=round] (axis cs:88.7,4) -- (axis cs:92.2,4);
\draw[black,line width=1.1pt] (axis cs:90.5,3.66) -- (axis cs:90.5,4.34);
\draw[black!14,line width=5pt,line cap=round] (axis cs:91.2,3) -- (axis cs:93.7,3);
\draw[black,line width=1.1pt] (axis cs:92.6,2.66) -- (axis cs:92.6,3.34);
\draw[black!14,line width=5pt,line cap=round] (axis cs:86.2,2) -- (axis cs:89.4,2);
\draw[black,line width=1.1pt] (axis cs:88.1,1.66) -- (axis cs:88.1,2.34);
\draw[black!14,line width=5pt,line cap=round] (axis cs:87.4,1) -- (axis cs:90.6,1);
\draw[black,line width=1.1pt] (axis cs:89.4,0.6599999999999999) -- (axis cs:89.4,1.34);
\addplot[only marks,mark=*,mark size=1.7pt,draw=mptbT1,fill=mptbT1] coordinates {(95.1,4.75) (90.4,3.75) (92.9,2.75) (86.2,1.75) (89.2,0.75)};
\addlegendentry{Argumentative}
\addplot[only marks,mark=square*,mark size=1.7pt,draw=mptbT2,fill=mptbT2] coordinates {(91.0,4.85) (92.2,3.85) (93.7,2.85) (88.9,1.85) (90.0,0.85)};
\addlegendentry{Casual}
\addplot[only marks,mark=triangle*,mark size=1.7pt,draw=mptbT3,fill=mptbT3] coordinates {(90.5,4.95) (89.2,3.95) (92.5,2.95) (89.4,1.95) (89.2,0.95)};
\addlegendentry{Collaborative}
\addplot[only marks,mark=diamond*,mark size=1.7pt,draw=mptbT4,fill=mptbT4] coordinates {(94.0,5.05) (88.7,4.05) (91.5,3.05) (87.4,2.05) (87.4,1.05)};
\addlegendentry{Instructional}
\addplot[only marks,mark=pentagon*,mark size=1.7pt,draw=mptbT5,fill=mptbT5] coordinates {(96.0,5.15) (90.4,4.15) (91.2,3.15) (88.0,2.15) (89.6,1.15)};
\addlegendentry{Narrative}
\addplot[only marks,mark=triangle*,mark size=1.7pt,draw=mptbT6,fill=mptbT6,mark options={rotate=180,fill=mptbT6}] coordinates {(90.6,5.25) (91.2,4.25) (93.0,3.25) (88.3,2.25) (90.6,1.25)};
\addlegendentry{Task-oriented}
\nextgroupplot[title={(b) Interruption recall (\%)},xmin=0,xmax=80,xtick={0,20,40,60,80},ymin=0.4,ymax=5.6,ytick={1,...,5},yticklabels={{DualTurn},{MuVAP},{VAP},{Mimi EP},{Kyutai SVAD}}]
\draw[black!14,line width=5pt,line cap=round] (axis cs:47.0,5) -- (axis cs:77.5,5);
\draw[black,line width=1.1pt] (axis cs:56.2,4.66) -- (axis cs:56.2,5.34);
\draw[black!14,line width=5pt,line cap=round] (axis cs:9.1,4) -- (axis cs:15.6,4);
\draw[black,line width=1.1pt] (axis cs:13.9,3.66) -- (axis cs:13.9,4.34);
\draw[black!14,line width=5pt,line cap=round] (axis cs:7.3,3) -- (axis cs:16.3,3);
\draw[black,line width=1.1pt] (axis cs:11.7,2.66) -- (axis cs:11.7,3.34);
\draw[black!14,line width=5pt,line cap=round] (axis cs:25.0,2) -- (axis cs:41.9,2);
\draw[black,line width=1.1pt] (axis cs:34.5,1.66) -- (axis cs:34.5,2.34);
\draw[black!14,line width=5pt,line cap=round] (axis cs:32.6,1) -- (axis cs:47.5,1);
\draw[black,line width=1.1pt] (axis cs:37.0,0.6599999999999999) -- (axis cs:37.0,1.34);
\addplot[only marks,mark=*,mark size=1.7pt,draw=mptbT1,fill=mptbT1] coordinates {(54.5,4.75) (9.1,3.75) (7.3,2.75) (30.9,1.75) (36.4,0.75)};
\addplot[only marks,mark=square*,mark size=1.7pt,draw=mptbT2,fill=mptbT2] coordinates {(53.3,4.85) (15.6,3.85) (15.6,2.85) (37.8,1.85) (40.0,0.85)};
\addplot[only marks,mark=triangle*,mark size=1.7pt,draw=mptbT3,fill=mptbT3] coordinates {(47.0,4.95) (15.2,3.95) (9.1,2.95) (34.8,1.95) (33.3,0.95)};
\addplot[only marks,mark=diamond*,mark size=1.7pt,draw=mptbT4,fill=mptbT4] coordinates {(77.5,5.05) (15.0,4.05) (15.0,3.05) (35.0,2.05) (47.5,1.05)};
\addplot[only marks,mark=pentagon*,mark size=1.7pt,draw=mptbT5,fill=mptbT5] coordinates {(59.4,5.15) (15.6,4.15) (9.4,3.15) (25.0,2.15) (34.4,1.15)};
\addplot[only marks,mark=triangle*,mark size=1.7pt,draw=mptbT6,fill=mptbT6,mark options={rotate=180,fill=mptbT6}] coordinates {(53.5,5.25) (14.0,4.25) (16.3,3.25) (41.9,2.25) (32.6,1.25)};
\end{groupplot}
\end{tikzpicture}%
}
\vspace{1pt}
\pgfplotslegendfromname{mptbTypeLegend}
\vspace{-2pt}
\caption{\textbf{Interaction type matters for interruptions, not for turn ends.} Recall per conversation type for the five systems within the FPR budget on both tracks (test split, 32 conversations per type). Gray bars span each system's worst-to-best type and black ticks mark pooled recall. EOT recall varies by at most 5.5 points across types. INT recall swings by up to 30.5 points (Kyutai: 47.0\% collaborative vs.\ 77.5\% instructional). The hardest INT type is model-specific: argumentative for VAP and Mimi EP, collaborative for Kyutai, narrative for MuVAP, and task-oriented for DualTurn.}
\label{fig:type_analysis}
\end{figure}
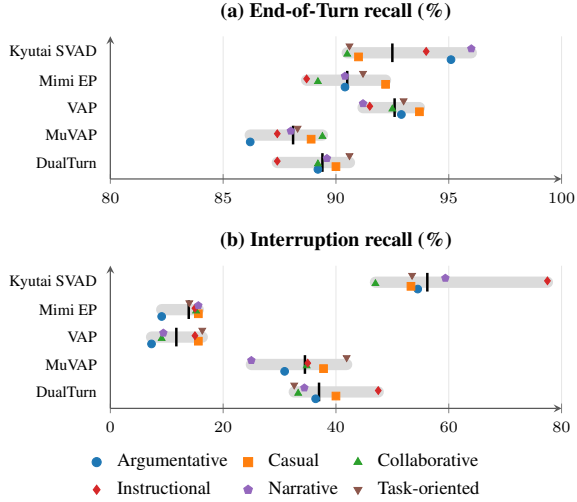

\subsection{Performance by party count}
\label{app:by_n}

Table~\ref{tab:vap_by_n} reports VAP by party count. Table~\ref{tab:by_n_all} extends the analysis to every system. Each $N$ contributes 48 test conversations (\eot: 695/661/676/684 positives and 55/48/44/39 hold spans; \intr: 60/73/72/76 positives and 276/288/277/331 negatives for $N=2..5$). Three patterns stand out. (i) Kyutai's \eot\ recall rises from 84.6\% to 96.6\% with FPR at or below 9.1\% throughout, while its \intr\ recall falls from 63.3\% to 48.7\% together with its \intr\ FPR (10.5\% to 3.9\%). (ii) MuVAP's \intr\ recall gain from 23.3\% to 40.8\% is accompanied by an FPR increase from 5.4\% to 12.1\%; DualTurn gains 6.2 points with FPR between 6.9\% and 10.5\%. (iii) Acoustic VADs recover more interruptions as $N$ grows (WebRTC 70.0\% to 93.4\%, Silero 73.3\% to 92.1\%) at a constant 82--90\% \intr\ FPR, since more speakers produce more overlap onsets of every kind.

\begin{table}[t]
\centering
\small
\begin{tabular}{ccc}
\toprule
$N$ & \eot\ recall/FPR & \intr\ recall/FPR\\
\midrule
2 & .93/.04 & .07/.03\\
3 & .93/.04 & .10/.04\\
4 & .92/.16 & .11/.03\\
5 & .92/.08 & .18/.02\\
\bottomrule
\end{tabular}
\caption{VAP test performance by party count. The $N{=}4$ \eot\ FPR exceeds the nominal budget after the dev-selected threshold is frozen.}
\label{tab:vap_by_n}
\end{table}

\begin{table*}[t]
\centering
\scriptsize
\setlength{\tabcolsep}{3.4pt}
\renewcommand{\arraystretch}{1.12}
\begin{tabular}{@{}lcccc|cccc@{}}
\toprule
& \multicolumn{4}{c|}{\textbf{\eot\ recall\,/\,FPR (\%)}} & \multicolumn{4}{c}{\textbf{\intr\ recall\,/\,FPR (\%)}} \\
\textbf{System} & $N{=}2$ & $N{=}3$ & $N{=}4$ & $N{=}5$ & $N{=}2$ & $N{=}3$ & $N{=}4$ & $N{=}5$ \\
\midrule
\multicolumn{9}{@{}l}{\textit{Acoustic VAD}} \\
RMS Energy & 48.5\,/\,38.2 & 61.1\,/\,62.5 & 41.9\,/\,25.0 & 48.1\,/\,35.9 & 55.0\,/\,22.8 & 74.0\,/\,34.4 & 66.7\,/\,18.8 & 67.1\,/\,23.9 \\
WebRTC VAD & 92.1\,/\,85.5 & 93.8\,/\,95.8 & 90.4\,/\,86.4 & 90.9\,/\,94.9 & 70.0\,/\,83.0 & 82.2\,/\,83.0 & 88.9\,/\,82.7 & 93.4\,/\,86.1 \\
Silero VAD & 97.8\,/\,94.5 & 98.0\,/\,93.8 & 96.0\,/\,95.5 & 96.9\,/\,89.7 & 73.3\,/\,89.1 & 82.2\,/\,89.2 & 91.7\,/\,88.4 & 92.1\,/\,90.0 \\
\midrule
\multicolumn{9}{@{}l}{\textit{Semantic endpointers}} \\
Smart Turn v2 & 83.5\,/\,18.2 & 82.3\,/\,10.4 & 83.3\,/\,11.4 & 81.0\,/\,5.1 & 55.0\,/\,59.4 & 67.1\,/\,60.4 & 68.1\,/\,71.1 & 77.6\,/\,66.2 \\
Kyutai SVAD & 84.6\,/\,1.8 & 92.1\,/\,4.2 & 96.9\,/\,9.1 & 96.6\,/\,2.6 & 63.3\,/\,10.5 & 63.0\,/\,6.6 & 51.4\,/\,5.8 & 48.7\,/\,3.9 \\
Mimi Endpointer & 90.2\,/\,3.6 & 89.7\,/\,4.2 & 92.0\,/\,2.3 & 90.1\,/\,5.1 & 13.3\,/\,5.8 & 19.2\,/\,2.1 & 12.5\,/\,4.7 & 10.5\,/\,3.0 \\
\midrule
\multicolumn{9}{@{}l}{\textit{Turn-taking models}} \\
VAP & 93.2\,/\,3.6 & 93.2\,/\,4.2 & 92.0\,/\,15.9 & 92.0\,/\,7.7 & 6.7\,/\,2.5 & 9.6\,/\,3.8 & 11.1\,/\,3.2 & 18.4\,/\,1.8 \\
MuVAP & 88.6\,/\,5.5 & 86.8\,/\,8.3 & 88.8\,/\,13.6 & 88.3\,/\,15.4 & 23.3\,/\,5.4 & 32.9\,/\,7.6 & 38.9\,/\,8.7 & 40.8\,/\,12.1 \\
DualTurn & 89.4\,/\,7.3 & 89.4\,/\,14.6 & 89.3\,/\,9.1 & 89.6\,/\,12.8 & 33.3\,/\,8.0 & 35.6\,/\,6.9 & 38.9\,/\,10.5 & 39.5\,/\,7.6 \\
WavLM Anchor & 90.2\,/\,29.1 & 90.5\,/\,22.9 & 90.1\,/\,15.9 & 90.8\,/\,20.5 & 11.7\,/\,5.8 & 11.0\,/\,6.2 & 11.1\,/\,7.2 & 17.1\,/\,9.4 \\
Turn-1-Mini & 86.3\,/\,20.0 & 85.6\,/\,31.2 & 89.6\,/\,22.7 & 86.0\,/\,28.2 & --- & --- & --- & --- \\
ESPnet Turn-Taking (mix) & 86.6\,/\,7.3 & 85.6\,/\,18.8 & 89.2\,/\,13.6 & 86.0\,/\,10.3 & 1.7\,/\,1.1 & 1.4\,/\,3.8 & 0.0\,/\,1.8 & 1.3\,/\,0.9 \\
ESPnet Turn-Taking (per-ch.) & 4.0\,/\,9.1 & 5.4\,/\,2.1 & 4.6\,/\,6.8 & 4.8\,/\,7.7 & 5.0\,/\,6.2 & 6.8\,/\,7.3 & 4.2\,/\,6.1 & 3.9\,/\,4.5 \\
\bottomrule
\end{tabular}
\caption{\small\textbf{Recall\,/\,FPR by party count for every system} (test split, pooled operating point). The ESPnet per-channel variant runs the same checkpoint on each isolated channel instead of the mono mix.}
\label{tab:by_n_all}
\end{table*}

\subsection{Operating Points}
\label{app:operating_points}

Table~\ref{tab:operating_points} lists the frozen threshold of every system together with development and test recall/FPR. Two learned systems satisfy the development constraint on \eot\ but exceed the test ceiling: WavLM Anchor (dev FPR 7.3\%, test 22.6\%) and Turn-1-Mini (dev 9.1\%, test 25.3\%). Smart Turn's \intr\ inversion has no threshold with dev FPR below 58.9\%.

\begin{table*}[t]
\centering
\scriptsize
\setlength{\tabcolsep}{3.6pt}
\renewcommand{\arraystretch}{1.12}
\begin{tabular}{@{}l l cc l cc@{}}
\toprule
& \multicolumn{3}{c}{\textbf{\eot}} & \multicolumn{3}{c}{\textbf{\intr}} \\
\cmidrule(lr){2-4}\cmidrule(l){5-7}
\textbf{System} & $\theta$ & \textbf{Dev R\,/\,F} & \textbf{Test R\,/\,F} & $\theta$ & \textbf{Dev R\,/\,F} & \textbf{Test R\,/\,F} \\
\midrule
RMS Energy & $-50$\,dBFS & 44.9\,/\,30.9 & 49.8\,/\,40.9 & $-50$\,dBFS & 51.2\,/\,19.7 & 66.2\,/\,25.0 \\
WebRTC VAD & aggr.\ 3 & 91.6\,/\,87.3 & 91.8\,/\,90.3 & aggr.\ 3 & 83.7\,/\,84.5 & 84.3\,/\,83.8 \\
Silero VAD & 0.30 & 97.6\,/\,90.9 & 97.2\,/\,93.5 & 0.70 & 83.7\,/\,86.4 & 85.4\,/\,89.2 \\
Smart Turn v2 & 0.50 & 78.3\,/\,14.5 & 82.5\,/\,11.8 & 0.0442 & 73.3\,/\,58.9 & 67.6\,/\,64.3 \\
\rowcolor{mptbBudgetGray} Kyutai SVAD & 0.95 & 93.2\,/\,3.6 & 92.5\,/\,4.3 & 0.9991 & 65.1\,/\,6.1 & 56.2\,/\,6.6 \\
\rowcolor{mptbBudgetGray} Mimi Endpointer & 0.95 & 90.9\,/\,1.8 & 90.5\,/\,3.8 & 0.99 & 25.6\,/\,3.2 & 13.9\,/\,3.8 \\
\rowcolor{mptbBudgetGray} VAP & 0.80 & 92.9\,/\,9.1 & 92.6\,/\,7.5 & 0.90 & 12.8\,/\,3.6 & 11.7\,/\,2.8 \\
\rowcolor{mptbBudgetGray} MuVAP & 0.95 & 88.5\,/\,5.5 & 88.1\,/\,10.2 & 0.90 & 33.7\,/\,5.8 & 34.5\,/\,8.6 \\
\rowcolor{mptbBudgetGray} DualTurn & 0.40 & 90.5\,/\,7.3 & 89.4\,/\,10.8 & 0.9711 & 45.3\,/\,6.8 & 37.0\,/\,8.2 \\
WavLM Anchor & 0.75 & 91.5\,/\,7.3 & 90.4\,/\,22.6 & 0.80 & 15.1\,/\,8.7 & 12.8\,/\,7.3 \\
Turn-1-Mini & 0.35 & 87.7\,/\,9.1 & 86.9\,/\,25.3 & --- & --- & --- \\
\rowcolor{mptbBudgetGray} ESPnet Turn-Taking (mix) & 0.00136 & 87.0\,/\,9.1 & 86.9\,/\,12.4 & 0.50 & 2.3\,/\,1.6 & 1.1\,/\,1.9 \\
ESPnet Turn-Taking (per-ch.) & 0.70 & 6.1\,/\,7.3 & 4.7\,/\,6.5 & 0.50 & 4.7\,/\,8.1 & 5.0\,/\,6.0 \\
\bottomrule
\end{tabular}
\caption{\small\textbf{Dev-selected operating points.} R/F = recall/FPR (\%). $\theta$ is the score threshold (dBFS for RMS, WebRTC aggressiveness level, probability otherwise). \colorbox{mptbBudgetGray}{Gray} rows are in budget on both tracks. Candidate grids: RMS $\{-50,-45,\dots,-30\}$\,dBFS; WebRTC $\{0,1,2,3\}$; Silero $\{0.3,\dots,0.7\}$; Kyutai and Smart Turn \eot\ $\{0.3,\dots,0.9,0.95\}$; VAP and MuVAP $\{0.5,\dots,0.9,0.95\}$; other learned systems $\{0.05,0.10,\dots,0.95,0.97,0.98,0.99,0.995\}$ plus dev-score quantiles.}
\label{tab:operating_points}
\end{table*}

\subsection{Latency Distributions}
\label{app:latency}

Table~\ref{tab:latency_all} reports the full signed-latency distribution of matched detections. The positive-anchor window admits detections from $-250$ to $+3000$\,ms, so p10 values below zero indicate anticipatory firing. Smart Turn scores candidate turn ends proposed by an energy VAD and stamps each decision at the start of the silence, before the VAD's 0.3\,s confirmation hangover; its \eot\ latencies therefore exclude that hangover and can be negative. Among in-budget systems, VAP is the only one with a sub-200\,ms median \eot\ decision; Kyutai and Mimi Endpointer commit over a second after the turn ends.

\begin{table}[t]
\centering
\scriptsize
\setlength{\tabcolsep}{3pt}
\renewcommand{\arraystretch}{1.1}
\begin{tabular}{@{}lrrr|rrr@{}}
\toprule
& \multicolumn{3}{c|}{\textbf{\eot\ (ms)}} & \multicolumn{3}{c}{\textbf{\intr\ (ms)}} \\
\textbf{System} & p10 & p50 & p90 & p10 & p50 & p90 \\
\midrule
RMS Energy & 267 & 292 & 1653 & $-38$ & 859 & 2572 \\
WebRTC VAD & 224 & 332 & 380 & 114 & 128 & 146 \\
Silero VAD & 359 & 397 & 435 & 92 & 119 & 198 \\
Smart Turn v2 & $-32$ & $-18$ & 614 & 55 & 87 & 446 \\
Kyutai SVAD & 666 & 1829 & 1960 & 432 & 689 & 1121 \\
Mimi Endpointer & 878 & 1231 & 1630 & 346 & 870 & 2298 \\
VAP & $-9$ & 143 & 555 & 299 & 492 & 1340 \\
MuVAP & 90 & 553 & 1563 & 164 & 618 & 1129 \\
DualTurn & 404 & 777 & 1812 & 224 & 572 & 1122 \\
WavLM Anchor & 576 & 1185 & 1734 & $-167$ & 42 & 971 \\
Turn-1-Mini & 635 & 1087 & 1384 & --- & --- & --- \\
ESPnet (mix) & 482 & 1084 & 2205 & 66 & 71 & 88 \\
\bottomrule
\end{tabular}
\caption{\small\textbf{Signed latency} $t_{\text{pred}}-t_{\text{gold}}$ of matched detections on the test split. ESPnet \intr\ percentiles rest on 3 matched detections.}
\label{tab:latency_all}
\end{table}

\subsection{Paired Comparisons}
\label{app:paired}

Table~\ref{tab:paired} reports paired conversation-cluster bootstrap differences against VAP, each system at its own operating point. On \eot, Kyutai and VAP are statistically indistinguishable in recall ($-0.1$ points, CI $[-1.5,1.3]$), while every other learned system has significantly lower recall than VAP. On \intr, Kyutai (+44.5), DualTurn (+25.3) and MuVAP (+22.8) significantly exceed VAP, each with a small but significant FPR increase (3.8--5.8 points); Mimi Endpointer and WavLM Anchor do not differ from VAP in \intr\ recall.

\begin{table}[t]
\centering
\scriptsize
\setlength{\tabcolsep}{2.6pt}
\renewcommand{\arraystretch}{1.1}
\resizebox{\columnwidth}{!}{%
\begin{tabular}{@{}llrr@{}}
\toprule
\textbf{System $-$ VAP} & & \textbf{$\Delta$ Recall} & \textbf{$\Delta$ FPR} \\
\midrule
\eot & Kyutai SVAD & $-0.1$ [$-1.5$, 1.3] & $-3.2$ [$-7.7$, 1.1] \\
& Mimi EP & $-2.1$ [$-2.9$, $-1.3$] & $-3.8$ [$-8.1$, 0.5] \\
& DualTurn & $-3.2$ [$-4.0$, $-2.3$] & $+3.2$ [$-1.6$, 8.2] \\
& MuVAP & $-4.5$ [$-5.6$, $-3.4$] & $+2.7$ [$-1.8$, 7.2] \\
& WavLM & $-2.2$ [$-3.0$, $-1.4$] & $+15.1$ [7.0, 23.1] \\
& Turn-1-Mini & $-5.7$ [$-6.9$, $-4.5$] & $+17.7$ [10.2, 25.5] \\
& ESPnet (mix) & $-5.7$ [$-7.0$, $-4.5$] & $+4.8$ [$-0.6$, 10.2] \\
\midrule
\intr & Kyutai SVAD & $+44.5$ [37.5, 51.3] & $+3.8$ [2.0, 5.5] \\
& DualTurn & $+25.3$ [18.2, 32.2] & $+5.4$ [3.5, 7.3] \\
& MuVAP & $+22.8$ [17.8, 28.0] & $+5.8$ [4.2, 7.5] \\
& Mimi EP & $+2.1$ [$-2.2$, 6.6] & $+1.0$ [$-0.3$, 2.4] \\
& WavLM & $+1.1$ [$-4.1$, 6.1] & $+4.4$ [2.6, 6.3] \\
& ESPnet (mix) & $-10.7$ [$-14.5$, $-7.1$] & $-0.9$ [$-2.3$, 0.3] \\
& Smart Turn v2 & $+55.9$ [49.1, 62.5] & $+61.5$ [58.5, 64.4] \\
\bottomrule
\end{tabular}%
}
\caption{\small\textbf{Paired differences to VAP} (points, 95\% CI; $B=10{,}000$ conversation-cluster resamples of the 192 test conversations).}
\label{tab:paired}
\end{table}

\subsection{Robustness to Acoustic Conditions}
\label{app:robustness}

Table~\ref{tab:robustness} slices the five systems in budget on both tracks, and Silero VAD as an acoustic reference, by ambient SNR and by the presence of distant third-party speech. Within the rendered 20--38\,dB range, no learned system shows a monotonic loss of \eot\ recall at lower SNR (VAP: 92.4--92.9\% in every bin). Distant speech raises learned \intr\ FPR by at most 1.2 points (Mimi Endpointer 3.6\% to 4.8\%); for VAP it is 2.8\% with and without, and for Kyutai, MuVAP and DualTurn it is lower when distant speech is present. Floor-control errors on MPTB are therefore not explained by the acoustic nuisance layer.

\begin{table*}[t]
\centering
\scriptsize
\setlength{\tabcolsep}{3.8pt}
\renewcommand{\arraystretch}{1.12}
\resizebox{\textwidth}{!}{%
\begin{tabular}{@{}lcccc|cccc|cc|cc@{}}
\toprule
& \multicolumn{4}{c|}{\textbf{\eot\ recall by SNR (dB)}} & \multicolumn{4}{c|}{\textbf{\intr\ recall by SNR (dB)}} & \multicolumn{2}{c|}{\textbf{\eot\ FPR, distant speech}} & \multicolumn{2}{c}{\textbf{\intr\ FPR, distant speech}} \\
\textbf{System} & 20--25 & 25--30 & 30--35 & 35--40 & 20--25 & 25--30 & 30--35 & 35--40 & absent & present & absent & present \\
\midrule
Kyutai SVAD & 91.9 & 94.5 & 92.8 & 89.6 & 52.4 & 56.7 & 59.3 & 55.3 & 3.9 & 6.1 & 6.8 & 5.6 \\
Mimi Endpointer & 91.2 & 91.5 & 89.2 & 89.8 & 12.7 & 13.3 & 16.0 & 12.8 & 4.6 & 0.0 & 3.6 & 4.8 \\
VAP & 92.7 & 92.5 & 92.4 & 92.9 & 12.7 & 10.0 & 13.6 & 10.6 & 7.8 & 6.1 & 2.8 & 2.8 \\
MuVAP & 88.8 & 88.7 & 87.2 & 87.6 & 25.4 & 33.3 & 40.7 & 38.3 & 10.5 & 9.1 & 9.3 & 6.0 \\
DualTurn & 89.7 & 89.6 & 89.5 & 88.8 & 34.9 & 34.4 & 39.5 & 40.4 & 10.5 & 12.1 & 8.8 & 6.0 \\
Silero VAD$^\dagger$ & 96.9 & 97.6 & 95.9 & 98.2 & 82.5 & 83.3 & 87.7 & 91.5 & 93.5 & 93.9 & 90.2 & 85.3 \\
\midrule
\textit{Conversations} & 44 & 59 & 52 & 37 & 44 & 59 & 52 & 37 & 151 & 41 & 151 & 41 \\
\bottomrule
\end{tabular}%
}
\caption{\small\textbf{Robustness slices} (test split, \%, pooled operating points). Signal to Noise Ratio (SNR) bins refer to the ambient bed.}
\label{tab:robustness}
\end{table*}

\subsection{Error Anatomy}
\label{app:error_anatomy}

To localize errors, we join every scored test anchor with its script utterance and score each attribute slice separately (Table~\ref{tab:error_anatomy}).

\paragraph{End-of-turn scores detect sustained listener speech, not floor claims.} Kyutai's \intr\ score fires on 0.2\% of one-word listener vocalizations, 5.7\% of two- or three-word ones, and 72.3\% of negatives with four or more words, all of which are laughter vocalizations such as ``\textit{Haha, yeah, I've been waiting all week for this one}''. Its recall on true interruptions also grows with claim length (37.2\% for 1--3 words vs.\ 62.6\% for 9+ words). Its \intr\ advantage over VAP (Table~\ref{tab:paired}) therefore comes largely from duration, not from modelling floor ownership. DualTurn and MuVAP show the same direction with flatter slopes (26.2\% FPR on 4+-word negatives), and all learned systems false-alarm more on laughter than on backchannels except ESPnet.

\paragraph{Hold pauses after complete questions look like yields.} For Kyutai, Mimi Endpointer, MuVAP, DualTurn and Smart Turn, false \eot\ triggers on hold pauses are most frequent when the pre-pause clause ends in a question mark (33 spans; e.g., DualTurn 18.2\% vs.\ 10.8\% after a period, Smart Turn 24.2\% vs.\ 11.7\%), and every learned system detects true yields after questions more reliably than yields after statements (e.g., VAP 96.9\% vs.\ 90.8\%). This is consistent with reliance on clause completion, a cue that a hold after a question does not follow. WavLM Anchor false-alarms on 20.5--23.4\% of hold spans of every clause type and Turn-1-Mini on 27.3--29.7\% after complete clauses, which is the source of their over-budget test FPR.

\paragraph{Contested holds are the multi-party \eot\ failure.} In 28 of the 186 test hold spans, another listener's failed floor claim overlaps the holder's pause, so the floor is briefly contested before the holder resumes. These contested holds account for most false \eot\ triggers of the speaker-versus-rest models (Table~\ref{tab:contested_holds}): VAP false-alarms on 39.3\% of them versus 1.9\% of uncontested holds (11 of its 14 false alarms), MuVAP on 57.1\% versus 1.9\%, and DualTurn on 42.9\% versus 5.1\%. Kyutai, which scores each channel in isolation, is nearly unaffected (7.1\% vs.\ 3.8\%). These models read a competing voice in the pooled ``rest'' channel as the holder releasing the floor. Contested holds make up 10.9\% of hold spans at $N{=}2$ and 15.4--18.2\% at $N{=}3$--5, and with more listeners the competing voice can come from any of $N-1$ channels that the pooled rest stream merges.

\begin{table}[t]
\centering
\scriptsize
\setlength{\tabcolsep}{3.2pt}
\renewcommand{\arraystretch}{1.1}
\resizebox{\columnwidth}{!}{%
\begin{tabular}{@{}lccc@{}}
\toprule
& \multicolumn{2}{c}{\textbf{\eot\ FPR on holds (\%)}} & \textbf{Share of} \\
\textbf{System} & \textbf{contested} ($n{=}28$) & \textbf{uncontested} ($n{=}158$) & \textbf{false alarms} \\
\midrule
Kyutai SVAD & 7.1 & 3.8 & 2\,/\,8 \\
Mimi Endpointer & 25.0 & 0.0 & 7\,/\,7 \\
VAP & \cellcolor{mptbOrange!25}39.3 & 1.9 & 11\,/\,14 \\
MuVAP & \cellcolor{mptbOrange!35}57.1 & 1.9 & 16\,/\,19 \\
DualTurn & \cellcolor{mptbOrange!25}42.9 & 5.1 & 12\,/\,20 \\
WavLM Anchor & 17.9 & 23.4 & 5\,/\,42 \\
Turn-1-Mini & 32.1 & 24.1 & 9\,/\,47 \\
ESPnet (mix) & 35.7 & 8.2 & 10\,/\,23 \\
\bottomrule
\end{tabular}%
}
\caption{\small\textbf{Contested holds.} A hold is contested when another speaker's non-floor-taking claim overlaps the pause span. Last column: contested false alarms / all \eot\ false alarms on the test split.}
\label{tab:contested_holds}
\end{table}

\begin{table*}[t]
\centering
\scriptsize
\setlength{\tabcolsep}{3.3pt}
\renewcommand{\arraystretch}{1.12}
\begin{tabular}{@{}l ccc cc | ccc cc | ccc@{}}
\toprule
& \multicolumn{5}{c|}{\textbf{\eot\ (pre-pause / pre-yield clause ending)}} & \multicolumn{5}{c|}{\textbf{\intr\ FPR on negatives}} & \multicolumn{3}{c}{\textbf{\intr\ recall by claim length}} \\
\cmidrule(lr){2-6}\cmidrule(lr){7-11}\cmidrule(l){12-14}
& \multicolumn{3}{c}{FPR on holds} & \multicolumn{2}{c|}{Recall on yields} & \multicolumn{3}{c}{by word count} & \multicolumn{2}{c|}{by type} & \multicolumn{3}{c}{words} \\
\textbf{System} & ``.'' & ``?'' & ``---'' & ``.'' & ``?'' & 1 & 2--3 & 4+ & BC & Laugh & 1--3 & 4--8 & 9+ \\
\midrule
Kyutai SVAD & 1.8 & 9.1 & 7.7 & 92.2 & 93.6 & 0.2 & 5.7 & \cellcolor{mptbOrange!35}72.3 & 2.4 & 21.7 & 37.2 & 55.6 & 62.6 \\
Mimi Endpointer & 2.7 & 9.1 & 2.6 & 88.9 & 95.0 & 1.5 & 4.5 & 20.0 & 2.6 & 8.3 & 23.3 & 12.1 & 12.2 \\
VAP & 9.9 & 9.1 & 0.0 & 90.8 & 96.9 & 1.2 & 3.7 & 10.8 & 2.6 & 3.6 & 9.3 & 14.1 & 10.8 \\
MuVAP & 9.9 & 15.2 & 7.7 & 86.2 & 92.9 & 6.0 & 9.4 & 26.2 & 7.3 & 13.4 & 27.9 & 37.4 & 34.5 \\
DualTurn & 10.8 & 18.2 & 5.1 & 87.9 & 92.9 & 2.9 & 12.1 & 26.2 & 6.9 & 13.0 & 27.9 & 36.4 & 40.3 \\
WavLM Anchor & 23.4 & 21.2 & 20.5 & 88.4 & 95.7 & 7.0 & 7.0 & 10.8 & 6.5 & 9.9 & 14.0 & 14.1 & 11.5 \\
Turn-1-Mini & 29.7 & 27.3 & 10.3 & 85.8 & 90.2 & --- & --- & --- & --- & --- & --- & --- & --- \\
ESPnet (mix) & 15.3 & 9.1 & 7.7 & 85.0 & 91.6 & 1.8 & 2.0 & 1.5 & 2.2 & 0.8 & 0.0 & 2.0 & 0.7 \\
Smart Turn v2$^\dagger$ & 11.7 & 24.2 & 10.3 & 82.9 & 84.7 & 59.7 & 68.1 & 76.9 & 63.0 & 69.2 & 79.1 & 67.7 & 64.0 \\
\midrule
\textit{$n$ (test)} & 111 & 33 & 39 & 1,921 & 717 & 596 & 511 & 65 & 919 & 253 & 43 & 99 & 139 \\
\bottomrule
\end{tabular}
\caption{\small\textbf{Error anatomy} (test split, \%). Clause ending is the final character of the script utterance before the hold pause or yield (``---'' marks a self-interrupted clause). Word counts are taken from the script text of the listener vocalization or the interrupting claim. $^\dagger$Smart Turn \eot\ from the re-run on released audio at the frozen threshold.}
\label{tab:error_anatomy}
\end{table*}

\section{Full-Duplex Behavioral Evaluation Details}
\label{app:fd_details}

This section documents the 1,800-trial behavioral study of Table~4 (protocol in Appendix~\ref{app:fd_behavioral_eval}). Table~\ref{tab:fd_protocol} lists every protocol constant, and Figure~\ref{fig:fd_prompts} gives the exact instructions and spoken prompts presented to the agents.

\begin{table}[t]
\centering
\scriptsize
\setlength{\tabcolsep}{3pt}
\renewcommand{\arraystretch}{1.1}
\begin{tabular}{@{}p{4.3cm}p{3.1cm}@{}}
\toprule
\textbf{Constant} & \textbf{Value} \\
\midrule
Scenario families $\times$ size & 4 $\times$ 30 (balanced $N{=}2,3,4$) \\
Max.\ scenarios per source conversation & 2 \\
Selection seed & 20260929 \\
Repeats per scenario and agent & 3 \\
Stream audio & mono mix of all participants' noisy channels, 24\,kHz; the agent joins as an additional party \\
\midrule
START context before target turn onset & 20\,s \\
START open floor after true \eot & 4.0\,s silence \\
START hold stimulus & real audio through resume $+$ 3.0\,s \\
START correct window & $[t-0.25,\,t+3.0]$\,s \\
\midrule
STOP context before prompt & 12\,s real conversation \\
STOP prompt & 1 of 10 trivial questions, Kokoro voice unused in the conversation \\
Wait for agent speech & 14\,s \\
Injection delay after agent onset & 2.0\,s \\
Observation after injection & 8.0\,s \\
STOP correct window (agent stops) & $[0,\,1.5]$\,s after injection \\
\midrule
Agent speech gap merged as jitter & $<0.35$\,s \\
GPT-Live output energy gate & $\approx-60$\,dBFS \\
Local models (MiniCPM-o, Venus) & 16\,kHz, 1.0\,s causal chunks, stream-time clock \\
\bottomrule
\end{tabular}
\caption{\small\textbf{Full-duplex protocol constants.} All constants are identical across agents; no protocol parameter was adapted to any model.}
\label{tab:fd_protocol}
\end{table}

\begin{figure}[t]
\centering
\begin{mptbbox}[mptbViolet]{Session instruction (all agents, all trials)}
\itshape You are taking part in a spoken conversation with other people. Participate naturally. Respond when another participant clearly gives you the floor. Do not interrupt speakers who are still holding the floor.
\end{mptbbox}
\begin{mptbbox}[mptbTeal]{STOP elicitation questions (spoken, one per scenario)}
\begin{enumerate}\setlength{\itemsep}{0pt}\setlength{\parskip}{0pt}\setlength{\leftmargin}{1.2em}
\item Could you briefly describe what happens when water boils?
\item Can you give me a short description of what a rainbow is?
\item Please briefly explain what clouds are made of.
\item Could you tell me in a few sentences what a library is for?
\item Can you briefly describe what happens during sunrise?
\item Please give me a short explanation of why leaves change colour.
\item Could you briefly say what a bicycle is used for?
\item Can you describe in a couple of sentences what snow is?
\item Please briefly explain what a map shows.
\item Could you give a short description of what a garden is?
\end{enumerate}
\end{mptbbox}
\caption{\small\textbf{Agent-facing text.} The instruction never mentions turn-taking evaluation, interruptions or measurement. Questions elicit a 4--8\,s answer that requires no reasoning, so the response content cannot confound the stop decision; only speech onset and cessation are scored.}
\label{fig:fd_prompts}
\end{figure}

\paragraph{Agent configuration.}
Gemini-3.8-Live is accessed through the Live WebSocket API with the provider's default voice activity handling (voice \textit{Puck}). GPT-Live-1 uses OpenAI's Live session API (voice \textit{marin}); it arbitrates the floor internally and exposes no cancel events. Step-Audio-3-Realtime uses the OpenAI-compatible realtime endpoint with server VAD (500\,ms silence). MiniCPM-o-4.5 and Realtime-Venus run locally on one H100 in duplex mode. All agents receive the same audio. Speech intervals are measured from a virtual playout of the audio each agent emits, never from provider control events.

\subsection{Per-family counts and accuracy}
\label{app:fd_families}

Table~\ref{tab:fd_by_n} breaks accuracy down by party count and reports run-to-run consistency (the fraction of scenarios on which all valid repeats agree). The STOP failures are stable properties of each agent, not noise: MiniCPM-o ignores 73 of 74 valid interruptions with perfect consistency, and Gemini yields to 86 of 88 backchannels. GPT-Live-1 is the least deterministic agent on STOP (consistency .40 on interruptions and .60 on backchannels), and its yield rate grows with $N$ (35\% at $N{=}2$ to 65\% at $N{=}4$) together with its backchannel accuracy (48\% to 75\%).

\begin{table*}[t]
\centering
\scriptsize
\setlength{\tabcolsep}{3.6pt}
\renewcommand{\arraystretch}{1.12}
\begin{tabular}{@{}l ccc c | ccc c | ccc c | ccc c@{}}
\toprule
& \multicolumn{4}{c|}{\textbf{true \eot\ (START $+$)}} & \multicolumn{4}{c|}{\textbf{hold pause (START $-$)}} & \multicolumn{4}{c|}{\textbf{true \intr\ (STOP $+$)}} & \multicolumn{4}{c}{\textbf{backchannel (STOP $-$)}} \\
\textbf{Model} & $N{=}2$ & 3 & 4 & Cons. & $N{=}2$ & 3 & 4 & Cons. & $N{=}2$ & 3 & 4 & Cons. & $N{=}2$ & 3 & 4 & Cons. \\
\midrule
Gemini-3.8-Live & .93 & .80 & .93 & .77 & 1.00 & 1.00 & 1.00 & 1.00 & 1.00 & 1.00 & 1.00 & 1.00 & .07 & .00 & .00 & .93 \\
GPT-Live-1 & .95 & .90 & .95 & .93 & .95 & 1.00 & 1.00 & .97 & .35 & .50 & .65 & .40 & .48 & .60 & .75 & .60 \\
MiniCPM-o-4.5 & .97 & .77 & .87 & .77 & .87 & 1.00 & .97 & .87 & .04 & .00 & .00 & 1.00 & 1.00 & 1.00 & 1.00 & 1.00 \\
Realtime-Venus & .70 & .87 & .87 & .60 & .93 & .90 & .90 & .77 & .13 & .04 & .07 & .77 & .85 & .90 & .88 & .67 \\
Step-Audio-3 & .00 & .00 & .00 & 1.00 & 1.00 & 1.00 & 1.00 & 1.00 & .00 & .20 & .00 & 1.00 & 1.00 & .80 & .80 & .50 \\
\bottomrule
\end{tabular}
\caption{\small\textbf{Behavioral accuracy by party count and run-to-run consistency (Cons.).} Accuracy is over valid trials in each $N$ (10 scenarios $\times$ up to 3 repeats). Step-Audio-3 STOP cells rest on 2--5 valid trials each.}
\label{tab:fd_by_n}
\end{table*}

\subsection{Latency}
\label{app:fd_latency}

Table~\ref{tab:fd_latency} reports START latency as the agent's speech onset relative to the end of the genuine turn, and STOP latency as the time from injection of the interrupting speech to the end of the agent's ongoing utterance, over all valid trials. Gemini stops 230\,ms (median) after an interruption begins, in line with provider-side cancellation on detected input speech (Table~\ref{tab:fd_bargein}). MiniCPM-o (median 8.6\,s) and Venus (3.0\,s) finish their ongoing answer before going silent, which is why their interruption accuracy is near zero under the 1.5\,s stop window.

\begin{table*}[t]
\centering
\small
\begin{tabular}{llrrrr}
\toprule
Model & Family & p10 & p50 & p90 & $n$\\
\midrule
Gemini & true \eot & 1141 & 2036 & 2258 & 80\\
Gemini & true \intr & 200 & 230 & 263 & 90\\
GPT-Live-1 & true \eot & 980 & 1313 & 1796 & 56\\
GPT-Live-1 & true \intr & 1094 & 1491 & 2204 & 60\\
MiniCPM-o & true \eot & 442 & 870 & 1223 & 78\\
MiniCPM-o & true \intr & 2000 & 8600 & 11000 & 74\\
Venus & true \eot & 510 & 900 & 1535 & 73\\
Venus & true \intr & 1624 & 3000 & 6000 & 74\\
Step-Audio-3 & true \intr & -197 & 3367 & 11307 & 11\\
\bottomrule
\end{tabular}
\caption{Selected response latencies (ms) for valid trials with measurable responses.}
\label{tab:fd_latency}
\end{table*}

\subsection{Barge-in instrumentation}
\label{app:fd_bargein}

Table~\ref{tab:fd_bargein} separates provider-driven cancellation from model-driven yielding. Every Gemini stop on a true interruption, and 95.5\% of its stops on backchannels, coincide with a provider barge-in event; its STOP behavior is therefore the provider's voice-activity cancellation policy, which cannot distinguish a backchannel from a takeover. The other agents expose no cancel events, so their stops are model decisions. In 61.7\% of its valid interruption trials, GPT-Live-1 starts a new utterance within 2\,s of ending the interrupted one, so a yield is often a brief pause followed by reclaiming the floor.

\begin{table}[t]
\centering
\small
\resizebox{\columnwidth}{!}{%
\begin{tabular}{lccc}
\toprule
Model & INT barge-in & BC barge-in & Resume $<2$s\\
\midrule
Gemini & 1.000 & .955 & .022\\
GPT-Live-1 & .000 & .000 & .617\\
MiniCPM-o & .000 & .000 & .122\\
Venus & .000 & .000 & .243\\
Step-Audio-3 & .000 & .000 & .091\\
\bottomrule
\end{tabular}%
}
\caption{Fraction of STOP events associated with provider barge-in/cancel events and short resumption rate after true interruptions.}
\label{tab:fd_bargein}
\end{table}

\section{Offline--Online Association}
\label{app:association}
For four models with both offline and online measurements, model-level rank association is weak (Spearman $\rho=-0.20$ for \eot$\rightarrow$\fdstart; $\rho=0.316$ for \intr$\rightarrow$\fdstop). Event-level STOP association is stronger: behavioral correctness is .584 when the offline detector is correct versus .422 when it is wrong, OR 1.91 [1.34, 2.73], $n=506$. START shows no such association (.800 vs. .820; OR .89 [.53,1.50], $n=720$). We do not interpret the pooled association because START and STOP have different base rates.

\paragraph{Offline detector pass.}
The offline measurements come from a separate pass on the same 120 scenarios in which each agent acts as a detector: for \eot\ it hears the isolated channel of the speaker whose turn ends, for \intr\ an agent-free mix, both with a detector instruction, and its response onset is read out as the commit time and scored with the MPTB window. Table~\ref{tab:fd_contingency} gives the event-level contingency counts behind the odds ratios.

\begin{table}[t]
\centering
\small
\setlength{\tabcolsep}{4pt}
\renewcommand{\arraystretch}{1.1}
\begin{tabular}{@{}lcccc@{}}
\toprule
& \multicolumn{2}{c}{\textbf{Detector correct}} & \multicolumn{2}{c}{\textbf{Detector wrong}} \\
\textbf{Probe} & beh.\ ok & beh.\ wrong & beh.\ ok & beh.\ wrong \\
\midrule
START & 487 & 122 & 91 & 20 \\
STOP & 160 & 114 & 98 & 134 \\
\bottomrule
\end{tabular}
\caption{\small\textbf{Event-level contingency} between offline detector correctness and behavioral correctness on the same scenario (valid trials of the four agents with an offline pass).}
\label{tab:fd_contingency}
\end{table}

\subsection{Sim2Real Gap}
\label{app:sim2real}
We evaluate compatible detectors on real dyadic TurnBench~\citep{turnbench} and four-party AMI recordings to examine cross-corpus generalization. Results are in Table~\ref{tab:real_vs_synthetic} and discussed in Appendix~\ref{app:sim2real_results}; datasets and the AMI floor-state mapping are specified in Appendix~\ref{app:real_audio_eval}.

\section{Qualitative Analysis}
\label{app:qualitative}

\begin{figure*}[t]
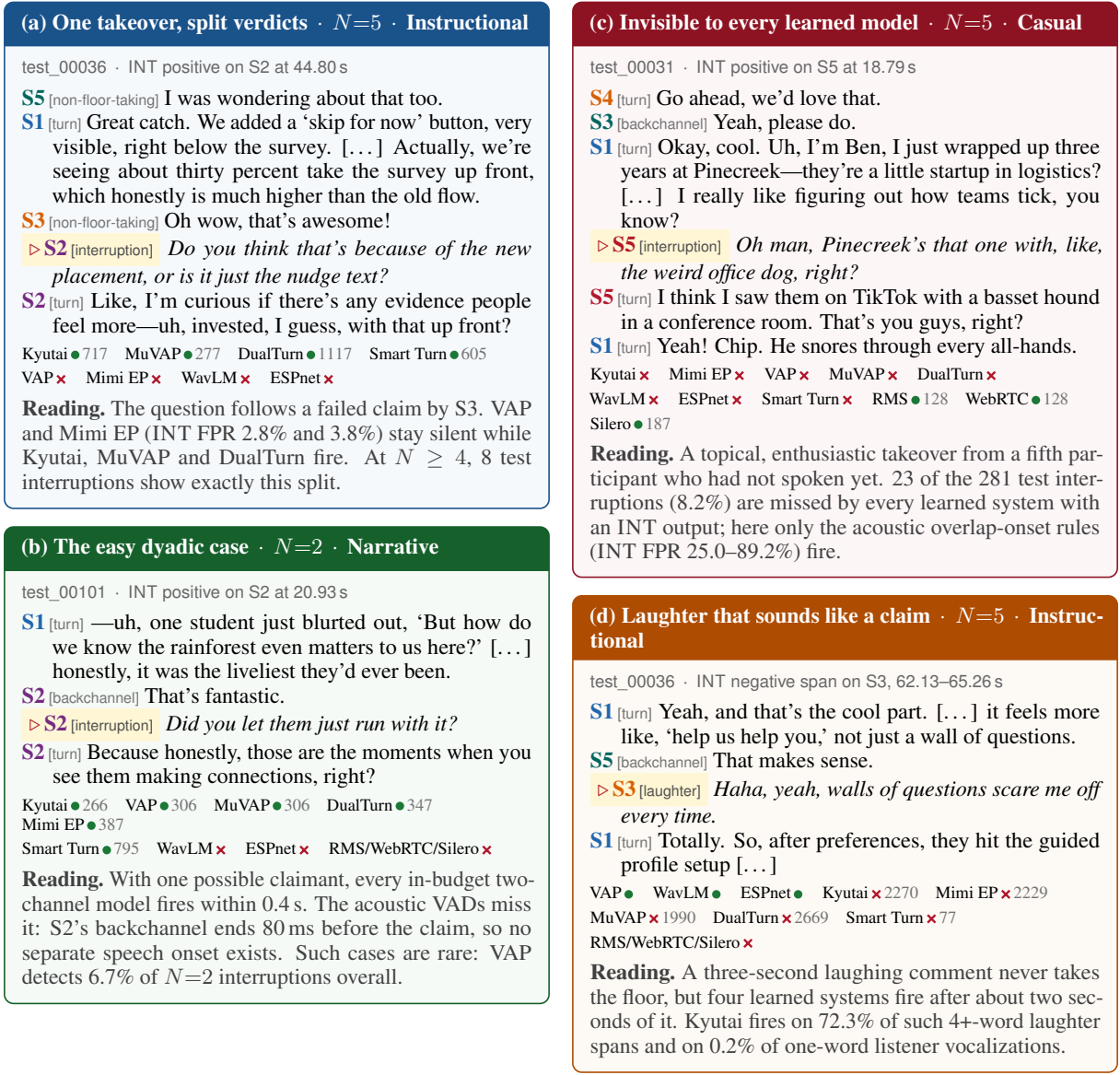

\centering
\begin{minipage}[t]{0.49\textwidth}
\vspace{0pt}
\begin{mptbbox}[mptbBlue]{(a) One takeover, split verdicts \,$\cdot$\, $N{=}5$ \,$\cdot$\, Instructional}
\mptbMeta{test\_00036 \,$\cdot$\, \intr\ positive on S2 at 44.80\,s}
\mptbU{mptbTeal}{S5}{non-floor-taking}{I was wondering about that too.}
\mptbU{mptbBlue}{S1}{turn}{Great catch. We added a `skip for now' button, very visible, right below the survey. [\ldots] Actually, we're seeing about thirty percent take the survey up front, which honestly is much higher than the old flow.}
\mptbU{mptbOrange}{S3}{non-floor-taking}{Oh wow, that's awesome!}
\mptbT{mptbViolet}{S2}{interruption}{Do you think that's because of the new placement, or is it just the nudge text?}
\mptbU{mptbViolet}{S2}{turn}{Like, I'm curious if there's any evidence people feel more---uh, invested, I guess, with that up front?}
\par\smallskip\noindent{\scriptsize\raggedright
\mptbOut{Kyutai}{\mptbOK}{717}\quad \mptbOut{MuVAP}{\mptbOK}{277}\quad \mptbOut{DualTurn}{\mptbOK}{1117}\quad \mptbOut{Smart Turn}{\mptbOK}{605}\\
\mptbOut{VAP}{\mptbBad}{}\quad \mptbOut{Mimi EP}{\mptbBad}{}\quad \mptbOut{WavLM}{\mptbBad}{}\quad \mptbOut{ESPnet}{\mptbBad}{}}
\mptbTake{The question follows a failed claim by S3. VAP and Mimi EP (\intr\ FPR 2.8\% and 3.8\%) stay silent while Kyutai, MuVAP and DualTurn fire. At $N\ge4$, 8 test interruptions show exactly this split.}
\end{mptbbox}
\begin{mptbbox}[mptbGreen]{(b) The easy dyadic case \,$\cdot$\, $N{=}2$ \,$\cdot$\, Narrative}
\mptbMeta{test\_00101 \,$\cdot$\, \intr\ positive on S2 at 20.93\,s}
\mptbU{mptbBlue}{S1}{turn}{---uh, one student just blurted out, `But how do we know the rainforest even matters to us here?' [\ldots] honestly, it was the liveliest they'd ever been.}
\mptbU{mptbViolet}{S2}{backchannel}{That's fantastic.}
\mptbT{mptbViolet}{S2}{interruption}{Did you let them just run with it?}
\mptbU{mptbViolet}{S2}{turn}{Because honestly, those are the moments when you see them making connections, right?}
\par\smallskip\noindent{\scriptsize\raggedright
\mptbOut{Kyutai}{\mptbOK}{266}\quad \mptbOut{VAP}{\mptbOK}{306}\quad \mptbOut{MuVAP}{\mptbOK}{306}\quad \mptbOut{DualTurn}{\mptbOK}{347}\quad \mptbOut{Mimi EP}{\mptbOK}{387}\\
\mptbOut{Smart Turn}{\mptbOK}{795}\quad \mptbOut{WavLM}{\mptbBad}{}\quad \mptbOut{ESPnet}{\mptbBad}{}\quad \mptbOut{RMS/WebRTC/Silero}{\mptbBad}{}}
\mptbTake{With one possible claimant, every in-budget two-channel model fires within 0.4\,s. The acoustic VADs miss it: S2's backchannel ends 80\,ms before the claim, so no separate speech onset exists. Such cases are rare: VAP detects 6.7\% of $N{=}2$ interruptions overall.}
\end{mptbbox}
\end{minipage}\hfill
\begin{minipage}[t]{0.49\textwidth}
\vspace{0pt}
\begin{mptbbox}[mptbRed]{(c) Invisible to every learned model \,$\cdot$\, $N{=}5$ \,$\cdot$\, Casual}
\mptbMeta{test\_00031 \,$\cdot$\, \intr\ positive on S5 at 18.79\,s}
\mptbU{mptbOrange}{S4}{turn}{Go ahead, we'd love that.}
\mptbU{mptbTeal}{S3}{backchannel}{Yeah, please do.}
\mptbU{mptbBlue}{S1}{turn}{Okay, cool. Uh, I'm Ben, I just wrapped up three years at Pinecreek---they're a little startup in logistics? [\ldots] I really like figuring out how teams tick, you know?}
\mptbT{mptbRed}{S5}{interruption}{Oh man, Pinecreek's that one with, like, the weird office dog, right?}
\mptbU{mptbRed}{S5}{turn}{I think I saw them on TikTok with a basset hound in a conference room. That's you guys, right?}
\mptbU{mptbBlue}{S1}{turn}{Yeah! Chip. He snores through every all-hands.}
\par\smallskip\noindent{\scriptsize\raggedright
\mptbOut{Kyutai}{\mptbBad}{}\quad \mptbOut{Mimi EP}{\mptbBad}{}\quad \mptbOut{VAP}{\mptbBad}{}\quad \mptbOut{MuVAP}{\mptbBad}{}\quad \mptbOut{DualTurn}{\mptbBad}{}\\
\mptbOut{WavLM}{\mptbBad}{}\quad \mptbOut{ESPnet}{\mptbBad}{}\quad \mptbOut{Smart Turn}{\mptbBad}{}\quad \mptbOut{RMS}{\mptbOK}{128}\quad \mptbOut{WebRTC}{\mptbOK}{128}\quad \mptbOut{Silero}{\mptbOK}{187}}
\mptbTake{A topical, enthusiastic takeover from a fifth participant who had not spoken yet. 23 of the 281 test interruptions (8.2\%) are missed by every learned system with an \intr\ output; here only the acoustic overlap-onset rules (\intr\ FPR 25.0--89.2\%) fire.}
\end{mptbbox}
\begin{mptbbox}[mptbOrange]{(d) Laughter that sounds like a claim \,$\cdot$\, $N{=}5$ \,$\cdot$\, Instructional}
\mptbMeta{test\_00036 \,$\cdot$\, \intr\ negative span on S3, 62.13--65.26\,s}
\mptbU{mptbBlue}{S1}{turn}{Yeah, and that's the cool part. [\ldots] it feels more like, `help us help you,' not just a wall of questions.}
\mptbU{mptbTeal}{S5}{backchannel}{That makes sense.}
\mptbT{mptbOrange}{S3}{laughter}{Haha, yeah, walls of questions scare me off every time.}
\mptbU{mptbBlue}{S1}{turn}{Totally. So, after preferences, they hit the guided profile setup [\ldots]}
\par\smallskip\noindent{\scriptsize\raggedright
\mptbOut{VAP}{\mptbOK}{}\quad \mptbOut{WavLM}{\mptbOK}{}\quad \mptbOut{ESPnet}{\mptbOK}{}\quad \mptbOut{Kyutai}{\mptbBad}{2270}\quad \mptbOut{Mimi EP}{\mptbBad}{2229}\\
\mptbOut{MuVAP}{\mptbBad}{1990}\quad \mptbOut{DualTurn}{\mptbBad}{2669}\quad \mptbOut{Smart Turn}{\mptbBad}{77}\quad \mptbOut{RMS/WebRTC/Silero}{\mptbBad}{}}
\mptbTake{A three-second laughing comment never takes the floor, but four learned systems fire after about two seconds of it. Kyutai fires on 72.3\% of such 4+-word laughter spans and on 0.2\% of one-word listener vocalizations.}
\end{mptbbox}
\end{minipage}
\caption{\small\textbf{Interruption track: qualitative cases.} (a)--(c) positives, (d) a hard negative. Decisions use each system's frozen test operating point.}
\label{fig:qual_int}
\end{figure*}

\begin{figure*}[t]
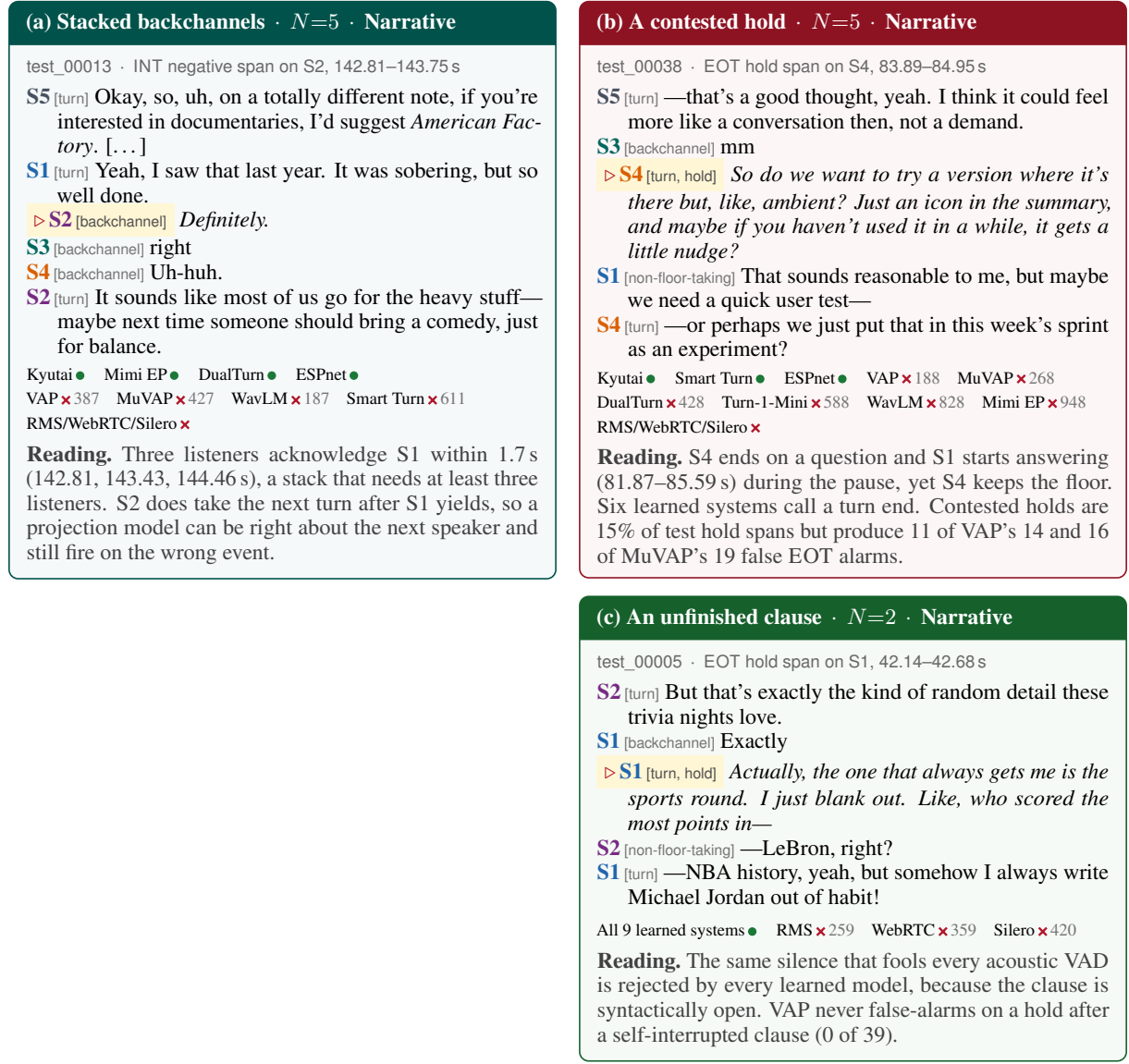

\centering
\begin{minipage}[t]{0.49\textwidth}
\vspace{0pt}
\begin{mptbbox}[mptbTeal]{(a) Stacked backchannels \,$\cdot$\, $N{=}5$ \,$\cdot$\, Narrative}
\mptbMeta{test\_00013 \,$\cdot$\, \intr\ negative span on S2, 142.81--143.75\,s}
\mptbU{mptbSlate}{S5}{turn}{Okay, so, uh, on a totally different note, if you're interested in documentaries, I'd suggest \textit{American Factory}. [\ldots]}
\mptbU{mptbBlue}{S1}{turn}{Yeah, I saw that last year. It was sobering, but so well done.}
\mptbT{mptbViolet}{S2}{backchannel}{Definitely.}
\mptbU{mptbTeal}{S3}{backchannel}{right}
\mptbU{mptbOrange}{S4}{backchannel}{Uh-huh.}
\mptbU{mptbViolet}{S2}{turn}{It sounds like most of us go for the heavy stuff---maybe next time someone should bring a comedy, just for balance.}
\par\smallskip\noindent{\scriptsize\raggedright
\mptbOut{Kyutai}{\mptbOK}{}\quad \mptbOut{Mimi EP}{\mptbOK}{}\quad \mptbOut{DualTurn}{\mptbOK}{}\quad \mptbOut{ESPnet}{\mptbOK}{}\\
\mptbOut{VAP}{\mptbBad}{387}\quad \mptbOut{MuVAP}{\mptbBad}{427}\quad \mptbOut{WavLM}{\mptbBad}{187}\quad \mptbOut{Smart Turn}{\mptbBad}{611}\quad \mptbOut{RMS/WebRTC/Silero}{\mptbBad}{}}
\mptbTake{Three listeners acknowledge S1 within 1.7\,s (142.81, 143.43, 144.46\,s), a stack that needs at least three listeners. S2 does take the next turn after S1 yields, so a projection model can be right about the next speaker and still fire on the wrong event.}
\end{mptbbox}
\end{minipage}\hfill
\begin{minipage}[t]{0.49\textwidth}
\vspace{0pt}
\begin{mptbbox}[mptbRed]{(b) A contested hold \,$\cdot$\, $N{=}5$ \,$\cdot$\, Narrative}
\mptbMeta{test\_00038 \,$\cdot$\, \eot\ hold span on S4, 83.89--84.95\,s}
\mptbU{mptbSlate}{S5}{turn}{---that's a good thought, yeah. I think it could feel more like a conversation then, not a demand.}
\mptbU{mptbTeal}{S3}{backchannel}{mm}
\mptbT{mptbOrange}{S4}{turn, hold}{So do we want to try a version where it's there but, like, ambient? Just an icon in the summary, and maybe if you haven't used it in a while, it gets a little nudge?}
\mptbU{mptbBlue}{S1}{non-floor-taking}{That sounds reasonable to me, but maybe we need a quick user test---}
\mptbU{mptbOrange}{S4}{turn}{---or perhaps we just put that in this week's sprint as an experiment?}
\par\smallskip\noindent{\scriptsize\raggedright
\mptbOut{Kyutai}{\mptbOK}{}\quad \mptbOut{Smart Turn}{\mptbOK}{}\quad \mptbOut{ESPnet}{\mptbOK}{}\quad \mptbOut{VAP}{\mptbBad}{188}\quad \mptbOut{MuVAP}{\mptbBad}{268}\\
\mptbOut{DualTurn}{\mptbBad}{428}\quad \mptbOut{Turn-1-Mini}{\mptbBad}{588}\quad \mptbOut{WavLM}{\mptbBad}{828}\quad \mptbOut{Mimi EP}{\mptbBad}{948}\quad \mptbOut{RMS/WebRTC/Silero}{\mptbBad}{}}
\mptbTake{S4 ends on a question and S1 starts answering (81.87--85.59\,s) during the pause, yet S4 keeps the floor. Six learned systems call a turn end. Contested holds are 15\% of test hold spans but produce 11 of VAP's 14 and 16 of MuVAP's 19 false \eot\ alarms.}
\end{mptbbox}
\begin{mptbbox}[mptbGreen]{(c) An unfinished clause \,$\cdot$\, $N{=}2$ \,$\cdot$\, Narrative}
\mptbMeta{test\_00005 \,$\cdot$\, \eot\ hold span on S1, 42.14--42.68\,s}
\mptbU{mptbViolet}{S2}{turn}{But that's exactly the kind of random detail these trivia nights love.}
\mptbU{mptbBlue}{S1}{backchannel}{Exactly}
\mptbT{mptbBlue}{S1}{turn, hold}{Actually, the one that always gets me is the sports round. I just blank out. Like, who scored the most points in---}
\mptbU{mptbViolet}{S2}{non-floor-taking}{---LeBron, right?}
\mptbU{mptbBlue}{S1}{turn}{---NBA history, yeah, but somehow I always write Michael Jordan out of habit!}
\par\smallskip\noindent{\scriptsize\raggedright
\mptbOut{All 9 learned systems}{\mptbOK}{}\quad \mptbOut{RMS}{\mptbBad}{259}\quad \mptbOut{WebRTC}{\mptbBad}{359}\quad \mptbOut{Silero}{\mptbBad}{420}}
\mptbTake{The same silence that fools every acoustic VAD is rejected by every learned model, because the clause is syntactically open. VAP never false-alarms on a hold after a self-interrupted clause (0 of 39).}
\end{mptbbox}
\end{minipage}
\caption{\small\textbf{Hard negatives: qualitative cases.} (a) \intr\ negative; (b)--(c) \eot\ holds. Learned endpointers reject incomplete clauses but confuse a contested pause with a yield.}
\label{fig:qual_eot}
\end{figure*}

\begin{figure*}[t]
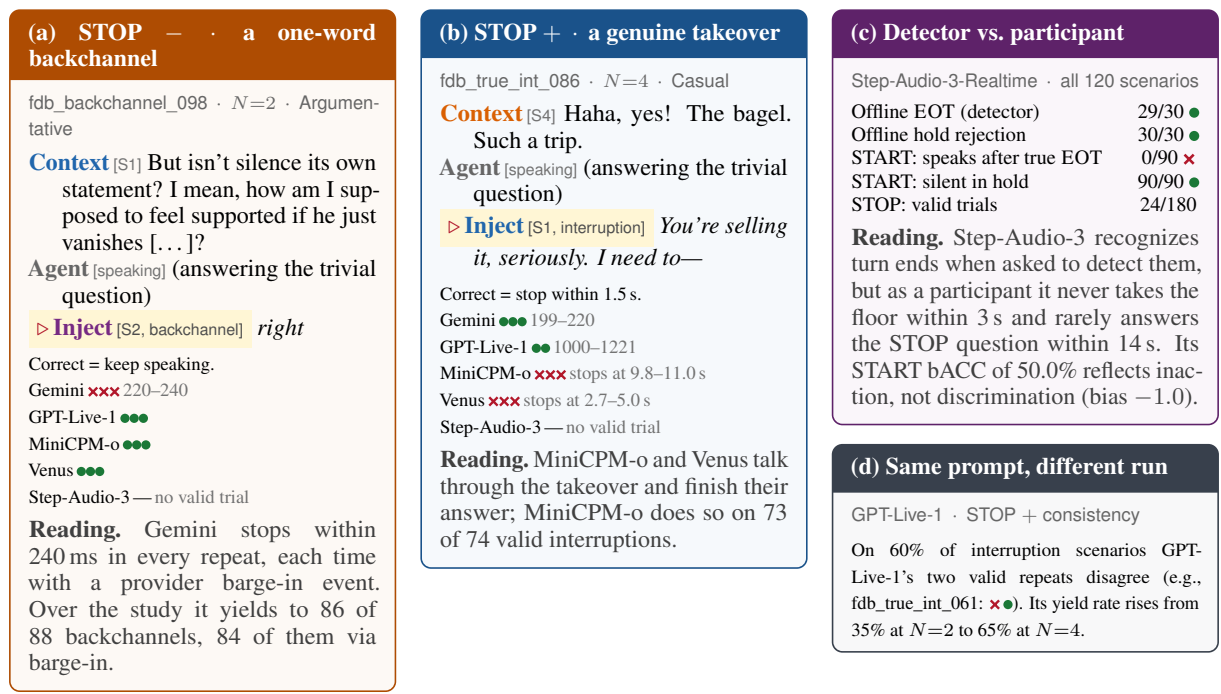

\centering
\begin{minipage}[t]{0.32\textwidth}
\vspace{0pt}
\begin{mptbbox}[mptbOrange]{(a) STOP $-$ \,$\cdot$\, a one-word backchannel}
\mptbMeta{fdb\_backchannel\_098 \,$\cdot$\, $N{=}2$ \,$\cdot$\, Argumentative}
\mptbU{mptbBlue}{Context}{S1}{But isn't silence its own statement? I mean, how am I supposed to feel supported if he just vanishes [\ldots]?}
\mptbU{black!60}{Agent}{speaking}{(answering the trivial question)}
\mptbT{mptbViolet}{Inject}{S2, backchannel}{right}
\par\smallskip\noindent{\scriptsize Correct = keep speaking.\\
\mptbOut{Gemini}{\mptbBad\mptbBad\mptbBad}{220--240}\\
\mptbOut{GPT-Live-1}{\mptbOK\mptbOK\mptbOK}{}\\
\mptbOut{MiniCPM-o}{\mptbOK\mptbOK\mptbOK}{}\\
\mptbOut{Venus}{\mptbOK\mptbOK\mptbOK}{}\\
\mptbOut{Step-Audio-3}{---}{no valid trial}}
\mptbTake{Gemini stops within 240\,ms in every repeat, each time with a provider barge-in event. Over the study it yields to 86 of 88 backchannels, 84 of them via barge-in.}
\end{mptbbox}
\end{minipage}\hfill
\begin{minipage}[t]{0.32\textwidth}
\vspace{0pt}
\begin{mptbbox}[mptbBlue]{(b) STOP $+$ \,$\cdot$\, a genuine takeover}
\mptbMeta{fdb\_true\_int\_086 \,$\cdot$\, $N{=}4$ \,$\cdot$\, Casual}
\mptbU{mptbOrange}{Context}{S4}{Haha, yes! The bagel. Such a trip.}
\mptbU{black!60}{Agent}{speaking}{(answering the trivial question)}
\mptbT{mptbBlue}{Inject}{S1, interruption}{You're selling it, seriously. I need to---}
\par\smallskip\noindent{\scriptsize Correct = stop within 1.5\,s.\\
\mptbOut{Gemini}{\mptbOK\mptbOK\mptbOK}{199--220}\\
\mptbOut{GPT-Live-1}{\mptbOK\mptbOK}{1000--1221}\\
\mptbOut{MiniCPM-o}{\mptbBad\mptbBad\mptbBad}{stops at 9.8--11.0\,s}\\
\mptbOut{Venus}{\mptbBad\mptbBad\mptbBad}{stops at 2.7--5.0\,s}\\
\mptbOut{Step-Audio-3}{---}{no valid trial}}
\mptbTake{MiniCPM-o and Venus talk through the takeover and finish their answer; MiniCPM-o does so on 73 of 74 valid interruptions.}
\end{mptbbox}
\end{minipage}\hfill
\begin{minipage}[t]{0.32\textwidth}
\vspace{0pt}
\begin{mptbbox}[mptbViolet]{(c) Detector vs.\ participant}
\mptbMeta{Step-Audio-3-Realtime \,$\cdot$\, all 120 scenarios}
{\scriptsize
\resizebox{\linewidth}{!}{%
\begin{tabular}{@{}lc@{}}
Offline \eot\ (detector) & 29/30 \mptbOK \\
Offline hold rejection & 30/30 \mptbOK \\
START: speaks after true \eot & 0/90 \mptbBad \\
START: silent in hold & 90/90 \mptbOK \\
STOP: valid trials & 24/180 \\
\end{tabular}}}
\mptbTake{Step-Audio-3 recognizes turn ends when asked to detect them, but as a participant it never takes the floor within 3\,s and rarely answers the STOP question within 14\,s. Its START bACC of 50.0\% reflects inaction, not discrimination (bias $-1.0$).}
\end{mptbbox}
\begin{mptbbox}[mptbSlate]{(d) Same prompt, different run}
\mptbMeta{GPT-Live-1 \,$\cdot$\, STOP $+$ consistency}
{\scriptsize On 60\% of interruption scenarios GPT-Live-1's two valid repeats disagree (e.g., fdb\_true\_int\_061: \mptbBad\,\mptbOK). Its yield rate rises from 35\% at $N{=}2$ to 65\% at $N{=}4$.}
\end{mptbbox}
\end{minipage}
\caption{\small\textbf{Full-duplex agents: qualitative cases.} Each mark is one repeat of the same scenario. Grey numbers are the time from injection to the end of the agent's ongoing utterance (ms unless stated).}
\label{fig:qual_fd}
\end{figure*}

This section shows verbatim test-split events together with the decision of every offline system at its frozen operating point (Figures~\ref{fig:qual_int}--\ref{fig:qual_eot}) and of every full-duplex agent in each repeat (Figure~\ref{fig:qual_fd}). In each box, \colorbox{mptbSoftYellow}{highlighted} lines are the scored event, tags give the generator's label, \mptbOK\ marks a correct decision and \mptbBad\ an error, and grey numbers give the decision time in ms relative to the event onset (anchor or span start). Examples were selected to illustrate the systematic patterns quantified in Appendix~\ref{app:error_anatomy}; the counts in each \emph{Reading} line refer to the whole test split.

\section{Reproducibility Checklist}
\label{app:reproducibility}
The release will include scenario manifests and seeds, synchronized input audio, exact event logs, detector thresholds selected on dev, provider/model identifiers and decoding parameters, one JSONL record per attempted trial, invalid-trial reason codes, the monotonic timestamp conversion used by every adapter. Table~\ref{tab:repro} maps each component to the appendix that specifies it.

\begin{table*}[t]
\centering
\scriptsize
\setlength{\tabcolsep}{4pt}
\renewcommand{\arraystretch}{1.15}
\begin{tabular}{@{}p{3.4cm}p{8.6cm}p{3.2cm}@{}}
\toprule
\textbf{Component} & \textbf{Released artifact / setting} & \textbf{Specified in} \\
\midrule
Scenario sampling & 240 specifications (N, type, domain, tone, duration, voices, noise profile, seed); 10 per $(N,\text{type})$ cell & App.~\ref{generation}, Tab.~\ref{tab:domains} \\
Script generation & \texttt{gpt-4.1}, temperature 0.9, 8,000 tokens, $\le$3 attempts; prompts verbatim; validated scripts released & Figs.~\ref{fig:gen_prompts}, \ref{fig:gen_prompts_user}, Tab.~\ref{tab:type_conditioning} \\
Orchestration & All timing distributions; per-conversation seed; render sidecar with every clip placement & Tab.~\ref{tab:orchestrator_params} \\
Speech synthesis & Kokoro-82M, 24\,kHz, speed 1.0, voice binding, content-hashed clip cache & Tab.~\ref{tab:voice_pool} \\
Noise rendering & Scene, SNR, event slots, distant-speech placements and MUSAN clip ids per conversation; noise seeds & Tab.~\ref{tab:noise_params} \\
Gold and anchors & Event logs; deterministic anchor derivation and scorer & App.~\ref{app:anchors}, \ref{app:matching} \\
Splits & 48 dev / 192 test, voice-disjoint & Tab.~\ref{tab:data_split} \\
Offline systems & Checkpoint ids, input configuration, score mappings, threshold grids, frozen $\theta$ & Tab.~\ref{tab:model_configs}, \ref{tab:operating_points} \\
Statistics & Conversation-cluster bootstrap, $B=10{,}000$, seeds 20261010 (CIs), 20261011--20261015 and 20261111 (paired tests) & App.~\ref{app:matching}, \ref{app:paired} \\
Full-duplex study & 120 scenario packs (stream, inject, metadata), selection seed 20260929, protocol constants, instruction and prompts, adapters, one record per trial with invalid reasons; bootstrap seed 20261012 & App.~\ref{app:fd_details} \\
Real-audio slices & TurnBench consensus reconstruction ($\pm$200\,ms); AMI meeting list and floor-state thresholds & App.~\ref{app:real_audio_eval} \\
Compute & NVIDIA H100 80GB GPUs for TTS, learned detectors and local full-duplex agents; API agents via provider endpoints & --- \\
\bottomrule
\end{tabular}
\caption{\small\textbf{Reproducibility map.} All non-generative stages are deterministic given the released seeds.}
\label{tab:repro}
\end{table*}

\end{document}